\documentclass[11pt]{article}
\usepackage{citesort,latexsym,graphics,color}

\newcounter{Diagrams}
\Alph{Diagrams}
\newcounter{cancellation}
\addtocounter{cancellation}{1}

\newtheorem{Diag}{}[Diagrams]
\newtheorem{DC}[Diag]{\{}
\newtheorem{DC-B}[Diag]{$|$}
\newtheorem{cancel}{Cancellation}

\newtheorem{Prop}{Proposition}
\newtheorem{D-ID}{Diagrammatic Identity}[section]

\newcommand{\ie}{{\it i.e.}\ }
\newcommand{\cf}{{\it cf.}\ }
\newcommand{\eg}{{\it e.g.}\ }
\newcommand{\etal}{{\it et al.}\ }
\newcommand{\aka}{{\it a.k.a.}\ }
\newcommand{\etc}{{\it etc.}\ }

\newcommand{\wrt}{with respect to}
\newcommand{\CC}{charge conjugation}
\newcommand{\CCI}{charge conjugation invariance}
\newcommand{\lhs}{left-hand side}
\newcommand{\rhs}{right-hand side}
\newcommand{\sic}{sic}
\newcommand{\BPZ}{\ensuremath{\mathsf{B}'(0)}}

\newcommand{\role}{r\^{o}le}

\newcommand{\be}{\begin{equation}}
\newcommand{\ee}{\end{equation}}
\newcommand{\bea}{\begin{eqnarray}}
\newcommand{\eea}{\end{eqnarray}}
\newcommand{\beas}{\begin{eqnarray*}}
\newcommand{\eeas}{\end{eqnarray*}}

\newcommand{\bear}{\begin{array}{l}}
\newcommand{\eear}{\end{array}}

\newcommand{\bcf}{\begin{center}\begin{figure}}
\newcommand{\ecf}{\end{figure}\end{center}}

\newcommand{\bct}{\begin{center}\begin{table}}
\newcommand{\ect}{\end{table}\end{center}}

\newcommand{\ds}{\displaystyle}

\def\eq#1{(\ref{#1})}
\def\eqs#1#2{(\ref{#1},\ref{#2})}
\def\sec#1{sec.~\ref{#1}}
\def\secs#1#2{secs.~\ref{#1} and~\ref{#2}}
\def\fig#1{fig.~\ref{#1}}
\def\Fig#1{Fig.~\ref{#1}}
\def\figs#1#2{figs.~\ref{#1} and~\ref{#2}}

\newcommand{\inte}{\! \int \!\!}

\newcommand{\Z}{ {\cal Z}}
\def\phi{ \varphi }
\def\A{{\cal A}}
\def\C{{\cal C}}
\def\D{{\cal D}}
\def\F{{\cal F}}
\def\Bb{\bar{B}}
\def\Db{\bar{D}}
\def\P{{\cal P}}

\def\dd{\dot{\Delta}}
\def\ker#1{\!\cdot\! #1 \!\cdot\!}
\def\hS{\hat{S}}
\def\e#1{\,{\rm e}^{\displaystyle #1}}
\def\one{\hbox{1\kern-.8mm l}}
\def\str{\mathrm{str}}

\newcommand{\tr}{{\mathrm{tr}}}

\newcommand{\Op}[1]{\mathcal{O}(p^{#1})}
\newcommand{\Oep}{\ensuremath{\mathcal{O}(\epsilon)}}
\newcommand{\Oepz}{\ensuremath{\mathcal{O}(\epsilon^0)}}
\newcommand{\OepPow}[1]{\ensuremath{\mathcal{O}\left(\epsilon^{#1}\right)}}

\newlength{\epminusone}
\newlength{\epzero}
\newcommand{\Int}[1]{\int \!\! d^D \! #1 \,}

\newcommand{\volume}[1]{d^D \! #1 \,}

\newcommand{\pder}[2]{\ensuremath{\frac{\partial #1}{\partial #2}}}
\newcommand{\fder}[2]{\ensuremath{\frac{\delta #1}{\delta #2}}}

\newcommand{\hf}{\frac{1}{2}}

\newlength{\PFheight}
\newcommand{\PF}[1]{
	 \stackrel{\rightarrow}{\vspace{\PFheight} #1}
}

\newcommand{\PB}[1]{
	\stackrel{\leftarrow}{\vspace{\PFheight} #1}
}

\newcommand{\DoubleCirc}{\ensuremath{\cdeps{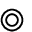}}}
\newcommand{\ReplaceAwC}{\ensuremath{\cdeps{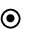}}}
\newcommand{\ReplaceAwSigma}{\ensuremath{\cdeps{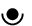}}}
\newcommand{\DiagSigma}{\ensuremath{\cdeps{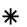}}}
\newcommand{\DummyKernel}{\ensuremath{\stackrel{\bullet}{\mbox{\rule{1cm}{.2mm}}}}}
\newcommand{\FullGR}{\ensuremath{\cdeps{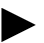}}}

\newcommand{\flow}{\Lambda \partial_\Lambda}
\newcommand{\flowConstAl}{\ensuremath{\Lambda \partial_\Lambda|_\alpha}}
\newcommand{\dec}[3][0]{\ensuremath{\left[ #2 \hspace{#1in} \right]^{#3}}}
\newcommand{\EP}[2]{\Delta^{#1}_{#2}}
\newcommand{\OLDs}{\mathcal{D}_1}

\newcommand{\GRk}{\rhd}
\newcommand{\GRkpr}{>}
\newcommand{\DiagDot}{\scriptstyle \bullet}

\newcommand{\CovKer}[2]{\{#1\}_{\!\!{}_{#2}}}
\newcommand{\TiedCovKer}[4]{#1 \, \CovKer{#2}{#3} #4}

\newlength{\LabLength}
\newlength{\ProcessRefLength}
\newlength{\ProcessLength}
\newlength{\CancelRefLength}
\newlength{\CancelRefLengthB}
\newlength{\CancelLength}
\newlength{\scriptboldcurlybracket}
\newlength{\strutheight}

\newcommand{\blankstrut}{\rule{0em}{\strutheight}}

\newcommand{\cdeps}[1]{\ensuremath{\begin{array}{c}\includegraphics{./eps/#1.eps} \end{array}}} 

\newcommand{\LD}[1]{
	\settowidth{\LabLength}{\scriptsize \textbf{\ref{#1}}}
	\addtolength{\LabLength}{0.8em}
	\begin{minipage}{\LabLength}
		\scriptsize
		\begin{Diag}\label{#1}\end{Diag}
	\end{minipage}
}

\newcommand{\PD}[2]{
	\settowidth{\LabLength}{\scriptsize\textbf{\ref{#1}}}
	\settowidth{\ProcessRefLength}{\scriptsize\ref{#2}}
	\addtolength{\LabLength}{\ProcessRefLength}
	\addtolength{\LabLength}{\ProcessLength}
	\addtolength{\LabLength}{0.8em}
	\begin{minipage}{\LabLength}
		\scriptsize
		\begin{Diag}\label{#1}$\rightarrow \ref{#2}$\end{Diag}
	\end{minipage}
}

\newcommand{\CD}[2]{
	\settowidth{\LabLength}{\scriptsize\textbf{\ref{#1}}}
	\settowidth{\CancelRefLength}{\scriptsize$\ref{#2}$}
	\addtolength{\LabLength}{\CancelRefLength}
	\addtolength{\LabLength}{\CancelLength}
	\begin{minipage}{\LabLength}
		\scriptsize
		\begin{DC}\label{#1} \ref{#2} \textbf{\}} \end{DC}
	\end{minipage}
}

\newcommand{\DCD}[3]{
\settowidth{\LabLength}{\scriptsize\textbf{\ref{#1}}}
\settowidth{\CancelRefLength}{\scriptsize$\ref{#2}$}
\settowidth{\CancelRefLengthB}{\scriptsize$\ref{#3}$}
\addtolength{\LabLength}{\CancelRefLength}
\addtolength{\LabLength}{\CancelRefLengthB}
\addtolength{\LabLength}{\CancelLength}
\addtolength{\LabLength}{0.5em}
\begin{minipage}{\LabLength}
\scriptsize
\begin{DC}\label{#1} \ref{#2}, \ref{#3} \textbf{\}}
\end{DC}
\end{minipage}
}

\newcommand{\CDB}[2]{
\settowidth{\LabLength}{\scriptsize\textbf{\ref{#2}}}
\settowidth{\CancelRefLength}{\scriptsize$\ref{#1}$}
\addtolength{\LabLength}{\CancelRefLength}
\addtolength{\LabLength}{\CancelLength}
\begin{minipage}{\LabLength}
\scriptsize
\begin{DC}\label{#1} \ref{#2} \textbf{$|$}
\end{DC}
\end{minipage}
}

\newcommand{\CDC}[2]{
\settowidth{\LabLength}{\scriptsize\textbf{\ref{#2}}}
\settowidth{\CancelRefLength}{\scriptsize$\ref{#1}$}
\addtolength{\LabLength}{\CancelRefLength}
\addtolength{\LabLength}{\CancelLength}
\begin{minipage}{\LabLength}
\scriptsize
\begin{DC-B}\label{#1} \ref{#2} \textbf{\}}
\end{DC-B}
\end{minipage}
}

\newcommand{\Discard}[1]{
	\settowidth{\LabLength}{\scriptsize\textbf{\ref{#1}}}
	\settowidth{\ProcessRefLength}{\scriptsize $\rightarrow 0$}
	\addtolength{\LabLength}{\ProcessRefLength}
	\addtolength{\LabLength}{\ProcessLength}
	\addtolength{\LabLength}{0.8em}
	\begin{minipage}{\LabLength}
	\scriptsize
		\begin{Diag}\label{#1}$\rightarrow 0$\end{Diag}
	\end{minipage}
}

\newcommand{\PO}[4][1]{
	\begin{array}{c}
		\PD{#3}{#4}
	\\[#1ex]
		#2
	\end{array}
}

\newcommand{\PDi}[4][1]{
	\PO[#1]{\ensuremath{\begin{array}{c}\input{pstex/#2.pstex_t} \end{array}}}{#3}{#4}
}

\newcommand{\LO}[3][1]{
	\begin{array}{c}
		\LD{#3}
	\\[#1ex]
		#2
	\end{array}
}

\newcommand{\LDi}[3][1]{\LO[#1]{\ensuremath{\begin{array}{c}\input{pstex/#2.pstex_t} \end{array}}}{#3}}

\newcommand{\CDi}[4][1]{
	\begin{array}{c}
		\CD{#3}{#4}
	\\[#1ex]
		\ensuremath{\begin{array}{c}\input{pstex/#2.pstex_t} \end{array}}
	\end{array}
}

\newcommand{\LHlabs}[3][0]{
\begin{array}{l}
\vspace{#1ex}
	#2\\
	#3		
\end{array}
}

\newcommand{\PDCD}[5][-0.5]{
	\LHlabs[#1]{\hspace{\scriptboldcurlybracket} \PD{#2}{#3}}{\CD{#4}{#5}}
}

\newcommand{\CDCD}[5][-0.5]{
	\LHlabs[#1]{\CD{#2}{#3}}{\CD{#4}{#5}}
}

\newcommand{\CDBl}[3][-0.5]
{
\LHlabs[#1]{\CD{#2}{#3}}{\blankstrut}
}

\newcommand{\Cancel}[2]{
\begin{cancel}
	Diagram~\ref{#1}  exactly cancels diagram~\ref{#2}.
	\label{cancel:#1}
\end{cancel}
}

\newcommand{\CancelCom}[3]{
\begin{cancel}
Diagram~\ref{#1}  exactly cancels diagram~\ref{#2}#3
\label{cancel:#1}
\end{cancel}
}

\newcommand{\CancelOpCom}[3]{
\begin{cancel}
Diagram~\ref{#1} cancels diagram~\ref{#2} at $\Op{2}$#3
\label{cancel:#1}
\end{cancel}
}

\newcommand{\jhep}[3]{\emph{JHEP} #1 (#2) #3}
\newcommand{\NuclPhys}[4]{\emph{Nucl.\ Phys.\ }\textbf{#1 #2} (#3) #4}
\newcommand{\PhysRev}[4]{\emph{Phys.\ Rev.\ }\textbf{#1 #2} (#3) #4}
\newcommand{\IntJModPhys}[4]{\emph{Int.\ J.\ Mod.\ Phys.\ }\textbf{#1 #2} (#3) #4}
\newcommand{\PhysRep}[4]{\emph{Phys.\ Rep.\ }\textbf{#1 #2} (#3) #4}

\newcommand{\TheorMathPhys}[3]{\emph{Theor.\ Math.\ Phys.\ }\textbf{#1} (#2) #3}

\newcommand{\AnnPhys}[3]{\emph{Ann.\ Phys.\ }\textbf{#1} (#2) #3}
\newcommand{\PhysLett}[4]{\emph{Phys.\ Lett.\ }\textbf{#1 #2} (#3) #4}

\newcommand{\Acta}[3]{\emph{Acta Phys.\ Slov.\ }\textbf{#1} (#2) #3}

\newcommand{\arxiv}[1]{[arXiv:#1]}
\newcommand{\hepth}[1]{hep-th/#1}
\newcommand{\hepph}[1]{hep-ph/#1}

\newcommand{\CommMathPhys}[3]{\emph{Comm.\ Math.\ Phys.\ }\textbf{#1}, #2 (#3)}
\newcommand{\Zeit}[3]{\emph{Zeitschr.\ f.\ Phys.\ }\textbf{#1} (#2) #3}
\newcommand{\Polon}[3]{\emph{Acta Phys.\ Polon\ }\textbf{#1} (#2) #3}

\newcommand{\http}[1]{http://#1}

\title{A Generalised Manifestly Gauge Invariant Exact Renormalisation Group for $SU(N)$ Yang-Mills}

\author{
Stefano Arnone,${}^\P$ Tim R. Morris${}^\ddag$ and Oliver J.
Rosten${}^\ddag$\\ 
${}^\P$Dipartimento di Fisica,
Universit\`a degli Studi di Roma ``La Sapienza''\\ P.le Aldo Moro,
2 - 00185 Roma, Italy\\ 
${}^\ddag$School of Physics and
Astronomy,  University of Southampton, \\ Highfield, Southampton
SO17 1BJ, U.K.\\ 
E-mails: {\tt
stefano.arnone@roma1.infn.it, T.R.Morris@soton.ac.uk,}\\ {\tt
ojr@phys.soton.ac.uk} }

\date{}

\begin{document}

\maketitle

\begin{abstract}
We take the manifestly gauge invariant exact renormalisation group
previously used to compute the one-loop $\beta$ function in
$SU(N)$ Yang-Mills without gauge fixing, and generalise it so that
it can be renormalised straightforwardly at any loop order. The
diagrammatic computational method is developed to cope with
general group theory structures, and new methods are introduced to
increase its power, so that much more can be done simply by
manipulating diagrams. The new methods allow the standard two-loop
$\beta$ function coefficient for $SU(N)$ Yang-Mills to be
computed, for the first time without fixing the gauge or
specifying the details of the regularisation scheme.
\end{abstract}

\vspace{-80ex}
\hfill SHEP 05-21

\newpage
\tableofcontents

\section{Introduction}
\label{Introduction}

In 1929, in pursuit of a manifestly relativistic Quantum
Electrodynamics, Pauli and Heisenberg discovered the famous
obstruction to its canonical quantisation, which in modern terms
is the lack of an inverse for the gauge field two-point 
vertex~\cite{Pauli}.\footnote{Of course they followed Hamiltonian
quantisation, where the problem manifests itself in the vanishing
of $\pi_0$, the momentum conjugate to the time component of the
gauge field.} Their solution in the same paper, effectively gauge
fixing, is still followed to this day. Much later, Feynman's
unitarity argument for Faddeev-Popov ghosts in non-Abelian gauge
theory~\cite{Feynman}, and the elegance and power of the resulting
BRST symmetries~\cite{BRST}, strengthened the case for gauge
fixing to such an extent that it is now often taken for granted
that it is a necessary first step to make sense of a quantum gauge
theory or, more extremely, that the original manifestly gauge
invariant formulation is nothing but a sort of slight of hand,
only the gauge fixed version having any real claim to existence
(this despite the fact that lattice gauge theory simulations are
routinely made without gauge fixing).

Nevertheless, in a series of 
works~\cite{u1,ym,ymi,ymii,Arnone:2000bv,sunn,Morris:2000jj,Arnone:2001iy,one,Arnone:2002ai,Arnone:2002qi,Antonio'sThesis,aprop,two,quarks,qed,Thesis},
we have been developing techniques that allow computations
directly in the continuum, in particular perturbative computations
in $SU(N)$ Yang-Mills, to proceed without any gauge fixing.
How
can we avoid the well-known obstructions above? We do not compute
contributions to the $S$ matrix (where at the perturbative level,
Feynman's arguments would necessarily apply~\cite{Feynman}) but
instead compute local objects: the vertices of a gauge invariant
Wilsonian effective action.\footnote{Correlators of gauge
invariant operators can be computed by introducing appropriate
sources~\cite{ymii}. To consider on-shell gluons, one can gauge
fix after the computation~\cite{aprop}.} The construction of a
real gauge invariant cutoff $\Lambda$, using spontaneously broken
$SU(N|N)$ gauge theory~\cite{sunn}, allows this for the first time
to be properly defined~\cite{ymii}. To compute the effective
action without fixing the gauge, we use the fact that there are an
infinity of possible exact renormalisation groups\footnote{
For alternative approaches---which gauge fix at some stage---see~\cite{Bonini:1993kt,Bonini:1993sj,Bonini:1994kp,Bonini:2001,Becchi:1993,Pernici:1998,Simionato:2000,Simionato:2000-B,Panza:2000,Reuter:1994sg,Reuter:1993kw,Freire:2000bq,Ellwanger:1994iz,D'Attanasio:1996jd,Litim:1998qi,Litim:1998nf,Pawlowski:2002eb,Litim:2002ce,Veneziano,Pawlowski-VDW}.
}
 (ERGs) that
specify its flow as modes are integrated out~\cite{jose} (the
continuum equivalent to the infinite number of ways of blocking on
a lattice~\cite{Morris:2000jj}---we expand on this 
in~\sec{Review}) and that out of these there are infinite number that
\emph{manifestly} preserve the gauge invariance, their weak
coupling expansion at no stage requiring the introduction of an
inverse for the gauge field two-point vertex~\cite{ym,ymi}.

Although it took some time to clarify and develop the ideas in the
initial works~\cite{u1,ym} to a point where we had a coherent
`calculus'---with which the one-loop $\beta$ function
coefficient, $\beta_1$, was computed without gauge fixing---we 
knew that significant additional development would be required to apply
these ideas further~\cite{aprop}. 
Indeed, although the analysis of~\cite{aprop} generalises the
original $N=\infty$ calculation to one that holds at finite $N$,
the Wilsonian effective action is restricted
to single supertrace terms only.
Removing this restriction
is one of the key elements in generalising
the formalism such that it is suitable for further computation.
The period between~\cite{aprop} and the present publication is in
part a measure of the scale of the difficulties we still had to
overcome.

The incorporation of multiple supertrace terms is necessary
for the complete renormalisation of the gauge
invariant Wilsonian effective action at one-loop and beyond,
as we discuss in~\sec{Flow}. Importantly, this
must be done in such a way as to respect the central term in the
$SU(N|N)$ algebra, implemented via the so-called no-$\A^0$
symmetry~\cite{aprop}. This then allows us to properly distinguish
the coupling $g(\Lambda)$ associated with the original
$SU(N)$ Yang-Mills from the coupling $g_2(\Lambda)$ associated
with the unphysical copy, which arises due to the $SU(N|N)$
regulating structure~\cite{Arnone:2001iy}.

Since the manifest preservation of gauge invariance ensures that the 
gauge fields have no
wavefunction renormalisation~\cite{ym},
we can ensure that $g$ and $g_2$ are the only quantities which
run.\footnote{For technical reasons, 
a superscalar field is given
zero mass dimension~\cite{aprop}, and thus is associated by the
usual dimensional reasoning with an infinite number of
dimensionless couplings. These couplings do not
require renormalisation, as we will see explicitly in~\cite{mgierg2}.}
We can then expect to recover the standard
two-loop $\beta$ function coefficient, $\beta_2$, by taking the
limit $g_2/g \rightarrow 0$, at the end of the calculation. 
As we will see in
ref.~\cite{mgierg2}, subject only to very general conditions, this
expectation is confirmed.
For future convenience, we define
\be
	\alpha := \frac{g^2_2}{g^2}.
\label{eq:alpha-defn}
\ee

However, introducing multiple supertrace terms rapidly increases
the number of Wilson-loop-like diagrams~\cite{ym,ymi,aprop},
even at one-loop. (Such diagrams simply correspond to drawing
explicitly the [super]colour flow, in common with other
diagrammatics inspired by `t Hooft's large $N$ diagrams~\cite{thooft}.)

The first step in bringing this complication under control, as we
describe in~\sec{sec:NewDiags}, is to replace these diagrams by
`group theory blind' diagrams. Just as in Feynman diagrams, which
they now very closely resemble, the result of combining the
structure constants is then implicit at the diagrammatic level.
This step is also the group theory analogue of leaving the
momentum dependence in the higher-point vertices of the flow
equation implicit, which underlies the diagrammatic computational
methods developed in refs.~\cite{one,aprop,two}. The result is a
streamlined computation, allowing many diagrams of different
group-theory structure to be processed in parallel.

Recall that, central to the diagrammatic methods of 
ref.~\cite{aprop}, is the introduction of an `effective propagator'.
Like ordinary propagators these are inverses of the classical
two-point vertex. However, since no gauge fixing has been done they
are inverses only in the transverse space. Equivalently we can say
that they are inverses up to remainder terms. These were called
`gauge remainder' terms because they are forced by gauge
invariance to be there; moreover, if a remainder
strikes a vertex, then they can be processed via
gauge invariance identities, which we recall in \sec{Review}.
In fact, the
full spontaneously broken $SU(N|N)$ acts in this way~\cite{aprop}.

In ref.~\cite{aprop} we used integration by parts of
$\Lambda\partial/\partial\Lambda$ (the generator of renormalisation group
flow) and these effective propagator relations to iteratively
simplify the expressions, purely by manipulating diagrams. We were
left with terms which were
algebraically completely determined, the gauge remainder terms
described above and total $\Lambda$-derivative terms (which,
in the formalism of this paper, become performed at constant $\alpha$).

Whilst all these terms can be further manipulated algebraically to
arrive at $\beta_1$, at two-loops this is no
longer possible, at least not without algebraic computing
facilities and even then this would be a major task. Instead, one
of us was inspired to generalise
these ideas~\cite{Thesis,oliver1} so that all contributions to $\beta_1$
can be reduced to diagrammatic $\Lambda$-derivative terms, as we will see
in \sec{sec:beta1}. It is from these terms that the numerical
value of $\beta_1$ can be straightforwardly extracted in a manifestly universal way.

The strategy at two-loops is essentially to repeat the same
diagrammatic steps. There are, however, a number of complications.
The first is that the procedure generates an almost prohibitively
large number of diagrams (we return to this point, and its
resolution~\cite{Thesis,oliver2}, in the conclusion).
Secondly, certain diagrammatic structures,
which look to be different, turn out to be algebraically the same.
Diagrams in which these structures are 
embedded as a sub-diagram cannot be manipulated
and so should be collected together. At two loops (and beyond~\cite{Thesis,oliver2})
all such terms cancel amongst themselves. The diagrammatic
identities required at two loops, for these cancellations,
are given in appendix~\ref{app:D-IDs}.
Thirdly, there is a subtlety associated with the
 Taylor expansions of a small number of
diagrams, necessary for the $\beta_2$ diagrammatics.
At the one-loop level, the Taylor expansions can be
performed blindly; at the two-loop level, however,
this process is not always trivial as it can generate IR
divergences in individual terms. The key is to consider
sets of terms together, whence these divergences
cancel out. There is a particularly elegant way of
thus  organising the calculation---using 
`subtraction techniques'~\cite{Thesis}---the 
explanation of which we defer until~\cite{mgierg2}.

Despite these complications, the two-loop and
one-loop diagrammatics are very similar. Although a very
large number of diagrams are generated at two-loops,
the vast majority cancel out to leave behind a manageable set of
$\Lambda$-derivative terms\footnote{And a set of terms that vanish
in the $\alpha \to 0$ limit.}.
The extraction of the numerical coefficient from
the  $\Lambda$-derivative terms at two-loops 
is subtle, heavily involving the use of the 
subtraction techniques.

In this paper, we focus on the construction of
the formalism and furnish the set of diagrammatic techniques
which allow, in principle, the reduction of $\beta_2$ to $\Lambda$-derivative
terms.\footnote{Though we note that this procedure is
made easier by the subtraction techniques.} A subset of these
 techniques are illustrated in the context of the
$\beta_1$ diagrammatics; a calculation which, at any rate, is necessary
for the two-loop calculation. (For further illustration
of the diagrammatic techniques, the reader is referred to~\cite{Thesis,oliver1}.)
In the partner paper~\cite{mgierg2}, we give the
expression for $\beta_2$ in terms of $\Lambda$-derivatives,
describe the subtraction techniques and extract the
universal coefficient.

The paper is organised as follows. In the next section we explain
why we need new terms in the flow equation to allow
straightforward renormalisation at one-loop and beyond. Having
introduced these new terms we first appropriately 
adapt the `old-style' diagrammatics of~\cite{aprop}
in~\sec{Diagrammatics}, before refining the diagrammatics in
\sec{sec:NewDiags}. The weak coupling flow equations are
phrased in this new notation, and some of their
properties are discussed in \sec{sec:WeakCoupling}.
The new diagrammatic techniques
are described in \sec{sec:FurtherDiagrammatics}.
In \sec{sec:beta1}, we use these techniques to redo the
diagrammatics for $\beta_1$, setting the stage for
the numerical evaluation of $\beta_2$, in~\cite{mgierg2}.
We conclude in \sec{sec:Conclusions}.

\section{Gauge invariance and the ERG}
\label{Flow}

We start with a quick review of the basic elements and then
explain why it is desirable to further modify the flow equation
for computations at arbitrary loop order. In all that follows,
we work in $D$ dimensional Euclidean space.

\subsection{Review}
\label{Review}


Our starting point is the 
recognition 
that there are an infinity of unrelated ERGs corresponding to the
infinite number of ways of blocking on the lattice~\cite{Morris:2000jj,jose}. 
It may help to make this observation
more concrete. Thus consider a general Kadanoff blocking for the
single, real scalar field $\phi$, which in continuum notation takes the
form $\phi(x) = b_x[\phi_0]$, $\phi$ being the blocked field and
$b_x$ the blocking function (a functional of the microscopic field
$\phi_0$). The blocking function could be linear $b_x[\phi_0] =
\int_y K(x-y)\phi_0(y)$, for some kernel $K(z)$ which is steeply
decaying once $z\Lambda>1$, or even something
non-linear.\footnote{The particular blocking that yields the
Legendre flow equation, \aka effective average action approach, is
given in ref.~\cite{wet}. This is directly related to Polchinski's
version through a Legendre transformation~\cite{erg}.} In the
standard way we define
\be
\label{emS}
\e{-S[\phi]} = \int\D\phi_0\ \delta\Big[ \phi - b[\phi_0]\Big]
\e{-S_{bare}[\phi_0]},
\ee
so that equality of microscopic and blocked partition functions
follows:
\[
\Z =  \inte{\D\phi}\e{-S[\phi]} = \int{\D\phi_0}
\e{-S_{bare}[\phi_0]}.
\]
We get an ERG by computing the flow of~\eq{emS}. Defining
\[
\Psi_x[\phi] =  - \e{S[\phi]}\, \int\D\phi_0\ \delta\Big[ \phi -
b[\phi_0]\Big] \Lambda{\partial b_x\over\partial\Lambda}\,
\e{-S_{bare}[\phi_0]},
\]
we see that
\be
\label{blockflow}
\Lambda{\partial\over\partial\Lambda} \e{-S[\phi]} =  \int_x
{\delta\over\delta\phi(x)}\left(\Psi_x \e{-S[\phi]}\right),
\ee
which is the starting point in refs.~\cite{ym,ymi,jose}, arrived
at by different arguments.

For the single scalar field, $\phi$, we can use~\cite{two}
\[
	\Psi_x = \hf \int_y \dd_{xy} \fder{\Sigma_1}{\phi(y)},
\]
where $\dd$  is an
ERG kernel. This choice yields:
\bea
	\dot{S} \equiv - \Lambda\partial_\Lambda S 
	& =	& a_0[S,\Sigma_1] - a_1[\Sigma_1]
\nonumber
\\[1ex]
	& =	& {1\over 2} {\delta S\over\delta\varphi} \ker{\dd} {\delta \Sigma_1 \over\delta\varphi} 
	- {1\over 2}{\delta \over\delta\varphi} \ker{\dd} {\delta \Sigma_1 \over\delta\varphi},
\label{poflo}
\eea
which will be referred to as the scalar flow equation.
$S$ is the Wilsonian effective action and
$\Sigma_1 = S - 2\hS$, where $\hat{S}$ is
the `seed action'~\cite{one,aprop,two}. 
We define
\[
f \ker{W} g = 
\int_{x,y}
f(x)\, W_{xy}\,g(y) = 
\int_x f(x) W(-\partial^2/\Lambda^2)\, g(x)
\]
which holds for any momentum space kernel $W(p^2/\Lambda^2)$ and
functions of spacetime $f$, $g$, using
\[
W_{x y} = W(-\partial^2/\Lambda^2)\,\delta(x-y) = \inte {d^D p
\over (2\pi)^D} \, W(p^2/\Lambda^2) \, {\rm e}^{i p \cdot (x-y)}.
\]

The seed
action is a generally non-universal input functional
which controls the flow and which satisfies the
same symmetries as $S$. 
$\hS$ must possess a kinetic term; in the case that $\hat{S}$
comprises just a kinetic term, eqn.~\eq{poflo}
reduces to Polchinski's version~\cite{Pol} of
Wilson's ERG~\cite{Wil}, up to a discarded
vacuum energy term. The seed action, like all other
ingredients of the flow equation, must be infinitely 
differentiable in momenta. This property, referred to
as quasilocality~\cite{TRM-MassiveScalar} guarantees that each RG step
$\Lambda \rightarrow \Lambda - \delta \Lambda$ does not
generate IR singularities~\cite{Wil}.

For a particular choice of $\hS$, which we will
discuss shortly, the integrated ERG kernel, $\Delta$, is 
the inverse of the classical two-point vertex. In recognition
of this, we henceforth refer to $\Delta$ as an 
`effective propagator'~\cite{two,aprop}. We use the word effective
advisably since its equivalence to a propagator
even in scalar field theory is down to choice. Moreover, when
we generalise to gauge theory, where we will not fix the
gauge, we cannot even define a propagator
in the usual sense. However, even in this case, $\Delta$
plays a diagrammatic \role\ somewhat analogous to a real propagator, and we
recognise this similarity.

Returning to~\eq{poflo}, we refer to $a_0$ as the classical term
of the flow equation, since this generates tree level diagrams,
and $a_1$ as the quantum term, since this generates loop corrections.
With this in mind, let us analyse the flow of the classical Wilsonian
effective action two-point vertex, $S_0^{\ \phi \phi}$. We can
consistently choose $\hS_0^{\ \phi \phi} = S_0^{\ \phi \phi}$
and, doing so, we have:
\be
	\dot{S}^{\ \phi \phi}_0 = -S_0^{\ \phi \phi} \dd \, S_0^{\ \phi \phi}.
\label{eq:Scalar-TLTP-flow}
\ee
Integrating up and setting the integration constant to zero
yields the effective propagator relation:
\be
	S_0^{\ \phi \phi} \Delta = 1.
\label{eq:Scalar-EP}
\ee

It is easy to generalise this analysis 
to pure $U(1)$ gauge theory~\cite{u1}: 
we simply replace $\phi$ by $A_\mu$ in~\eq{poflo} to give
\be
\label{u1flo} \dot{S}  = {1\over 2}
 {\delta S\over\delta A_\mu} \ker{\dd}
{\delta \Sigma_1 \over\delta A_\mu} - {1\over 2}{\delta
\over\delta A_\mu} \ker{\dd} {\delta \Sigma_1 \over\delta A_\mu}.
\ee
Because the gauge field only transforms by a shift under a $U(1)$
transformation $A_\mu\mapsto A_\mu +
\partial_\mu\Omega$, the functional derivatives $\delta/\delta
A_\mu$ are gauge \emph{invariant}. Thus~\eq{u1flo} 
is also gauge invariant. Once again, we choose the classical seed
action two-point vertex equal to its Wilsonian effective action
counterpart. Eqn.~\eq{eq:Scalar-TLTP-flow}
becomes
\be
\label{twoflo}
\dot{S}_{0\, \mu\nu} = -\, S_{0\, \mu\alpha} \,\dd\, S_{0\,\alpha\nu}.
\ee
Gauge invariance forces $ S_{0\,\alpha\nu}$ to be transverse. 
Utilising this and integrating up, we now find that the integration
constant is constrained to be zero, if we demand that $\Delta$ is
well behaved as $p \rightarrow \infty$. Eqn.~\eq{eq:Scalar-EP}
becomes
\be
\label{effpropid}
S_{0\, \mu\nu}\,\Delta = \delta_{\mu\nu} - {p_\mu p_\nu\over p^2}.
\ee
Thus, the effective propagator is the inverse of the classical
two-point vertex only in the transverse space. The object
$p_\mu p_\nu / p^2$ is our first example of a gauge remainder.

Of course, pure $U(1)$ gauge theory is not very interesting, but
nevertheless the flow equation we have written down is 
\emph{manifestly gauge invariant}. No gauge fixing is necessary to
define it and no gauge fixing is needed to calculate with it,
perturbatively or nonperturbatively~\cite{u1}.

The next step is to generalise our analysis to gauge theories with
interesting interactions and clearly the most important, and
challenging, direction is to generalise this to non-Abelian gauge
theories. (However, for application to QED,  see ref.~\cite{qed}.)
The first, obvious, difficulty is that if $A_\mu$ is a non-Abelian
gauge field, eqn.~\eq{u1flo} is no longer gauge invariant: the
functional derivatives $\delta/\delta A_\mu$ now transform
homogeneously in the adjoint representation of the gauge group. It is
easy to overcome this problem: we simply choose
\be
\label{Psi}
\Psi = {1\over2} \{\dd\}{\delta\Sigma_{g}\over\delta A_\mu},
\ee
where $\{\dd\}$ is some covariantisation of the kernel. For
example, we could replace $\dd(-\partial^2/\Lambda^2)$ by
$\dd(-D^2/\Lambda^2)$, where $D_\mu =
\partial -i A_\mu$ is the covariant derivative, but there are
infinitely many other possible covariantisations~\cite{ym,ymi}. 
We have scaled $g$ out of its normal place in the
covariant derivative, which is why in~\eq{Psi} we have~\cite{ym,ymi}
\be
\label{Sigma}
\Sigma_g := g^2 S - 2\hS.
\ee
Replacing $\phi$ by $A_\mu$ and
substituting $\Psi$ in~\eq{blockflow}, we get the manifestly gauge
invariant flow equation
\be
\label{sunflo}
\dot{S}  =
 {1\over2}{\delta S\over\delta A_\mu} \{\dd\}
{\delta \Sigma_g \over\delta A_\mu} - {1\over2}{\delta \over\delta
A_\mu} \{\dd\} {\delta \Sigma_g \over\delta A_\mu}.
\ee

Although this equation allows computations in non-Abelian
Yang-Mills without gauge fixing, we run into a problem at one-loop
because covariant higher-derivatives provide insufficient
regularisation~\cite{ymi,cov}. The essentially unique 
solution~\cite{ym} to this problem is to embed the theory inside a
spontaneously broken supergauge theory~\cite{Arnone:2001iy}.

Hence we regularise $SU(N)$ Yang-Mills theory by instead working
with $SU(N|N)$ Yang-Mills. The gauge field is valued in the Lie
superalgebra and thus takes the form of a Hermitian supertraceless
supermatrix:
\be
\label{defA}
\A_\mu =  \left( \!\! \begin{array}{cc}
                   A^{1}_{\mu} & B_{\mu} \\
                   \bar{B}_{\mu} & A^{2}_{\mu}
                   \end{array} \!\!
            \right) + {\A}^{0}_{\mu} \one.
\ee
Here, $A^1_\mu(x)\equiv A^1_{a\mu}\tau^a_1$ is the
physical $SU(N)$ gauge field, $\tau^a_1$ being the $SU(N)$
generators orthonormalised to
$\tr(\tau^a_1\tau^b_1)=\delta^{ab}/2$, while $A^2_\mu(x)\equiv
A^2_{a\mu}\tau^a_2$ is an unphysical $SU(N)$ gauge field.
The $B$ fields are fermionic gauge fields which will gain a mass
of order $\Lambda$ from the spontaneous breaking; they play the
\role\ of gauge invariant Pauli-Villars (PV) fields, furnishing the
necessary extra regularisation to supplement the covariant higher
derivatives. 

To unambiguously define contributions which are finite
only by virtue of the PV regularisation, a preregulator must
be used in $D=4$~\cite{Arnone:2001iy}. This amounts to a prescription for 
discarding otherwise non-vanishing surface terms which
can be generated by shifting loop momenta. Hence, we work
in $D=4-2\epsilon$, so that such contributions are automatically
discarded. There are, however, very strong indications~\cite{Thesis,oliver2}
that an entirely diagrammatic prescription can instead be adopted, which
one might hope would be applicable to phenomena for which
one must work strictly in $D=4$.

$\A^0$ is the gauge field for the centre of the
$SU(N|N)$ Lie superalgebra. Equivalently, one can write
\be
\label{expandA}
\A_\mu = {\A}^{0}_{\mu} \one + \A^A_\mu T_A,
\ee
where the $T_A$ are a complete set of traceless and supertraceless
generators normalised as in ref.~\cite{Arnone:2001iy}. 

The theory is
subject to the local invariance:
\be
\label{Agauged}
\delta\A_\mu = [\nabla_\mu,\Omega(x)] +\lambda_\mu(x) \one,
\ee
where the first term generates supergauge transformations,
$\nabla_\mu = \partial_\mu -i\A_\mu$ being the covariant
derivative, and the second divides out by the centre of the
algebra. Indeed this `no-$\A^0$ shift symmetry' ensures that
nothing depends on $\A^0$ and that $\A^0$ has no degrees of 
freedom~\cite{aprop}. The spontaneous breaking is carried by a superscalar
field
\be
\label{defC}
\C = \left( \begin{array}{cc} C^1 & D \\
                  \bar{D} & C^2
           \end{array}
       \right),
\ee
which transforms covariantly:
\be
\label{Cgauged}
\delta\C = -i\,[\C,\Omega].
\ee

It can be shown that, at the classical level, the spontaneous
breaking scale (effectively the mass of $B$) tracks the covariant
higher derivative effective cutoff scale $\Lambda$, if $\C$ is
made dimensionless (by using powers of $\Lambda$) and  $\hS$ has
the minimum of its effective potential at:
\be
\label{sigma}
<\C>\ = \sigma \equiv \pmatrix{\one & 0\cr 0 & -\one}.
\ee
In this case the classical action $S_0$ also has a minimum 
at~\eq{sigma}. At the quantum level this can be imposed as a
constraint on $S$, which can be satisfied by a suitable choice of
$\hS$~\cite{aprop,Thesis}. When we shift to the broken phase, $D$ becomes
a super-Goldstone mode (eaten by $B$ in unitary gauge) whilst the
$C^i$ are Higgs bosons and can be given a running mass of order
$\Lambda$~\cite{ym,Arnone:2001iy,aprop}. Working in our manifestly
gauge invariant formalism, $B$ and $D$ gauge transform into each
other; in recognition of this, we define the composite fields\footnote{These
definitions are consistent with those in~\cite{Thesis} but differ,
for convenience,
from those in~\cite{aprop} by signs in the fifth component.}
\be
	F_R = (B_\mu, D), \qquad \bar{F}_R = (\bar{B}_\mu, -\bar{D}).
\label{eq:F}
\ee

It will be useful to define the projectors $\sigma_1 =
{1\over2}(\one+\sigma)$ and $\sigma_-={1\over2}(\one-\sigma)$.
With a slight abuse of notation we can then write the components
of the superfields in terms of full supermatrices: $\tilde{A}^1_\mu =
\sigma_+\A_\mu\sigma_+$,\ $\tilde{A}^2_\mu = \sigma_-\A_\mu\sigma_-$,\
$B_\mu = \sigma_+\A_\mu\sigma_-$,\ $\Bb_\mu =
\sigma_-\A_\mu\sigma_+$, and similarly for $C^i$, $D$ and $\Db$.
(Note that these $\tilde{A}^i$ thus contain $\A^0\sigma_i$.
We will see in \secs{sec:BrokenPhase}{sec:GRs} how we can effectively
remove $\A^0$ from our considerations, using it to map us into a
particular diagrammatic picture.)
As will become clear later, in general this is a more useful
notation in the broken phase than the one employed in 
ref.~\cite{aprop} where we split the superfields only into full block
(off)diagonal components such as $A_\mu = \tilde{A}^1_\mu + \tilde{A}^2_\mu$. 

We generalise~\eq{sunflo} by first working out a form for the
classical two-point vertices in the broken phase, in particular
consistent with spontaneously broken $SU(N|N)$ invariance~\cite{aprop}. 
Generalisations 
of~\eq{twoflo} then determine the kernels $\dd$, providing there are
sufficiently many different terms in the flow equation to furnish
different kernels for each different two-point vertex. This,
together with respect for~\eq{Agauged} and~\eq{Cgauged}, are the
main constraints on the choice of $\Psi$. The solution given in
ref.~\cite{aprop} amounts to:
\be
\label{sunnfl}
\dot{S}  = a_0[S,\Sigma_g] - a_1[\Sigma_g],
\ee
where
\be
\label{a0}
 a_0[S,\Sigma_g] ={1\over2}\,\frac{\delta S}{\delta {\cal
A}_{\mu}}\{\dd^{\!\A\A}\}\frac{\delta \Sigma_g}{\delta {\cal
A}_{\mu}}+{1\over2}\,\frac{\delta S}{\delta {\cal
C}}\{\dd^{\C\C}\} \frac{\delta \Sigma_g}{\delta {\cal C}},
\ee
and
\be
\label{a1}
a_1[\Sigma_g] = {1\over2}\,\frac{\delta }{\delta {\cal
A}_{\mu}}\{\dd^{\!\A\A}\}\frac{\delta \Sigma_g}{\delta {\cal
A}_{\mu}} + {1\over2}\,\frac{\delta }{\delta {\cal
C}}\{\dd^{\C\C}\} \frac{\delta \Sigma_g}{\delta {\cal C}},
\ee
where the natural definitions of functional derivatives of
$SU(N|N)$ matrices are used~\cite{ymii,Arnone:2001iy,aprop}:
\be
\label{dCdef}
{\delta \over {\delta\C}} := { \left(\!{\begin{array}{cc} {\delta
/ {\delta C^1}} & - {\delta / {\delta \bar{D}}} \\ {\delta /
{\delta D}} & - {\delta / {\delta C^2}} \end{array}} \!\!\right)},
\ee
and from~\eq{expandA}~\cite{Arnone:2001iy,aprop}:
\be
\label{dumbdef}
{\delta\over\delta\A_\mu}:=
2T_A{\delta\over\delta\A_{A\,\mu}}+{\sigma\over2N}
{\delta\over\delta\A^0_\mu}.
\ee

The crucial freedom we need to generalise~\eq{effpropid} is hidden
in the definition of what we now mean by a covariantised kernel.
For both $W=\dd^{\A\A}$ and $W=\dd^{\C\C}$ we set
$\{W\}=\CovKer{W}{\A\C}$, where
\be
\label{wev}
\TiedCovKer{u}{W}{\A\C}{v} = \TiedCovKer{u}{W}{\A}{v}
-{1\over4} \TiedCovKer{[\C,u]}{W_{m}}{\A}{[\C,v]},
\ee
and $\CovKer{W}{\A}$ is a supercovariantisation, extending
that introduced in~\eq{Psi} and which is defined precisely below,
and the $\C$ commutator terms are introduced to allow a difference
between $B$ and $A$ kernels, and $C$ and $D$ kernels, in the
broken phase. They do this because at the level of two-point flow
equations, $\C$ is replaced by $\sigma$ in~\eq{wev}, and $\sigma$
(anti)commutes with the (fermionic) bosonic elements of the
algebra. Thus, extracting the broken-phase two-point classical
flow equations from~\eq{a0}, we find that the $A^i$ kernels are
given by $\dd^{\!\A\A}$, the $C^i$ kernels by $\dd^{\C\C}$, but
the $B$ kernel is $\dd^{\!\A\A}+\dd^{\!\A\A}_m$ and the $D$ kernel
is $\dd^{\C\C}+\dd^{\C\C}_m$~\cite{aprop}. The $B$ and $D$ kernels can be
combined:
\begin{equation}
	\dd^{\,F\; \bar{F}}_{MN}(p) =
	\left(
		\begin{array}{cc}
			\dd_p^{B \bar{B}} \delta_{\mu \nu}	& 0
		\\
									0			& -\dd^{D\bar{D}}_p
		\end{array}
	\right).
\label{eq:F-Wine}
\end{equation}
Notice in~\eq{wev}
that we use $\{W\}_{\!\!{}_{\A\C}}$ to label a covariantisation
of two kernels, $W$ and $W_{m}$.

The other main constraint is invariance 
under~\eqs{Agauged}{Cgauged}. The general covariantisation encodes the
invariance by insisting that
\bea
\label{wv}
		\TiedCovKer{u}{W}{\A}{v} = 
		\sum_{m,n=0}^{\infty} \int_{x_1, \cdots, x_n;y_1,\cdots, y_m;x,y}
%
W_{\mu_1\cdots\mu_n,\nu_1\cdots\nu_m}(x_1,\ldots,x_n;y_1,\ldots,y_m;x,y) 
& &
\nonumber 
\\
	\str \left[u(x)\A_{\mu_1}(x_1)\cdots \A_{\mu_n}(x_n)v(y)\A_{\nu_1}(y_1)\cdots \A_{\nu_m}(y_m) \right].
&&
\eea
is invariant under~\eq{Agauged}, where $u$ and $v$ are supermatrix
representations transforming homogeneously as in~\eq{Cgauged} and
where, without loss of generality, we may insist that
$\CovKer{W}{\A}$ satisfies $\TiedCovKer{u}{W}{\A}{v} \equiv
\TiedCovKer{v}{W}{\A}{u}$. For simplicity's sake, we have chosen~\eq{wv}
to contain only a single supertrace. 
The $m=n=0$ term is just the original kernel, \ie
\be
\label{mno}
W_,(;;x,y)\equiv W_{xy}.
\ee
The requirement that~\eq{wv} is supergauge invariant enforces a
set of Ward identities on the vertices
$W_{\mu_1\cdots\mu_n,\nu_1\cdots\nu_m}$ which we describe
later. Note that, for the sake of brevity, we will often 
loosely refer to
vertices of the covariantised kernels simply as kernels.
The no-$\A^0$ symmetry is obeyed by requiring the coincident line
identities~\cite{ymi}.
These identities are equivalent to the requirement that the gauge
fields all act by commutation~\cite{ymii}, thus trivially the
$\A^0$ parts of~\eq{defA} disappear, ensuring that the no-$\A^0$
part of~\eq{Agauged} is satisfied. A consequence of
the coincident line identities, which also trivially follows from the
representation of~\eq{wv} in terms of commutators, is that if
$v(y)=\one g(y)$ for all $y$, \ie is in the scalar representation
of the gauge group, then the covariantisation collapses to
\be
\label{Acoline}
\TiedCovKer{u}{W}{\A}{v}
= (\str\, u)\ker{W}g.
\ee
Note that the same identity therefore holds for the extended
version~\eq{wev}.

Under~\eq{Cgauged}, the $\C$ functional derivative
transforms homogeneously:
\be
\label{dCgauged}
\delta \left({\delta\over\delta\C}\right) =
-i\left[{\delta\over\delta\C},\Omega\right],
\ee
and thus by~\eq{wev} and~\eq{wv}, the corresponding terms 
in~\eqs{a0}{a1} are invariant. The $\A$ functional derivative, however,
transforms as~\cite{aprop}:
\be
\label{dAgauged}
\delta \left({\delta\over\delta\A_\mu}\right) =
-i\left[{\delta\over\delta\A_\mu},\Omega\right]
+{i\one\over2N}\tr\left[{\delta\over\delta\A_\mu},\Omega\right].
\ee
The correction is there because~\eq{dumbdef} is traceless, which
in turn is a consequence of the supertracelessness of~\eq{defA}.
The fact that $\delta/\delta\A$ does not transform homogeneously
means that supergauge invariance is destroyed unless the
correction term vanishes for other reasons.

Here, no-$\A^0$ symmetry comes to the rescue. Using the invariance
of~\eq{wev} for homogeneously transforming $u$ and $v$, and the
invariance of $S$ and $\hS$, we have by~\eq{dAgauged} 
and~\eq{Acoline}, that the $\A$ term in~\eq{a0} transforms to
\be
\label{tlgaugetr}
\delta\left(\frac{\delta S}{\delta {\cal
A}_{\mu}}\{\dd^{\!\A\A}\}\frac{\delta \Sigma_g}{\delta {\cal
A}_{\mu}}\right) = {i\over2N}\, \tr\!\left[{\delta
S\over\delta\A_\mu},\Omega\right]\!\cdot
\dd^{\!\A\A}\!\cdot\str{\delta\Sigma_g\over\delta\A_\mu} +
(S\leftrightarrow\Sigma_g),
\ee
where $S\leftrightarrow\Sigma_g$ stands for the same term with $S$
and $\Sigma_g$ interchanged. But by~\eq{dumbdef} and no-$\A^0$
symmetry,
\[
\str{\delta\Sigma_g\over\delta\A_\mu}={\delta\Sigma_g\over\delta\A^0_\mu}=0,
\]
similarly for $S$, and thus the tree level terms are invariant
under~\eqs{Agauged}{Cgauged}. Likewise, the quantum terms 
in~\eq{a1} are invariant, since
\be
\label{qcgaugetr}
\delta\left(\frac{\delta }{\delta {\cal
A}_{\mu}}\{\dd^{\!\A\A}\}\frac{\delta }{\delta {\cal
A}_{\mu}}\Sigma_g\right) = {i\over N}\,\tr\!\left[{\delta
\over\delta\A_\mu}, \Omega\right]\!\cdot\dd^{\!\A\A}\!\cdot
\str{\delta\Sigma_g\over\delta\A_\mu} =0.
\ee
This completes the proof that the ERG~\eq{sunnfl} is both
supergauge and no-$\A^0$ invariant.

To state the Ward identities we first
modify the definition of the vertices of
the kernels to reflect the structure of the ERG equation:
\bea
	\lefteqn{
		\fder{}{Z_1^c} \left\{W^{Z_1Z_2}\right\} \fder{}{Z_2^c} = 
		\sum_{m,n=0}^{\infty} \int_{x_1, \cdots, x_n;y_1,\cdots, y_m;x,y}
	} \nonumber 
\\ [1ex]
	& 	& W^{X_1 \cdots X_n, Y_1 \cdots Y_m, Z_1 Z_2}_{\, a_1 \; \cdots \, a_n,\, b_1\, \cdots b_m}(x_1,\ldots,x_n;y_1,\ldots,y_m;x,y) \nonumber 
\\ [1ex]
	& 	& \str \left[\fder{}{Z_1^c(x)}X^{a_1}_1(x_1)\cdots X^{a_n}_n(x_n) \fder{}{Z_2^c(y)} Y^{b_1}_1 (y_1) \cdots Y^{b_m}_m(y_m) \right],
\label{eq:Wine-B}
\eea
where the $X_i$, $Y_i$ and $Z_i$ are any of the broken phase fields
(up to certain restrictions to be discussed)
and the indices $a_i$, $b_i$ and $c$ are Lorentz indices or null, 
as appropriate. The definition~\eq{wev} restricts the
appearances of the $C^i$, and the structure of the flow equation
insists that the $Z_i$
must be components of either $\A$ or $\C$.
The combined  $X_i$, $Y_i$ and $Z_i$
must be net-bosonic but  note, in particular, that 
the there is no reason for the $Z_i$
not to be net fermionic \eg $W^{F,A^1B}$. Finally,
we assign a value of zero to vertices which imply
an illegal supertrace structure \eg a vertex
in which an $A^1$ directly follows an $A^2$.

Now we
define the (broken phase) Wilsonian effective action / seed action vertices.
Using the former as a template, this is simple:
supergauge invariance demands that this expansion be in terms of supertraces and
products of supertraces~\cite{aprop}:
\bea
	S	& =	& 	\sum_{n=1}^\infty \frac{1}{s_n} \int_{x_1, \cdots, x_n}
				S^{X_1 \cdots X_n}_{\,a_1 \, \cdots \, a_n}(x_1, \cdots, x_n) \str X_1^{a_1}(x_1) \cdots X_n^{a_n} (x_n)
	\nonumber \\
		& +&	\frac{1}{2!} \sum_{m,n=0}^{\infty} \frac{1}{s_n s_m} 
				\int_{x_1, \cdots, x_n;y_1,\cdots, y_m}
				 S^{X_1 \cdots X_n, Y_1 \cdots Y_m}_{\, a_1 \, \cdots \, a_n \, , \, b_1\cdots \, b_m}(x_1, \cdots, x_n; y_1 \cdots y_m)
	\nonumber \\
		&	&	
				\hspace{6em} \str X_1^{a_1}(x_1) \cdots X_n^{a_n} (x_n) \, \str Y_1^{b_1}(y_1) \cdots Y_m^{b_m} (y_m)
	\nonumber \\
		&+	& \ldots \label{eq:ActionExpansion},
\eea
where the labels and indices are as before
and the vacuum energy is ignored. 
Due to the invariance of the supertrace under cyclic permutations of its
arguments, we take only one cyclic ordering for the lists $X_1 \cdots X_n$, $Y_1 \cdots Y_m$ in the sums over $n,m$.
Furthermore, if either list is invariant under some nontrivial
cyclic permutations, then $s_n$ ($s_m$) is the order of the cyclic
subgroup, otherwise $s_n=1$ ($s_m=1$). Charge conjugation invariance,
under which $\A \rightarrow -\A^T$ and $\C \rightarrow \C^T$~\cite{ymi,aprop}
provides relationships between various vertices, which will be exploited
throughout this paper.

We write the momentum space vertices as
\begin{eqnarray*}
	\lefteqn{
		S^{X_1 \cdots X_n}_{\,a_1 \, \cdots \, a_n}(p_1, \cdots, p_n) \left(2\pi \right)^D \delta \left(\sum_{i=1}^n p_i \right)
	}
\\
	& &
	=
	\Int{x_1} \!\! \cdots \volume{x_n} e^{-i \sum_i x_i \cdot p_i} S^{X_1 \cdots X_n}_{\,a_1 \, \cdots \, a_n}(x_1, \cdots, x_n),
\end{eqnarray*}
where all momenta are taken to point into the vertex, and
similarly for the multiple supertrace vertices. We will employ the shorthand
\[
	S^{X_1 X_2}_{\, a_1 \, a_2}(p) \equiv S^{X_1 X_2}_{\, a_1 \, a_2}(p,-p).
\]

The Ward identities, which follow from applying~\eq{Agauged}, \eq{Cgauged}, \eq{dCgauged}
and~\eq{dAgauged} to the flow eqns.~\eq{sunnfl}--\eq{a1} can now be expressed in terms of
the vertices of \eqs{eq:Wine-B}{eq:ActionExpansion}. 
From the unbroken bosonic gauge transformations 
follows~\cite{ymi,aprop}
\be
	q_\nu U^{\cdots X \! A^{1,2} Y \cdots}_{\cdots \, a \; \nu \ \ \  b \ \cdots} (\ldots, p,q,r, \ldots) 
	=
	U^{\cdots XY \cdots}_{\cdots \, a \; b \ \cdots}( \ldots, p, q+r, \ldots) - 	
	U^{\cdots XY \cdots}_{\cdots \, a \; b \ \cdots}( \ldots, p+q, r, \ldots),
\label{eq:WID-U}
\ee
where $U$ can be a vertex from either action or any of the kernels.
The field $X$ and / or $Y$ can, unlike $ A^{1,2}_\nu$, represent
the end of a vertex of a covariantised
kernel; in which case they are to be identified with
the $Z_1$ and / or $Z_2$  of eqn.~\eq{eq:Wine-B}, as appropriate.
There is an appealing geometrical picture of~\eq{eq:WID-U}: 
we can view the momentum of the 
field $A^{1,2}_\nu$ as being pushed forward on to
the next field with a plus and pulled back on to the previous field
with a minus~\cite{ym,ymi,aprop}.

The second Ward identity follows from  the (broken)
fermionic gauge transformations and is most neatly written in terms of the
composite fields $F$ and $\bar{F}$. To this end, we define
a five-momentum~\cite{aprop}\footnote{Again, this definition is different 
from~\cite{ymi}, where the two comes with a minus sign.}
\be
	q_M = (q_\mu, 2).
\label{eq:Intro:qM}
\ee
The Ward identity corresponding to the broken, fermionic gauge transformations
can now be written in the following compact form (there is an identical eqn.\
involving $\bar{F}_N$):
\be
	q_N U^{\cdots X \! F Y \cdots}_{\cdots \, a  N  b \ \cdots} (\ldots, p,q,r, \ldots) 
	=
	U^{\cdots X\PF{Y} \cdots}_{\cdots \, a \; b \ \cdots}( \ldots, p, q+r, \ldots) - 	
	U^{\cdots \PB{X}Y \cdots}_{\cdots \, a \; b \ \cdots}( \ldots, p+q, r, \ldots),
\label{eq:WID-B}
\ee
where the $\PF{X}$ \etc should be interpreted according to 
table~\ref{tab:GR:Flavour} and the index $N$ is summed over,
such that each product of components is weighted with unity.
The null entries are those for which the required ordering of
fields does not exist.
\renewcommand{\arraystretch}{1.5}
\begin{center}
\begin{table}[h]
	\[
	\begin{array}{l|l|c|c}
		U 		& 	V 		& \PF{V} 			& \PB{V}
\\ \hline\hline
		\bar{F} & (A^1,C^1) &(\bar{B}, \bar{D})	& \mbox{---}
\\
  				& (A^2,C^2) &\mbox{---} 		& (\bar{B},\bar{D}) 
\\
  				& F 		&(\tilde{A}^2, C^2) & (\tilde{A}^1,C^1) 
\\ \hline
		F 		& (A^2,C^2) & F 				& \mbox{---}
\\ 
  				& (A^1,C^1) & \mbox{---} 		& F 
\\
  				& \bar{F} 	&(\tilde{A}^1,-C^1) & (\tilde{A}^2,-C^2) 
	\end{array}
	\]
\caption{The flavour changing effect of pushing forward and / or pulling back the 
momentum of the fermionic field, $U$ on to the field $V$.}
\label{tab:GR:Flavour}
\end{table}
\end{center}
\renewcommand{\arraystretch}{1}

Apart from the general constraints already discussed, some weak
constraints necessary for regularisation as reviewed below, and of
course charge conjugation invariance and Poincar\'e invariance, the 
covariantisation~\eq{wv} and seed action $\hS$ in the flow eqn.~\eq{sunnfl} can
be left unspecified~\cite{aprop}: the calculational methods (reviewed and
extended in~secs.~\ref{Diagrammatics}--\ref{sec:MomentumExpansions}) 
are designed to turn this freedom to our
advantage, by acting as a guide to the most efficient method of
computation. In addition, this provides an automatic check on the
universality of the terms we are computing.

The ERG~\eq{sunnfl} is properly ultraviolet regularised by virtue
of the manifest spontaneously broken $SU(N|N)$ invariance and
covariant higher derivative regularisation. This latter is
determined by the functions in the general form of the two-point
vertices~\cite{aprop}, and corresponds to the
introduction of cutoff functions. For a specific choice of bare
action and power-law cutoff functions, the weak constraints that
these powers have to satisfy were determined in ref.~\cite{Arnone:2001iy}.
They are not necessarily, however, the correct inequalities to
ensure that the flow eqn.~\eq{sunnfl} itself is regularised,
for a specific choice of seed action and covariantisation~\eq{wv},
nor is it necessary to restrict to only power law cutoff
functions. Undoubtedly it is possible to work out some very
general constraints on $\hS$, the covariantisation~\eq{wv} and the
cutoff functions, which ensure that all quantities are properly
ultraviolet regularised at all stages of calculation. However
since all these details drop out of the calculation at the end,
this effort would be largely wasted. Instead we 
follow~\cite{aprop}, and simply assume that these constraints are
satisfied.

Finally, we will require that the covariantisation satisfies
\be
\label{noATailBiting}
\TiedCovKer{\fder{}{\A_\mu}}{W}{\A}{} = 0,
\ee
(where the functional derivative acts on all terms inside
$\CovKer{W}{\A}$ but not on the unspecified right hand
attachment) \ie that there are no diagrams  in which the kernel
bites its own tail~\cite{ym,ymi,ymii}. In general, certain such
diagrams do appear to be improperly regularised, and this is why we
apply this restriction; we will give an example in \sec{sec:WBT},
after we have introduced the necessary diagrammatics.

For general
solutions to this constraint see refs.~\cite{ym,aprop}. 
Eqn.~\eq{noATailBiting} leads to identities for the $W$ vertices which
again we do not need in practice. For the future we note that
there are indications that the restriction~\eq{noATailBiting} can
in fact be lifted: in any calculation of universal quantities, 
the ultraviolet divergences in kernel-biting-their-tail
diagrams are then cancelled by implicit divergences in other
terms, such diagrams always disappearing in the final 
answer~\cite{Thesis,oliver1,oliver2}.

\subsection{The need for a new flow equation}
\label{need}


The running coupling $g(\Lambda)$ in~\eq{sunnfl}, entering 
via~\eq{Sigma}, is identified with the original $SU(N)$ coupling via
the renormalisation condition
\be
\label{defg}
S[\A=A^1, \C=\sigma] ={1\over2g^2}\,\str\!\int\!\!d^D\!x\,
\left(F^1_{\mu\nu}\right)^2+\cdots,
\ee
where the ellipsis stands for higher dimension operators and the
ignored vacuum energy. (No wavefunction renormalisation is required
because exact preservation of gauge invariance forbids it once the
coupling is scaled out of the covariant derivative~\cite{ym,ymi}.)
This means that the coefficient for the two-point vertex with
structure $\str\, A^1_\mu(p) A^1_\nu(-p)$ must have the form
\be
\label{gcondn}
S^{11}_{\mu\nu}(p) = {2\over g^2}\Box_{\mu\nu}(p) + \Op{4},
\ee
where $\Box_{\mu \nu}(p) = p^2 \delta_{\mu \nu} - p_\mu p_\nu$.
In order to go beyond one-loop computations of purely $A^1$
vertices (for example in computing the two-loop beta 
function~\cite{Thesis,mgierg2}) we will need to take into account the running
coupling $g_2(\Lambda)$ associated with $A^2$. We define it via
the renormalisation condition
\be
\label{defg2}
S[\A=A^2, \C=\sigma] ={1\over2g^2_2}\,\str\!\int\!\!d^D\!x\,
\left(F^2_{\mu\nu}\right)^2+\cdots,
\ee
where the ellipsis has the same meaning as in~\eq{defg}. Note in
particular the unphysical wrong sign for $g^2_2$, forced by the
supertrace. (In turn this is a necessary consequence of global
invariance under the supergroup in the unbroken phase. Indeed here
the lowest dimension gauge field operator is proportional to
$\str\,\F^2_{\mu\nu}$, $\F_{\mu\nu}=i[\nabla_\mu,\nabla_\nu]$
being the superfieldstrength. The wrong sign in the $A^2$ sector
leads to unitarity violation which decouples in the continuum
limit~\cite{Arnone:2001iy,aprop}.) Eqn.~\eq{defg2} implies that the
two-point vertex with the structure $\str\, A^2_\mu(p)
A^2_\nu(-p)$ must have the form
\be
\label{g2condn}
S^{22}_{\mu\nu}(p) = {2\over g^2_2}\Box_{\mu\nu}(p) + \Op{4}.
\ee

In ref.~\cite{aprop} we restricted the classical effective action
to a single supertrace. Since $\str\,\sigma A_\mu A_\nu$ is not
no-$\A^0$ invariant, this requires that the classical two-point
$A$ vertices have the structure:
\[
\str\, A_\mu A_\nu = \str\, A^1_\mu A^1_\nu + \str\, A^2_\mu
A^2_\nu,
\]
which in turn enforces $g_2=g$ at the classical level (in common
with the unbroken phase).

However we also showed that the $A$ two-point vertex~\eq{gcondn} flows at
one-loop into~\cite{aprop}
\be
\label{aproponeloop}
{22\over3(4\pi)^2}\ \Box_{\mu\nu}(p)\ \str\,\sigma\ \str\,\sigma
A_\mu A_\nu +\Op{4},
\ee
where the $\Box_{\mu\nu}$ factor follows from manifest gauge invariance.
Utilising $\str\,\sigma = 2N$, we recognise 
that the numerical coefficient of~\eq{aproponeloop} is just
equal to 
$-4\beta_1$, where $\beta_1$ is the
famous one-loop $SU(N)$ Yang-Mills $\beta$ function coefficient.
Since
\[
\str\,\sigma A_\mu A_\nu = \str\, A^1_\mu A^1_\nu - \str\, A^2_\mu
A^2_\nu,
\]
this gives the expected flow of~\eq{gcondn} and the expected
wrong-sign $\beta$ function for~\eq{g2condn}.

Now an obvious question arises: how can~\eq{aproponeloop} be
consistent with no-$\A^0$ invariance? In fact it is
straightforward to repeat the computation of ref.~\cite{aprop} for
the case where a gauge field appears in each supertrace and thus
show that the one-loop term
\be
\label{apropplus}
-{22\over3(4\pi)^2}\ \Box_{\mu\nu}(p)\ \str\,\sigma A_\mu\
\str\,\sigma A_\nu
\ee
is also generated. Note that, from~\eq{defA} or~\eq{expandA}, this
contains only $\A^0$s, and it cancels the $\A^0$ terms 
in~\eq{aproponeloop}. Therefore, together these terms form a
combination invariant under the no-$\A^0$ symmetry of~\eq{Agauged}.

We thus confirm that~\eq{Agauged} is operating as expected, but we
see clearly that we have to abandon the single supertrace form for
$S$ in order to renormalise $g$ and $g_2$ differently, already at
one loop.

\section{The new flow equation}
\label{New}

\subsection{Modifying the flow equation}

The need to allow $g_2\ne g$ implies that the classical two-point
vertices $S^{\ \, 1\, 1}_{0\, \mu\nu}$ and $S^{\ \, 2 \, 2}_{0\, \mu\nu}$ 
must also
be free to differ. If we are to maintain the advantages of setting
two-point vertices in $S_0$ and $\hS$ equal, we must generalise
the flow equation so that the kernels for $A^1$ and $A^2$ can also
differ (\cf the arguments above~\eq{sunnfl}
and surrounding~\eq{wev}).

We therefore need to add new terms to the definition of the
covariantised kernel in~\eq{wev} for the case where this is used
to connect two functional $\A$ derivatives. (We do not need it for
the kernels connecting two functional $\C$ derivatives because we
do not need to allow the $C^i$ kernels to differ. This is because
there are no terms that require renormalising in
the $\C$ sector, as we will explicitly demonstrate in~\cite{mgierg2}.) 
Clearly we need an insertion of $\C$, since in
the broken phase this gives $\sigma$, the only algebraic object
available for distinguishing $A^1$ from $A^2$. However it cannot
be a commutator as in~\eq{wev} since $[\sigma,\delta/\delta
A^i]=0$. From the above discussion of no-$\A^0$ invariance
and the discussion of~\eq{dAgauged} (where
also the need is to cancel terms proportional to $\one$) it is
clear that we need to entertain multiple supertrace terms in order
to preserve~\eqs{Agauged}{Cgauged}. The simplest solution is to
define in~\eqs{a0}{a1},
\be
\label{weev}
	u\{\dd^{\A\A}\}v := \TiedCovKer{u}{\dd^{\A\A}}{\A\C}{v} 
		+ \TiedCovKer{u}{\dd^{\A\A}_\sigma}{\A}{\P(v)} 
		+ \TiedCovKer{\P(u)}{\dd^{\A\A}_\sigma}{\A}{v},
\ee
where
\be
\label{Pdef}
8N\,\P(X) = \{\C,X\}\,\str\,\C - 2\, \C\,\str\,\C X.
\ee
Note that $\P(X)$ has the following special properties:
$\str\,\P(X)=0$ and $\P(\one)=0$. Using~\eq{Acoline}, 
it is easy to see these ensure
that the extra terms in~\eq{weev} are supergauge invariant despite
the inhomogeneous part of~\eq{dAgauged}. The anticommutator
structure in~\eq{Pdef} arises without loss of generality from
charge conjugation invariance of~\eqs{a0}{a1}. $\P$ is added
symmetrically in~\eq{weev} to maintain the symmetry
$u\{\dd^{\A\A}\}v = v\{\dd^{\A\A}\}u$ purely for convenience in the
diagrammatic formalism to follow. Likewise the factor of $8N$ 
in~\eq{Pdef} is purely for convenience. Finally note that the
redefinition~\eq{weev} does not disturb the fact that~\eq{sunnfl}
can be written in the form~\eq{blockflow} (with $\phi$ replaced by
a sum over $\A$ and $\C$ fields). Thus~\eq{sunnfl} still
corresponds to a valid ERG.


\subsection{The old-style diagrammatics}
\label{Diagrammatics}

We introduce the diagrammatics of~\cite{aprop},
adapted to the new flow equation, in the symmetric
phase, as this allows us to describe the necessary
elements whilst minimising new notation.

\subsubsection{Diagrammatics for the action}
\label{sec:Diags:Action}

The diagrammatic representation of the action is given in \fig{fig:Diagrammatics:Action}.
\bcf[h]
	\[
		\ensuremath{\begin{array}{c}\begin{picture}(0,0)%
\includegraphics{pstex/GeneralSingleSupertrace.pstex}%
\end{picture}%
\setlength{\unitlength}{3947sp}%
\begingroup\makeatletter\ifx\SetFigFont\undefined%
\gdef\SetFigFont#1#2#3#4#5{%
  \reset@font\fontsize{#1}{#2pt}%
  \fontfamily{#3}\fontseries{#4}\fontshape{#5}%
  \selectfont}%
\fi\endgroup%
\begin{picture}(422,457)(2070,-1028)
\put(2217,-833){\makebox(0,0)[lb]{\smash{\SetFigFont{11}{13.2}{\rmdefault}{\mddefault}{\updefault}{\color[rgb]{0,0,0}$S$}%
}}}
\end{picture}
 \end{array}} +  \ensuremath{\begin{array}{c}\begin{picture}(0,0)%
\includegraphics{pstex/GeneralDoubleSupertrace.pstex}%
\end{picture}%
\setlength{\unitlength}{3947sp}%
\begingroup\makeatletter\ifx\SetFigFont\undefined%
\gdef\SetFigFont#1#2#3#4#5{%
  \reset@font\fontsize{#1}{#2pt}%
  \fontfamily{#3}\fontseries{#4}\fontshape{#5}%
  \selectfont}%
\fi\endgroup%
\begin{picture}(617,960)(2330,-1516)
\put(2761,-1135){\makebox(0,0)[lb]{\smash{\SetFigFont{11}{13.2}{\rmdefault}{\mddefault}{\updefault}{\color[rgb]{0,0,0}$S$}%
}}}
\end{picture}
 \end{array}} + \cdots
	\]
\caption{Diagrammatic representation of the action.}
\label{fig:Diagrammatics:Action}
\ecf

Each circle stands for a single supertrace containing any number of fields. The $S$
is to remind us that this is an expansion of the action, since the seed action, $\hS$,
has a similar expansion.
The dotted line reminds us that the two supertraces in the double supertrace term are
part of the same vertex.
The arrows
indicate the cyclic sense in which fields should be read off; henceforth, this
will always be done in the counterclockwise sense and so these arrows will generally
be dropped.\footnote{Arrows can always be dropped in complete diagrams formed by
the flow equation. However, if we look at the diagrammatic representation of a kernel,
in isolation, then it will be necessary to keep these arrows.} 
In turn, each of
these supertraces can now be expanded in terms of a sum over all explicit, cyclically
independent combinations of fields, as in \fig{fig:Diagrammatics:SingleStr},
where closed circles represent $\A$s and open circles represent $\C$s.
\bcf[h]
	\[
	\ensuremath{\begin{array}{c}\begin{picture}(0,0)%
\includegraphics{pstex/GeneralSingleSupertrace-B.pstex}%
\end{picture}%
\setlength{\unitlength}{3947sp}%
\begingroup\makeatletter\ifx\SetFigFont\undefined%
\gdef\SetFigFont#1#2#3#4#5{%
  \reset@font\fontsize{#1}{#2pt}%
  \fontfamily{#3}\fontseries{#4}\fontshape{#5}%
  \selectfont}%
\fi\endgroup%
\begin{picture}(422,420)(2070,-991)
\end{picture}
 \end{array}} =
		\left[
		\ensuremath{\begin{array}{c}\begin{picture}(0,0)%
\includegraphics{pstex/Vertex-SH.pstex}%
\end{picture}%
\setlength{\unitlength}{3947sp}%
\begingroup\makeatletter\ifx\SetFigFont\undefined%
\gdef\SetFigFont#1#2#3#4#5{%
  \reset@font\fontsize{#1}{#2pt}%
  \fontfamily{#3}\fontseries{#4}\fontshape{#5}%
  \selectfont}%
\fi\endgroup%
\begin{picture}(392,423)(2144,-999)
\end{picture}
 \end{array}}
		+ \
		{\renewcommand{\arraycolsep}{0.2em}
		\begin{array}{rl}
				& \ensuremath{\begin{array}{c}\begin{picture}(0,0)%
\includegraphics{pstex/Vertex-SFSF.pstex}%
\end{picture}%
\setlength{\unitlength}{3947sp}%
\begingroup\makeatletter\ifx\SetFigFont\undefined%
\gdef\SetFigFont#1#2#3#4#5{%
  \reset@font\fontsize{#1}{#2pt}%
  \fontfamily{#3}\fontseries{#4}\fontshape{#5}%
  \selectfont}%
\fi\endgroup%
\begin{picture}(392,458)(2144,-1034)
\end{picture}
 \end{array}}
		\\
			+	& \ensuremath{\begin{array}{c}\begin{picture}(0,0)%
\includegraphics{pstex/Vertex-SHSH.pstex}%
\end{picture}%
\setlength{\unitlength}{3947sp}%
\begingroup\makeatletter\ifx\SetFigFont\undefined%
\gdef\SetFigFont#1#2#3#4#5{%
  \reset@font\fontsize{#1}{#2pt}%
  \fontfamily{#3}\fontseries{#4}\fontshape{#5}%
  \selectfont}%
\fi\endgroup%
\begin{picture}(392,458)(2144,-1034)
\end{picture}
 \end{array}}
		\end{array}
		}
		+ \
		{\renewcommand{\arraycolsep}{0.2em}
		\begin{array}{rl}
				& \ensuremath{\begin{array}{c}\begin{picture}(0,0)%
\includegraphics{pstex/Vertex-SFSFSF.pstex}%
\end{picture}%
\setlength{\unitlength}{3947sp}%
\begingroup\makeatletter\ifx\SetFigFont\undefined%
\gdef\SetFigFont#1#2#3#4#5{%
  \reset@font\fontsize{#1}{#2pt}%
  \fontfamily{#3}\fontseries{#4}\fontshape{#5}%
  \selectfont}%
\fi\endgroup%
\begin{picture}(392,423)(2144,-999)
\end{picture}
 \end{array}}	
		\\
			+	& \ensuremath{\begin{array}{c}\begin{picture}(0,0)%
\includegraphics{pstex/Vertex-SFSFSH.pstex}%
\end{picture}%
\setlength{\unitlength}{3947sp}%
\begingroup\makeatletter\ifx\SetFigFont\undefined%
\gdef\SetFigFont#1#2#3#4#5{%
  \reset@font\fontsize{#1}{#2pt}%
  \fontfamily{#3}\fontseries{#4}\fontshape{#5}%
  \selectfont}%
\fi\endgroup%
\begin{picture}(392,423)(2144,-999)
\end{picture}
 \end{array}}
		\\
			+	& \ensuremath{\begin{array}{c}\begin{picture}(0,0)%
\includegraphics{pstex/Vertex-SFSHSH.pstex}%
\end{picture}%
\setlength{\unitlength}{3947sp}%
\begingroup\makeatletter\ifx\SetFigFont\undefined%
\gdef\SetFigFont#1#2#3#4#5{%
  \reset@font\fontsize{#1}{#2pt}%
  \fontfamily{#3}\fontseries{#4}\fontshape{#5}%
  \selectfont}%
\fi\endgroup%
\begin{picture}(392,423)(2144,-999)
\end{picture}
 \end{array}}
		\\
			+	& \ensuremath{\begin{array}{c}\begin{picture}(0,0)%
\includegraphics{pstex/Vertex-SHSHSH.pstex}%
\end{picture}%
\setlength{\unitlength}{3947sp}%
\begingroup\makeatletter\ifx\SetFigFont\undefined%
\gdef\SetFigFont#1#2#3#4#5{%
  \reset@font\fontsize{#1}{#2pt}%
  \fontfamily{#3}\fontseries{#4}\fontshape{#5}%
  \selectfont}%
\fi\endgroup%
\begin{picture}(392,423)(2144,-999)
\end{picture}
 \end{array}}
		\end{array}
		}
		+\cdots
		\right.
	\]
\caption{Expansion of a single supertrace in powers of $\A$ and $\C$.}
\label{fig:Diagrammatics:SingleStr}
\ecf

In this case, we have chosen not to indicate whether the supertraces we
are considering come from the Wilsonian effective action, the
seed action, or some linear combination of the two.
Explicit instances of fields are referred to as decorations. Note that there are
no supertraces containing a single $\A$, since $\str \A =0$. The
 $\str \A \C$ vertex vanishes by charge conjugation invariance.

As it stands, \figs{fig:Diagrammatics:Action}{fig:Diagrammatics:SingleStr}
provide representations of the action~\eq{eq:ActionExpansion}. However, as we will
see shortly, it is often useful to interpret the explicitly decorated terms as just the
vertex coefficient functions $S^{X_1 \cdots X_n}_{\,a_1 \, \cdots \, a_n}(p_1, \cdots, p_n)$ etc.,
the accompanying supertrace(s) and cyclic symmetry factors having been stripped off.

\subsubsection{Diagrammatics for the kernels}
\label{sec:Diag:kernels}

The kernels  have a very similar diagrammatic expansion~\cite{ymii,aprop}
to the action, as shown in \fig{fig:Diagrammatics:Wines}.
\bcf[h]
	\[
		\cdeps{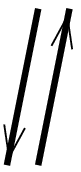} = \cdeps{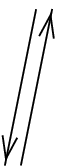} + \cdeps{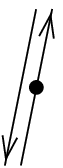} +\cdeps{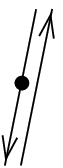} + \cdeps{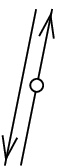} +\cdeps{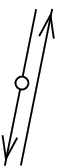} +\cdots
	\]
\caption{Diagrammatic representation of the kernels.}
\label{fig:Diagrammatics:Wines}
\ecf

The ellipsis represents terms with any number of fields distributed in 
all possible ways between the two sides of the kernel.
We should note that \fig{fig:Diagrammatics:Wines} is not
strictly a representation  of (the symmetric phase version of) eqn.~\eq{eq:Wine-B}.
Eqn.~\eq{eq:Wine-B} involves not
only the kernel vertices and the associated decorative fields, 
but also two functional derivatives which sit at the ends of
the kernel. Moreover, according to eqn.~\eq{weev},
the kernels can now attach via $\P$.

We can
directly include both the functional derivatives 
and instances of $\P$
in our
diagrammatics, as shown in
\fig{fig:Diagrammatics:Wines-True}; notice that, in combination,
they unambiguously label
the different kernels in the flow equation.
\bcf[h]
	\[
		\ensuremath{\begin{array}{c}\begin{picture}(0,0)%
\includegraphics{pstex/GeneralWine-Derivs.pstex}%
\end{picture}%
\setlength{\unitlength}{3947sp}%
\begingroup\makeatletter\ifx\SetFigFont\undefined%
\gdef\SetFigFont#1#2#3#4#5{%
  \reset@font\fontsize{#1}{#2pt}%
  \fontfamily{#3}\fontseries{#4}\fontshape{#5}%
  \selectfont}%
\fi\endgroup%
\begin{picture}(366,816)(2068,-1294)
\end{picture}
 \end{array}} +\frac{1}{8N} 
		\left[
			\cdeps{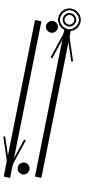} + \cdeps{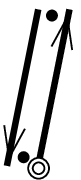} + \cdeps{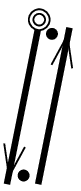} + \cdeps{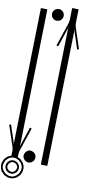}
		\right]
	\]
\caption{Diagrammatic representation of the kernels, 
which recognises the structure of the flow equation.}
\label{fig:Diagrammatics:Wines-True}
\ecf

In the first diagram, 
the grey circles can be either both $\A$s or both $\C$s;
since they sit at the end of the kernel
they represent functional derivatives and label the kernel.
The double circle notation, explicitly defined
in \fig{fig:DoubleCircle}, represents $\P$. Note that the presence of the
double circles tells us that the associated kernel is
of type $\dd^{\A\A}_{\sigma}$.
\bcf[h]
	\[
	\cdeps{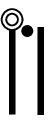} \equiv \cdeps{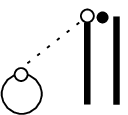} - \cdeps{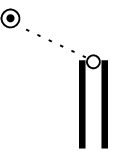}
	\]
\caption{Interpretation of the double circle notation.}
\label{fig:DoubleCircle}
\ecf

In both diagrams, the dotted line
is a `false' kernel in the sense that, if joining two
fields with position arguments $x$ and $y$,
it is just $\delta(x-y)$. 
Thus, in the first diagram on the \rhs, the $\C$ embedded at the end of the kernel
is attached, via a false kernel, to a separate $\str \C$. 
In the second diagram, the operator \ReplaceAwC\
tells us to replace an $\A$ with a $\C$. As for the kernel, it
is `plugged' by a $\C$, rather than ending in a functional
derivative; the functional derivative being linked to this $\C$
via a false kernel.
Note that one of the $\A$s
which labels the kernel may now be linked to the end of the kernel
by a false kernel.

\subsubsection{The symmetric phase flow equation}

To go from the diagrammatic representations of
the actions and kernels to the diagrammatic
flow equation, we require some properties of
the functional derivative \wrt\ $\A$ and $\C$.
Following~\cite{ymii,aprop} we introduce the constant
supermatrices $X$ and $Y$ and, neglecting the irrelevant spatial 
dependence,  schematically represent
first supersowing, whereby two supertraces are sown together:
\be
	\pder{}{\C} \str \C Y = Y \ \ \Longrightarrow \ \ \str X \pder{}{\C} \str \C Y = \str XY,
\label{eq:SuperSowing-C}
\ee
and secondly supersplitting, whereby a single supertrace is split into two supertraces:
\be
	\str \pder{}{\C} X \C Y = \str X \str Y.
\label{eq:SuperSplitting-C}
\ee

In the $\A$-sector, the analogue of these 
relationships, which can be deduced from the
completeness relation for the $T_A$~\cite{Arnone:2001iy}, receive
corrections, since $\A$ is constrained to be
supertraceless:
\bea
	\str X \pder{}{\A} \str \A Y	& = & \str XY - \frac{1}{2N} \str X \tr Y \label{eq:SuperSowing-A}
\\ [1ex]
	\str \pder{}{\A} X \A Y 		& = & \str X \str Y - \frac{1}{2N} \tr XY \label{eq:SuperSplitting-A}.
\eea
These corrections, containing $\tr \cdots \equiv \str \sigma \cdots$, generically
violate
$SU(N|N)$ invariance and, as proved in~\cite{aprop}, effectively vanish
as a consequence of the $SU(N|N)$ invariance of the flow equation (the modification
of the flow equation according to~\eq{weev} does nothing to change this
conclusion). Thus, the supersplitting and supersowing rules are
essentially exact for both fields, enabling us to straightforwardly
incorporate the gauge algebra into the diagrammatics, even at finite $N$~\cite{aprop}.

\Fig{fig:DiagFE} shows the diagrammatic representation
of the flow equation; the ellipses represent terms with
additional supertraces. The `dumbbell' terms of the top row are formed
by $a_0$ (which sews two supertraces together) whereas the `padlock' terms
of the bottom row are formed by $a_1$ (which splits a single supertrace
apart).
\bcf[h]
	\[
	\begin{array}{c}
		\dec{\ensuremath{\begin{array}{c}\begin{picture}(0,0)%
\includegraphics{pstex/Action.pstex}%
\end{picture}%
\setlength{\unitlength}{3947sp}%
\begingroup\makeatletter\ifx\SetFigFont\undefined%
\gdef\SetFigFont#1#2#3#4#5{%
  \reset@font\fontsize{#1}{#2pt}%
  \fontfamily{#3}\fontseries{#4}\fontshape{#5}%
  \selectfont}%
\fi\endgroup%
\begin{picture}(422,420)(2070,-991)
\put(2225,-827){\makebox(0,0)[lb]{\smash{\SetFigFont{11}{13.2}{\rmdefault}{\mddefault}{\updefault}{\color[rgb]{0,0,0}$S$}%
}}}
\end{picture}
 \end{array}} + \ensuremath{\begin{array}{c}\begin{picture}(0,0)%
\includegraphics{pstex/Action-DS.pstex}%
\end{picture}%
\setlength{\unitlength}{3947sp}%
\begingroup\makeatletter\ifx\SetFigFont\undefined%
\gdef\SetFigFont#1#2#3#4#5{%
  \reset@font\fontsize{#1}{#2pt}%
  \fontfamily{#3}\fontseries{#4}\fontshape{#5}%
  \selectfont}%
\fi\endgroup%
\begin{picture}(617,923)(2330,-1479)
\end{picture}
 \end{array}} + \cdots }{\bullet} = 
	\\[8ex]
		\ds
		\hf
		\left[
			\begin{array}{l}
				\ensuremath{\begin{array}{c}\input{pstex/ClassicalTerm-B.pstex_t} \end{array}} + \ds \frac{1}{8N}
				\left[
					\ensuremath{\begin{array}{c}\input{pstex/NewClassicalTerm-A.pstex_t} \end{array}}
					+\ensuremath{\begin{array}{c}\input{pstex/NewClassicalTerm-B.pstex_t} \end{array}}
					+\ensuremath{\begin{array}{c}\input{pstex/NewClassicalTerm-C.pstex_t} \end{array}}
					+\ensuremath{\begin{array}{c}\input{pstex/NewClassicalTerm-D.pstex_t} \end{array}}
				\right]
			\\[10ex]
				-\ensuremath{\begin{array}{c}\begin{picture}(0,0)%
\includegraphics{pstex/QuantumTerm-B.pstex}%
\end{picture}%
\setlength{\unitlength}{3947sp}%
\begingroup\makeatletter\ifx\SetFigFont\undefined%
\gdef\SetFigFont#1#2#3#4#5{%
  \reset@font\fontsize{#1}{#2pt}%
  \fontfamily{#3}\fontseries{#4}\fontshape{#5}%
  \selectfont}%
\fi\endgroup%
\begin{picture}(596,897)(2709,-1551)
\put(2913,-986){\makebox(0,0)[lb]{\smash{\SetFigFont{11}{13.2}{\rmdefault}{\mddefault}{\updefault}{\color[rgb]{0,0,0}$\Sigma_g$}%
}}}
\end{picture}
 \end{array}} - \ds \frac{1}{8N}
					\left[
						\ensuremath{\begin{array}{c}\begin{picture}(0,0)%
\includegraphics{pstex/NewQuantumTerm-A.pstex}%
\end{picture}%
\setlength{\unitlength}{3947sp}%
\begingroup\makeatletter\ifx\SetFigFont\undefined%
\gdef\SetFigFont#1#2#3#4#5{%
  \reset@font\fontsize{#1}{#2pt}%
  \fontfamily{#3}\fontseries{#4}\fontshape{#5}%
  \selectfont}%
\fi\endgroup%
\begin{picture}(615,897)(2690,-1551)
\put(2913,-986){\makebox(0,0)[lb]{\smash{\SetFigFont{11}{13.2}{\rmdefault}{\mddefault}{\updefault}{\color[rgb]{0,0,0}$\Sigma_g$}%
}}}
\end{picture}
 \end{array}}
						+\ensuremath{\begin{array}{c}\begin{picture}(0,0)%
\includegraphics{pstex/NewQuantumTerm-C.pstex}%
\end{picture}%
\setlength{\unitlength}{3947sp}%
\begingroup\makeatletter\ifx\SetFigFont\undefined%
\gdef\SetFigFont#1#2#3#4#5{%
  \reset@font\fontsize{#1}{#2pt}%
  \fontfamily{#3}\fontseries{#4}\fontshape{#5}%
  \selectfont}%
\fi\endgroup%
\begin{picture}(596,897)(2744,-1587)
\put(2955,-1024){\makebox(0,0)[lb]{\smash{\SetFigFont{11}{13.2}{\rmdefault}{\mddefault}{\updefault}{\color[rgb]{0,0,0}$\Sigma_g$}%
}}}
\end{picture}
 \end{array}}
						+\ensuremath{\begin{array}{c}\begin{picture}(0,0)%
\includegraphics{pstex/NewQuantumTerm-D.pstex}%
\end{picture}%
\setlength{\unitlength}{3947sp}%
\begingroup\makeatletter\ifx\SetFigFont\undefined%
\gdef\SetFigFont#1#2#3#4#5{%
  \reset@font\fontsize{#1}{#2pt}%
  \fontfamily{#3}\fontseries{#4}\fontshape{#5}%
  \selectfont}%
\fi\endgroup%
\begin{picture}(596,897)(2764,-1610)
\put(2956,-1057){\makebox(0,0)[lb]{\smash{\SetFigFont{11}{13.2}{\rmdefault}{\mddefault}{\updefault}{\color[rgb]{0,0,0}$\Sigma_g$}%
}}}
\end{picture}
 \end{array}}
						+\ensuremath{\begin{array}{c}\begin{picture}(0,0)%
\includegraphics{pstex/NewQuantumTerm-B.pstex}%
\end{picture}%
\setlength{\unitlength}{3947sp}%
\begingroup\makeatletter\ifx\SetFigFont\undefined%
\gdef\SetFigFont#1#2#3#4#5{%
  \reset@font\fontsize{#1}{#2pt}%
  \fontfamily{#3}\fontseries{#4}\fontshape{#5}%
  \selectfont}%
\fi\endgroup%
\begin{picture}(622,897)(2744,-1587)
\put(2955,-1024){\makebox(0,0)[lb]{\smash{\SetFigFont{11}{13.2}{\rmdefault}{\mddefault}{\updefault}{\color[rgb]{0,0,0}$\Sigma_g$}%
}}}
\end{picture}
 \end{array}}
					\right]
			\end{array}
		\right] + \cdots
	\end{array}
	\]
\caption{The diagrammatic flow equation.}
\label{fig:DiagFE}
\ecf

There are a number of important points to
make about the diagrams of \fig{fig:DiagFE}.
First is that we have discarded all diagrams in
which the kernels bite their tails (see \secs{Review}{sec:WBT}). Secondly, the
diagrammatic flow equation is naturally phrased
(as with all good Feynman diagrammatics) in 
terms of \emph{coefficient functions only}: all explicit
supertraces and symmetry factors have been stripped off.
Thus, the interpretation of the diagrammatic
elements in \fig{fig:DiagFE}, is not the same
as in figs.~\ref{fig:Diagrammatics:Action}--\ref{fig:DoubleCircle}, where both
the explicit supertraces and symmetry factors
are still present.

However, we can think of each diagram as having
an \emph{implied} supertrace structure, and this
must of course match between the diagrams on the
right-hand side and the term on the left-hand side of
the flow equation.
The supertrace structure on the left-hand side
is obvious, just being that naturally associated
with the vertex coefficient function whose flow
we are computing. To ensure that the diagrams
on the right-hand side are consistent with this,
we must understand how to read off the supertrace
structure from our diagrams. The rules are simple
and follow from exact supersowing / supersplitting.
Having expanded out the double circles, 
each closed circuit represents a supertrace, the
fields on the circuit representing the argument
of the supertrace. For each circuit, we should
sum over all independent cyclic permutations of the
fields. If a circuit is empty, it corresponds
to $\str \one$, which vanishes. There is a single
correction to this picture, which can be traced back
to eqn.~\eq{eq:SuperSplitting-A}~\cite{aprop}: if 
both ends of an undecorated $\dd^{\A\A}$ kernel
attach to the same $\str \A \A$ factor, then we supplement
the usual group theory factor with an additional -2.

In the unbroken phase, this actually causes a class of
diagram which would na\"ively vanish, to survive. These
diagrams can possess any number of supertraces;  the case in
which a
double supertrace is split into three supertraces 
is shown in \fig{fig:SuperSplittingCorrection}.
\begin{center}
\begin{figure}[h]
	\[
		\ensuremath{\begin{array}{c}\input{pstex/SuperSplitCorr-Ex.pstex_t} \end{array}}
	\]
\caption{A term which survives, despite na\"ively vanishing by group theory considerations.}
\label{fig:SuperSplittingCorrection}
\end{figure}
\end{center}

Now, the expected group theory factor associated with this diagram 
arises from the two empty
circuits in the padlock structure. These yield $(\str \one)^2 = 0$. 
However, according
to the above prescription, the group theory factor of this diagram 
is supplemented by $-2$,
causing it to survive.

\subsubsection{Kernels which bite their tails}
\label{sec:WBT}

As promised at the end of \sec{Review}
we give, in \fig{fig:WBT}, an explicit example of a portion of a
diagram, potentially generated by the flow equation, which 
is not properly UV regularised; such diagrams, in which the
kernel bites its tail are, as discussed already, to be discarded.
\bcf[h]
	\[
		\ensuremath{\begin{array}{c}\begin{picture}(0,0)%
\includegraphics{pstex/WBT-Ex.pstex}%
\end{picture}%
\setlength{\unitlength}{3947sp}%
\begingroup\makeatletter\ifx\SetFigFont\undefined%
\gdef\SetFigFont#1#2#3#4#5{%
  \reset@font\fontsize{#1}{#2pt}%
  \fontfamily{#3}\fontseries{#4}\fontshape{#5}%
  \selectfont}%
\fi\endgroup%
\begin{picture}(584,877)(2067,-1597)
\put(2651,-1012){\makebox(0,0)[lb]{\smash{\SetFigFont{8}{9.6}{\rmdefault}{\mddefault}{\updefault}{\color[rgb]{0,0,0}$k$}%
}}}
\end{picture}
 \end{array}}
	\]
\caption{An improperly UV regularised diagram.}
\label{fig:WBT}
\ecf

We take the explicitly shown fields, including the differentiated 
ones, to be components of supergauge fields, and the loop momentum 
to be $k$, as shown. The problem is that in the special case
exemplified by \fig{fig:WBT}, 
the integrand behaves at best as $1/k^3$ for large $k$, 
independent of the way the uncovariantised kernel decays 
with large momenta (and with a typically non-polynomial 
coefficient); this all follows from the Ward identities, 
\cf Appendix A of~\cite{ymi}. Although Lorentz invariance 
will then ensure that such a loop integral is only 
logarithmically divergent in $D=4$ dimensions, the only 
escape from this divergence is by 
imposing~\eq{noATailBiting} or by
cancellation from some 
other contribution. In the latter case, however, whenever the inner and outer 
supertraces are decorated with specific flavours, the $SU(N|N)$ 
group theory properties are insufficient to enforce a cancellation. 
(For example we can take the inner supertrace to trap two $A^1$ 
fields, and assume the outer supertrace is fixed to be in the 
$A^1$ sector by decorations at the base of the kernel or on the 
vertex itself---where as a result these extra decorations 
do not dampen the large $k$ behaviour.)

\subsubsection{The broken phase flow equation}
\label{sec:BrokenPhase}

To work in the broken phase we now decorate our diagrams
with the dynamical broken phase fields and $\sigma$, rather
than $\A$ and $\C$~\cite{aprop}. Functional derivatives
are performed \wrt\ these dynamical fields, leading to
corrections. These are easily computed by noting that 
(with a slight abuse of notation) \eg
\[
	\fder{}{\tilde{A^1}} = \sigma_+ \fder{}{\A} \sigma_+.
\]
The effect of differentiating \wrt\ partial, rather than full,
supermatrices is, therefore, simply insertions of $\sigma_\pm$. On a circuit
containing other fields, these insertions have no
effect, as they correspond to inserting unity in the appropriate
subspace. On a circuit devoid of any fields, though, they have a
profound effect: whereas, in the unbroken phase, such a circuit 
yields (up to the correction discussed earlier)
$\str \one = 0$ now, however, such a circuit can produce
$\str \sigma_\pm = \pm N$.
(Note that the correction to exact supersplitting discussed earlier now
manifests itself wherever there is an attachment of an undecorated $\dd^{AA}$
kernel to a separate $\str AA$.)

Let us illustrate the use of the diagrammatic flow equation
by considering the flow of the $S^{1\, 1}_{\mu \nu}(p)$ vertex.
In \fig{fig:ClassicalFlow-A1A1}
we focus on the classical part of the flow equation.
Filled circles represent $A$s and, if they are
tagged with a `1', then they are restricted to
the $A^1$ sector. The symbol $\DiagSigma$ represents
insertions of $\sigma$ and the symbol \ReplaceAwSigma\
tells us that an $A$ has been differentiated and
replaced with a $\sigma$.
We have suppressed the Lorentz
indices of the decorative fields. Note also that we
have attached our false kernel to instances of $\sigma$,
which do not carry a position argument. If the $\sigma$
has replaced an $A$, then the position argument of
the $A$ is the information carried by the appropriate end
of the false kernel. However, the other end may attach to
a $\sigma$ with no associated field to provide a position
argument. In this case, 
the false
kernel does not carry a position space $\delta$-function,
but just serves to remind us how the diagrams of
which they comprise a part were formed.
The ellipsis represents the un-drawn diagrams spawned
by the quantum term. 
\begin{center}
\begin{figure}[h]
	\[	
	\begin{array}{c}
	\vspace{1ex}
		\begin{array}{ccc}
		\vspace{0.1in}
									&	& \LD{Dumbbell-A1A1-Bars}{}
		\\
			-\flow \ensuremath{\begin{array}{c}\input{pstex/Vertex-A1A1.pstex_t} \end{array}} & = & \ensuremath{\begin{array}{c}\input{pstex/Dumbbell-A1A1-Bars.pstex_t} \end{array}}
		\end{array}
	\\
		\ds
		+
		\frac{1}{16N}
		\left[
			\begin{array}{cccccc}
			\vspace{0.1in}
					& \LD{Dumbbell-A1A1-Bars-New-A}	&	&\LD{Bottle-A1A1-Unity}	&	&\LD{Bottle-A1A1-Unity-B}
			\\
				8	&\ensuremath{\begin{array}{c}\input{pstex/Dumbbell-A1A1-Bars-New-A.pstex_t} \end{array}}	&-4	&\ensuremath{\begin{array}{c}\input{pstex/Bottle-A1A1-Unity.pstex_t} \end{array}}	&-4	&\ensuremath{\begin{array}{c}\input{pstex/Bottle-A1A1-Unity-B.pstex_t} \end{array}}
			\end{array}
		\right]
		+ \cdots
	\end{array}
	\]
\caption{Classical part of the flow of a vertex decorated
by two $A^1$s.}
\label{fig:ClassicalFlow-A1A1}
\end{figure}
\end{center}

There are a number of important points to make about the
diagrams of \fig{fig:ClassicalFlow-A1A1}.
Diagram~\ref{Dumbbell-A1A1-Bars} is the only
explicitly drawn diagram for which the kernel
attaches to both vertices directly, as opposed
to via $\P$. The factor of $1/2$ associated
with the original $a_0$ terms has been killed by the factor
of two coming from summing over the two possible locations of
the fields which decorate the dumbbell.
Notice that the differentiated fields are in the $A$-sector
only. Generically, the differentiated fields can also
be $F$s and $C$s. However, both of these
are forbidden in this case by our choice of
external fields: in the former case because $S^{AF}$
vertices do not exist and, in the latter case, because the
$S^{AC}$ vertex vanishes by \CCI.

Whilst we are free to choose the external
fields to be $A^1$s, as opposed to $\tilde{A}^1$s,
the internal fields necessarily possess an $\A^0$ component.
Indeed, for internal fields to properly label the $\dd^{AA}$
kernel, we should take them to be full $A$s, though noting
that they are effectively reduced to $\tilde{A}^1$s by the
choice of external fields. This feels somewhat unnatural
and we will shortly adopt a more pleasing prescription. First,
though, we will analyse the remaining diagrams of 
\fig{fig:ClassicalFlow-A1A1}.

Diagram~\ref{Dumbbell-A1A1-Bars-New-A}
is a combination of four diagrams produced by the 
contributions to the flow equation involving $\P$,
corresponding to the four ends of the kernel on which the
$\sigma$ can sit. Since these diagrams are equivalent,
we can add them. Summing over the two locations of the decorative fields
gives an overall relative factor of eight, as shown. The kernel is  
$\dot{\Delta}^{AA}_\sigma$. That it must be this, and not $\dot{\Delta}^{AA}$,
can be deduced by the presence of components of \DoubleCirc\ at one of its ends.
The final point to note about this diagram is that the extra supertrace,
attached to the kernel via a false kernel,
is just $\str \sigma = 2N$. Thus we see that this diagram has
the same overall factor as the first diagram.

Diagrams~\ref{Bottle-A1A1-Unity} and~\ref{Bottle-A1A1-Unity-B}
both vanish. The uppermost vertices have an associated
supertrace structure $\str \sigma^2 = 0$. 
In fact, we need never have drawn these diagrams. 
In the former case, the $S$ vertex
coefficient function is $S^{A \sigma}$; single $A$ vertices 
vanish by both \CC\ and Lorentz invariance (similarly in 
the latter case, but with $S \leftrightarrow \Sigma_g$). 
In anticipation of what is to come, though, we note
that multiple supertrace
terms can have separate $\str A \sigma$ factors. This will
play a key \role\ in what follows, since $\str A \sigma$
is none other than $2N \A^0$.

Thus, to recapitulate, only the first two diagrams survive. 
They come with the same relative factor, have exactly
the same vertices and so can be combined. The result
will be that the two vertices are now joined by the
sum of $\dot{\Delta}^{AA}$ and $\dot{\Delta}^{AA}_\sigma$.
It seems natural to define
\be
	\dot{\Delta}^{A^1A^1} \equiv \dot{\Delta}^{11} \equiv \dot{\Delta}^{AA} + \dot{\Delta}^{AA}_\sigma.
\label{eq:Wine-A1A1:Defn}
\ee

We can do an analogous analysis in the $A^2$ sector. The difference
here is that the embedded $\sigma$ of the analogue of 
diagram~\ref{Dumbbell-A1A1-Bars-New-A}
gives rise to a minus sign since, whilst $\str A^1 A^1 \sigma = \str A^1 A^1$,
$\str A^2 A^2 \sigma = -\str A^2 A^2$. In this case, we are led
to the definition
\be
	\dot{\Delta}^{A^2A^2} \equiv \dot{\Delta}^{22} \equiv \dot{\Delta}^{AA} - \dot{\Delta}^{AA}_\sigma.
\label{eq:Wine-A2A2:Defn}
\ee

However, neither~\eq{eq:Wine-A1A1:Defn} nor~\eq{eq:Wine-A2A2:Defn}
work quite as we would like: the kernels $\dd^{AA}$ and 
$\dd^{AA}_\sigma$
do not attach to just $A^1$ or $A^2$, as the labels of $\dd^{11}$ 
and $\dd^{22}$ seem to imply; 
we know that there is also attachment to $\A^0$.

Nonetheless, both the physics of the situation and the diagrammatics 
seem to be guiding us to a formalism
where we work with $A^1$s and $A^2$s---the so-called
$A^{1,2}$ basis. It turns out that the most
efficient way to proceed is to follow this lead. Recalling
eqn.~\eq{dumbdef},
we now split up the first term into derivatives \wrt\ $A^1$ and $A^2$:
\be
	\fder{}{A_\mu} = 2 \sigma_+ \tau_1^{a} \fder{}{A_{\mu a}^1}  - 2 \sigma_- \tau_2^a \fder{}{A_{\mu a}^2}  + \frac{\sigma}{2N} \fder{}{\A_\mu^0}.
\label{eq:SF-deriv-cmpts}
\ee
We now exploit no-$\A^0$ symmetry and so use the prescription that, since all 
\emph{complete} functionals
are independent of $\A^0$, we can take $\delta/ \delta \A^0$ not to act.

It is important to realise, though, that this does not mean that
we can simply dispense with $\A^0$ altogether. Were we to attempt
to do this, we could always regenerate it, via gauge transformations.
Thus, the only way to ensure that various actions are invariant
under no-$\A^0$ symmetry is to enforce constraints between various
vertices: though $\A^0$ is thus removed in one sense, its effects
are nonetheless present. We have seen an example of this
already in \sec{New}. The vertices $S^{AA\sigma}$
and $S^{A\sigma,A\sigma}$ are both individually associated with
$\A^0$ components. However, the set of
relationships of which  
\be
	S^{AA\sigma} + 2N S^{A\sigma,A\sigma} = 0 
\label{eq:No-A0-Ex}
\ee
is an example
ensure that the action as a whole is independent of $\A^0$~\cite{Thesis}.

Let us consider the effects of taking $\A^0$ not to act
in more detail. First, we note that we can no longer
use exact supersowing / supersplitting: there will
be additional corrections, which we will compute in
a moment. Temporarily neglecting this, let us see
what we have gained.

In diagrams~\ref{Dumbbell-A1A1-Bars} 
and~\ref{Dumbbell-A1A1-Bars-New-A}, the internal fields can
now be taken to be $A^1$s. Upon combining the diagrams,
and utilising~\eq{eq:Wine-A1A1:Defn}, the internal
fields now naturally label the kernel (similarly in the $A^2$
sector).

What of diagrams~\ref{Bottle-A1A1-Unity} and~\ref{Bottle-A1A1-Unity-B}?
 We have already noted that they vanish
anyway, but this is only an accidental feature of the particular
vertex whose flow we chose to compute. The real point is
that they both contain a $\str \sigma A = 2N \A^0$, which is
attached to a kernel. In our new picture, we simply never draw such
diagrams: we are not interested
in external $\A^0$s and need not consider internal
ones because we do not allow $\delta / \delta \A^0$ to act.

Moreover, when working in the $A^{1,2}$ basis,  all diagrams involving
\ReplaceAwSigma vanish. This is because the differentiated field
is now restricted to being an $A^1$ or $A^2$. However,
\[
	\str \sigma \fder{}{A} = \tr \sigma_+ \fder{}{A^1} - \tr \sigma_- \fder{}{A^2} \equiv 0.
\]
Had we still been using exact supersowing and supersplitting,
$\str \sigma \delta / \delta A$ terms could have survived~\cite{Thesis}.
(If supersowing and supersplitting are exact, we can 
think of the constrained superfield 
$\A$ behaving effectively like a `full' superfield $\A^e$~\cite{aprop}.
Now $\str \sigma \delta / \delta A^e \neq 0$.)

To compute the corrections arising from
the abandonment of exact supersowing / supersplitting,
consider 
attachment of a kernel via $\delta / \delta A^1$, as shown in
\fig{fig:Attachement-A1}. 
\bcf[h]
	\[		\ensuremath{\begin{array}{c}\input{pstex/Attach-A1.pstex_t} \end{array}} = 2 \ensuremath{\begin{array}{c}\input{pstex/Attach-A1-B.pstex_t} \end{array}}
	\]
\caption{Attachment of a kernel via $\delta / \delta A^1$.}
\label{fig:Attachement-A1}
\ecf

Note that, in the second diagram, we have retained the `1'
which previously labelled the $A^1$, to remind us that
the flavours of any fields which followed or preceded
the $A^1$ are restricted. 

Next, we use the completeness relation for $SU(N)$,
\begin{equation}
	2 \left( \tau_1^a \right)^i_{\ j} \left( \tau_1^a \right)^k_{\ l} = \delta^i_{\ l} \delta^k_{\ j} - \frac{1}{N} \delta^i_{\ j} \delta^k_{\ l},
\label{eq:SU(N)-Completeness}
\end{equation}
to obtain \fig{fig:Attachement-A1-B}, where attachments like those in
the first column of diagram on the right-hand side will be henceforth
referred to as direct.
\begin{center}
\begin{figure}[h]
	\begin{eqnarray*}
		\ensuremath{\begin{array}{c}\input{pstex/Attach-A1.pstex_t} \end{array}} 	& = 	& \ensuremath{\begin{array}{c}\input{pstex/Attach-A1-C.pstex_t} \end{array}} - \frac{1}{N} \ensuremath{\begin{array}{c}\input{pstex/Attach-A1-D.pstex_t} \end{array}}
	\\
						& \equiv& \ensuremath{\begin{array}{c}\input{pstex/Attach-A1-C.pstex_t} \end{array}} -\frac{1}{N} \ensuremath{\begin{array}{c}\input{pstex/Attach-A1-E.pstex_t} \end{array}}
	\end{eqnarray*}
\caption{A re-expression of \fig{fig:Attachement-A1}.}
\label{fig:Attachement-A1-B}
\end{figure}
\end{center}

There is a very similar expression for the $A^2$ sector. Now, however, $\sigma_+$s
are replaced with $\sigma_-$s and the sign of the $1/N$ contribution 
flips~\cite{Arnone:2001iy}.
We can also understand the 
sign flip heuristically because we are tying everything
back into supertraces and not traces; we recall that the supertrace 
yields the trace of the bottom block-diagonal of a supermatrix but picks
up a minus sign.

In the $B, D$ and $C$ sectors we do not get any $1/N$ 
attachment corrections. Derivatives \wrt\
the fields $B$ and $\bar{B}$,
can simply be written
\begin{eqnarray}
	\fder{}{B}		& = & \sigma_- \fder{}{\A}\sigma_+,
\label{eq:SigmaAlg-B}
\\
	\fder{}{\bar{B}}& = & \sigma_+ \fder{}{\A}\sigma_-,
\label{eq:SigmaAlg-Bbar}
\end{eqnarray}
and so derivatives \wrt\ these partial superfields just yield
insertions of $\sigma_\pm$.

In the $C$-sector, $\C^0$ does not play a privileged \role~\cite{Arnone:2001iy} 
and so has not been factored out of the definition of $C^1$ and $C^2$
(see eqn.~\eq{defC}). Hence, derivatives \wrt\ the components
of $\C$ simply yield insertions of $\sigma_\pm$.

Let us now reconsider the classical part of the flow of
a vertex decorated by two fields. These fields can be any
of $A^1, \ A^2, \ C^1, \ C^2, \ F$ and $\bar{F}$ (though
certain choices \eg $A^1A^2$ correspond to a vanishing vertex).
The diagrammatics is shown in \fig{fig:ClassicalFlow-TP},
where we have neglected not only the quantum terms but also
any diagrams in which the kernel is decorated.\footnote{
Such diagrams only exist in the $C$-sector, in this case, since
this is the only sector for which one-point (seed action) vertices
exist.}
\begin{center}
\begin{figure}[h]
	\[
	\begin{array}{cccc}
	\vspace{0.1in}
								&	& \LD{Dumbbell-TP} &
	\\
		-\flow\ensuremath{\begin{array}{c}\input{pstex/Vertex-TP.pstex_t} \end{array}} 	&=	& \ensuremath{\begin{array}{c}\input{pstex/Dumbbell-TP.pstex_t} \end{array}} &+ \cdots
	\end{array}
	\]
\caption{Classical part of the flow of a two-point vertex decorated
by the fields $A^1, \ A^2, \ C^1, \ C^2, \ F$ and $\bar{F}$.}
\label{fig:ClassicalFlow-TP}
\end{figure}
\end{center}

The differentiated fields now label the kernels, where
we take $\dd^{C^1C^1} = \dd^{C^2C^2} = \dd^{CC}$
and $\dd^{\bar{F}F} = \dd^{F \bar{F}}$. The only subtlety
comes in the $A^{1,2}$ sectors, where we know that
there are corrections, which have been implicitly absorbed
into the Feynman rules. We will be more specific about
the corrections in this case---\ie where the kernel is 
undecorated.

From \fig{fig:Attachement-A1-B} we obtain 
the relation of \fig{fig:Attachment-A1A1}.
\begin{center}
\begin{figure}[h]
	\[
		\ensuremath{\begin{array}{c}\input{pstex/Attach-A1A1.pstex_t} \end{array}} =	 \ensuremath{\begin{array}{c}\input{pstex/Attach-A1A1-A.pstex_t} \end{array}} -\frac{1}{N} \ensuremath{\begin{array}{c}\input{pstex/Attach-A1A1-B.pstex_t} \end{array}} -\frac{1}{N} \ensuremath{\begin{array}{c}\input{pstex/Attach-A1A1-C.pstex_t} \end{array}} +\frac{1}{N^2} \ensuremath{\begin{array}{c}\input{pstex/Attach-A1A1-D.pstex_t} \end{array}}
	\]
\caption{The attachment corrections for
 $\dd^{11}$.}
\label{fig:Attachment-A1A1}
\end{figure}
\end{center}

The expression in \fig{fig:Attachment-A1A1} now simplifies. The loop in
the middle of the final diagram is decorated only by $\sigma_+^2$ and so
yields $\str \sigma_+ = N$. This diagram then cancels either of those
with factor $-1/N$. We now redraw the remaining diagram with factor $-1/N$,
as shown in \fig{fig:Attachment-A1s,A2s}, together
with a similar expression in the $A^2$ sector.
\begin{center}
\begin{figure}[h]
	\begin{eqnarray}
		\ensuremath{\begin{array}{c}\input{pstex/Attach-A1A1.pstex_t} \end{array}} &=	& \ensuremath{\begin{array}{c}\input{pstex/Attach-A1A1-A.pstex_t} \end{array}} -\frac{1}{N} \ensuremath{\begin{array}{c}\input{pstex/Attach-A1-A1-N.pstex_t} \end{array}}
	\label{eq:Attach-A1A1}
	\\[2ex]
		\ensuremath{\begin{array}{c}\input{pstex/Attach-A2A2.pstex_t} \end{array}} &=	& \ensuremath{\begin{array}{c}\input{pstex/Attach-A2A2-A.pstex_t} \end{array}} +\frac{1}{N} \ensuremath{\begin{array}{c}\input{pstex/Attach-A2-A2-N.pstex_t} \end{array}}
	\label{eq:Attach-A2A2}
	\end{eqnarray}	
\caption{The attachment corrections for
$\dd^{11}$ and $\dd^{22}$.}
\label{fig:Attachment-A1s,A2s}
\end{figure}
\end{center}

Then meaning of the double dotted lines and the associated field
hiding behind the line of the supertrace should be clear: 
the double dotted lines stand for $\dd^{A^i A^i}$, and the sector of the
associated fields tells us whether $i=1$ or $2$. 

Returning now to diagram~\ref{Dumbbell-TP} (\fig{fig:ClassicalFlow-TP})
we should interpret the
kernel when the internal fields are in the $A^1$ or $A^2$ sectors
according to~\eqs{eq:Attach-A1A1}{eq:Attach-A2A2}.

In our analysis of the terms spawned by the classical part of the flow
equation, we have so far restricted ourselves to diagrams in which the
kernel is undecorated and for which there are no insertions of the dynamical
components of $\C$ via $\P$ (recall that such 
insertions occur as in  \fig{fig:DoubleCircle}).

Let us first relax just the former restriction and analyse
the mapping into the $A^1$, $A^2$ basis~\cite{Thesis}.  
The simplest case to deal with is where the 
decorations of the covariantised
kernels $\{\dd^{\A\A}\}$ and $\{\dd_\sigma^{\A\A}\}$
are, on each side \emph{independently}, net bosonic. In this
scenario, we are just mapped into the $A^1$, $A^2$ basis,
as before. 

The next case to examine is where the decorations on one side
of the kernel are net bosonic but those on the other side are net
(anti) fermionic. This immediately tells us that one of the functional
derivatives sitting at the end of the kernel must be 
(anti) fermionic. Now, as before, we would like to pair up diagrams
with a $\{\dd^{\A\A}\}$ vertex with those with a 
$\{\dot{\Delta}_\sigma^{\A\A}\}$ vertex.
However, there is a subtlety: $\sigma$ anticommutes with net fermionic
structures. Thus, of the four terms generated
 with a $\sigma$ at the end of one side of the kernel
attached to a separate $\str \sigma$, two cancel.
Hence, the vertex from $\{\dot{\Delta}_\sigma^{\A\A}\}$
has a factor of half relative to the vertex from
$\{\dd^{\A\A}\}$. 

We can still choose a prescription to map us into the $A^1$, $A^2$ basis
by absorbing this factor into our definition of 
\eg $\dd^{1B,F \cdots,\cdots}$, where both ellipses denote net bosonic decorations.\footnote{
Note that the $B$ denotes a functional derivative \wrt\
$B$; since this removes a fermion, the kernel decorations
must be net fermionic.}

The final case to examine is where the decorations on 
one side of the kernel are net fermionic whilst those on the
other side are net anti-fermionic.\footnote{If the decorations
on both sides are (anti) fermionic, then the vertex belongs
to $\{\dd^{F\bar{F}} \}$.}
The functional 
derivatives must both be bosonic and in 
separate sub-sectors.\footnote{
It is straightforward to check that one cannot construct a legal
kernel of this type for which the functional derivatives
are fermionic and anti-fermionic.
}
There are no 
surviving contributions involving $\{\dd_\sigma^{\A\A}\}$
vertices.
Nonetheless, we can still define an object $\{\dd^{A^1 A^2}\}$
though we note that it must have net fermionic decorations on both sides.

Let us now relax the second restriction above and so allow insertions
of the dynamical components of $\C$ at the ends of kernels, via $\P$.
Such diagrams occur only with $\dot{\Delta}_\sigma^{\A\A}$ kernels
and
there is no natural way of mapping such terms 
into the $A^1$, $A^2$ basis. There
is, of course, nothing to stop us performing functional derivatives
\wrt\ $A^1$ and $A^2$ only---indeed, this is what we will do. However,
the  differentiated fields do not now label the kernels in a natural
way. 

In fact, as we will see (see also~\cite{Thesis,oliver1,oliver2}), 
considerations such as this
ultimately do not bother us: we will find that, in our calculation
of $\beta$-function coefficients, all instances of these awkward terms 
cancel, amongst themselves. Hence, our strategy is to lump
all the unpleasant terms into our diagrammatic rules, so that
they essentially become hidden. We can, at any stage, unpack them
if necessary but, generally speaking, we will find that we do not
need to do so (see~\cite{Thesis} for an explicit example of this).

This completes the
analysis of the diagrammatics we have performed for the new flow equation.
We have not explicitly looked at the quantum term, but there are no new
considerations in this case. Thus, we can summarise the prescription
that we use in the broken phase.
\begin{enumerate}
	\item	All decorative fields are instances of
			$A^1$, $A^2$, $F$, $\bar{F}$, $C^1$, $C^2$ and $\sigma$;
			$\A^0$ is excluded.

	\item	Differentiation is \wrt\ all dynamical fields, above, where:
			\begin{enumerate}
				\item	differentiation \wrt\ $A^1$ or $A^2$ leads to 
						attachment corrections of the type
						shown in \fig{fig:Attachement-A1-B};
				
				\item	differentiation \wrt\ all other fields just involves
						insertions of $\sigma_\pm$;

				\item	diagrams involving $\str \sigma \delta / \delta A^{1,2}$
						vanish, identically.
			\end{enumerate}

	\item	Full diagrams without any insertions of the components of \DoubleCirc\
			at the ends of the kernels are naturally written in terms of
			the above fields and their corresponding kernels \ie $A^1$s
			attach to a (decorated) $\dd^{11}$ etc.
		
	\item	Full diagrams with insertions of the components of  \DoubleCirc\
			are restricted
			to those for which at least one component is not a $\sigma$
			(in the case that both components are $\sigma$s, these
			terms have already been used to map us into the $A^{1,2}$ basis). These
			diagrams involve the kernel $\dot{\Delta}^{AA}_\sigma$ but, for convenience,
			are packaged together with the decorated kernels of the previous item.
\end{enumerate}

\subsection{The new diagrammatics}
\label{sec:NewDiags}

\subsubsection{Construction}

The work of the previous section now guides us to
a more compact and intuitive diagrammatics, which
is considerably easier to deal with. By packaging up
the remaining $\dd^{AA}_\sigma$ kernels
with decorated instances of the other kernels, 
we have taken a step in the right
direction. In anticipation that these compact, packaged
objects cancel in their entirety when we perform actual
calculations, it clearly makes sense
to bundle together kernels of a different flavour. However,
by doing this, we have started to
combine diagrams with differing supertrace structures.
In this section, we extend this to its natural 
conclusion.

The basic idea is that, rather than considering diagrams
with a specific supertrace structure, we instead sum 
over all legal supertrace structures, consistent with
the decorative fields. Thus, let us suppose
that we wish to compute the flow of all vertices
which can be decorated by the set of fields $\{f\}$.
The new flow equation takes a very simple,
intuitive form, as shown in \fig{fig:New-Diags-1}.
\begin{center}
\begin{figure}[h]
	\[
	-\flow 
	\dec{
		\ensuremath{\begin{array}{c}\input{pstex/Vertex-S.pstex_t} \end{array}}
	}{\{f\}}
	=
	\frac{1}{2}
	\dec{
		\ensuremath{\begin{array}{c}\input{pstex/Dumbbell-S-Sigma_g.pstex_t} \end{array}} - \ensuremath{\begin{array}{c}\input{pstex/Vertex-Sigma_g.pstex_t} \end{array}}
	}{\{f\}}
	\]
\caption{The diagrammatic form of the flow equation, when we treat
single and multiple supertrace terms together.}
\label{fig:New-Diags-1}
\end{figure}
\end{center}

Let us now analyse each of the elements of \fig{fig:New-Diags-1} in turn.
On the \lhs, we have the set of vertices whose flow we are computing.
This set comprises all cyclically independent arrangements of the fields $\{f\}$,
over all possible (legal) supertrace structures.
When we specify
the fields $\{f\}$, we use a different notation from before. As an example,
consider $\{f\} = \{A^1_\mu, A^1_\nu, C^1\}$, shown in \fig{fig:NewVertices-Ex}.
Note that we have not drawn any vertices comprising a supertrace decorated only by an $A^1$,
since these vanish.
\begin{center}
\begin{figure}[h]
	\[
	\dec{
		\ensuremath{\begin{array}{c}\input{pstex/Vertex-S.pstex_t} \end{array}}
	}{A^1_\mu A^1_\nu C^1}
	\equiv
	\ensuremath{\begin{array}{c}\input{pstex/Vertex-S-A1A1C1.pstex_t} \end{array}}
	=
	\ensuremath{\begin{array}{c}\input{pstex/Vertex-S-A1_muA1_nuC1.pstex_t} \end{array}} + \ensuremath{\begin{array}{c}\input{pstex/Vertex-S-A1_nuA1_muC1.pstex_t} \end{array}} + \ensuremath{\begin{array}{c}\input{pstex/Vertex-S-A1_nuA1_mu-C1.pstex_t} \end{array}}
	\]
\caption{A new style vertex decorated by two $A^1$s and a $C^1$.}
\label{fig:NewVertices-Ex}
\end{figure}
\end{center}

It is apparent that we denote $A$s by wiggly lines and $C$s by dashed lines. 
A wildcard field will be denoted by a solid line.

Notice how, in the new style diagram, we explicitly indicate the sub-sector of
all the fields. This is because there is no need for them to be on the same supertrace
and so, for example, there is nothing to prevent an $A^1A^1C^2$ vertex. In the old
notation, however, all fields on the same circle are on the same supertrace and
so, once we know the sub-sector of one field, the sub-sectors of the remaining fields
on the same circle follow, uniquely.

To symbolically represent the new vertices, we will somewhat loosely write \eg
$S_{\mu \nu}^{1 \, 1 C^{1}}$. If we need to emphasise that we are using the
new style diagrammatics, as opposed to the old style diagrammatics, then
we will write $S_{\ \, \mu \nu}^{\{ 1 \, 1 C^{1} \}}$, reminding us that the
fields are arranged in all cyclically independent ways
over all possible (legal) supertrace structures.

With these points in mind, let us return to \fig{fig:New-Diags-1}. 
The diagrams
on the \rhs\ both involve the structure \DummyKernel. This is a dummy kernel which
attaches, at either end, to dummy fields. 
The fields at the ends can be any of $A^1$, $A^2$, $C^1$,
$C^2$, $\bar{F}$ or $F$, so long as the corresponding diagram actually exists. 
The dummy
kernel can be decorated by any subset of the fields $\{f\}$ where, if a
pair of decorative fields are both components of $\C$ 
(and one of them is dynamical),
then we include the possibility that the kernel can be of the type 
$\dd^{\A\A}_\sigma$.
In this case, we note that there are implicit factors of $1/16N$.

The relationship between the new diagrammatics for the kernels and the 
old diagrammatics is straightforward, and is illustrated in 
\fig{fig:NewKernel-Ex}
for the case of a new-style kernel decorated by a single $A^1$.
\begin{center}
\begin{figure}[h]
	\[
	\dec{
		\ensuremath{\begin{array}{c}\input{pstex/DummyKernel.pstex_t} \end{array}}
	}{A^1}
	\equiv
	\ensuremath{\begin{array}{c}\input{pstex/DummyKernel-A1.pstex_t} \end{array}}
	=
	\ensuremath{\begin{array}{c}\input{pstex/DummyKernel-A1-R.pstex_t} \end{array}} + \ensuremath{\begin{array}{c}\input{pstex/DummyKernel-A1-L.pstex_t} \end{array}}
	\]
\caption{A new style (dummy) kernel decorated by a single $A^1$.}
\label{fig:NewKernel-Ex}
\end{figure}
\end{center}

Having described the new diagrammatics for vertices and kernels,
we are nearly ready to complete our interpretation of 
\fig{fig:New-Diags-1}.
Before we do so, however, we use our new notation to hide one further detail:
instances of $\sigma$. Instances of $\sigma$ correspond
either to insertions of $\pm \one$ in the relevant subspace
or to factors of $2N$. Thus, instances of $\sigma$
can be replaced by numerical factors accompanying kernels / vertices.
We need only remember that the multiple supertrace decorations
of kernels exist.
Such terms require two instances of $\C$ (at least one of
which we take to be dynamical). 
In the case that one of these fields is a $\sigma$,
we must remember that it is now hidden.

With these implicit instance of $\sigma$ in mind, the interpretation
of the \rhs\ of \fig{fig:New-Diags-1} is simple:
the decorative fields $\{f\}$ are distributed around the two diagrams
in all possible, independent ways.

Whilst we will generally use the
new diagrammatics thus described, from
now on, it is occasionally useful to
flip back to the old style mentality
of specifying the supertrace structure.
It turns out that, in this paper, we
only ever have recourse to do
this for single supertrace terms and
so introduce the notation `Fields as Shown'
or FAS. An example of this is illustrated
in \fig{fig:FAS}.
\begin{center}
\begin{figure}[h]
	\[
		\left[\ensuremath{\begin{array}{c}\input{pstex/Vertex-S-A1A1C1.pstex_t} \end{array}}\right]_\mathrm{FAS} \equiv \ensuremath{\begin{array}{c}\input{pstex/Vertex-S-A1_muA1_nuC1.pstex_t} \end{array}}
	\]
\caption{An example of the meaning of FAS.}
\label{fig:FAS}
\end{figure}
\end{center}

\subsubsection{Constraints in the $C$-sector} \label{sec:ND-II:App}

We conclude our description of the
new diagrammatics with an explicit illustration
of their use.
Recall from \sec{Review}
that, in order to ensure quantum corrections do not
shift the minimum of the Higgs potential from the classical
choice, $\sigma$, $\hat{S}$ is constrained. 
Assuming that these quantum corrections vanish means that
all Wilsonian effective action one-point $C$ vertices
vanish. We can thus use the flow equation to give us a
constraint equation,
as shown in \fig{fig:C-secConstraint}. Note that
the external fields can be in either the $1$ or $2$ sub-sector.
\begin{center}
\begin{figure}[h]
	\begin{eqnarray}
	-\flow \ensuremath{\begin{array}{c}\input{pstex/Vertex-S-C.pstex_t} \end{array}} = 0 	&=				& \frac{1}{2} \left[ \ensuremath{\begin{array}{c}\input{pstex/Dumbbell-S-C-Sigma_g.pstex_t} \end{array}} + \ensuremath{\begin{array}{c}\input{pstex/Dumbbell-S-Sigma_g-C.pstex_t} \end{array}} - \ensuremath{\begin{array}{c}\input{pstex/Sigma_g-C-DEP.pstex_t} \end{array}} - \ensuremath{\begin{array}{c}\input{pstex/Sigma_g-W-C.pstex_t} \end{array}} \right] 
\nonumber
\\[2ex]
								&\Rightarrow 	& -2 \ensuremath{\begin{array}{c}\input{pstex/Dumbbell-S-C-Sh.pstex_t} \end{array}} = \ensuremath{\begin{array}{c}\input{pstex/Sigma_g-C-DEP.pstex_t} \end{array}} + \ensuremath{\begin{array}{c}\input{pstex/Sigma_g-W-C.pstex_t} \end{array}}
\label{eq:NPC-sec-Constraint}
	\end{eqnarray}
\caption{The constraint arising from ensuring that
the position of the minimum of the Higgs potential is
unaffected by quantum corrections.}
\label{fig:C-secConstraint}
\end{figure}
\end{center}

To go from the first line to the second line, we have 
used~\eq{Sigma} and have
discarded all one-point, Wilsonian effective action vertices.
To satisfy eqn.~\eq{eq:NPC-sec-Constraint}, we tune the
one-point, seed action vertex in the first diagram, which is something
we are free to do.

\subsection{The weak coupling expansion}
\label{sec:WeakCoupling}

In preparation for the computation of perturbative
$\beta$ function coefficients, we examine the form 
the flow equation takes in the limit of weak coupling
and investigate some of its properties.

\subsubsection{The flow equation}

Following~\cite{ymi,aprop}, the action has the weak
coupling expansion
\begin{equation}
	S = \sum_{i=0}^\infty \left( g^2 \right)^{i-1} S_i = \frac{1}{g^2}S_0 + S_1 + \cdots,
\label{eq:WeakCouplingExpansion-Action}
\end{equation} 
where $S_0$ is the classical effective action and the $S_{i>0}$
the $i$th-loop corrections. The seed action has a similar expansion:
\begin{equation}
	\hat{S} = \sum_{i=0}^\infty  g^{2i}\hat{S}_i.
\label{eq:WeakCouplingExpansion-SeedAction}
\end{equation} 
Note that these definitions are consistent with $\Sigma_g = g^2 S -2\hat{S}$; 
identifying powers of $g$ in the flow equation, it is clear that
$S_i$ and $\hat{S}_i$ will always appear together.
With this in mind, we now define
\begin{equation}
	\Sigma_i = S_i - 2\hat{S}_i.
\label{eq:WeakCouplingSigma}
\end{equation}

The $\beta$ functions for $g$ and $g_2$ are
\begin{eqnarray}
	\flow \frac{1}{g^2} 	& = & -2 \sum_{i=1}^\infty \beta_i(\alpha) g^{2(i-1)} 
\label{eq:WCE:flow-g}
\\
	\flow \frac{1}{g^2_2} 	& = & -2 \sum_{i=1}^\infty \tilde{\beta}_i(1/\alpha) g^{2(i-1)}_2,
\label{eq:WCE:flow-g_2}
\end{eqnarray}
where the $\beta_i(\alpha)$ are determined through the
renormalisation condition~\eq{defg} and the 
$\tilde{\beta}_i(1/\alpha)$ are determined through~\eq{defg2}.
The coefficient $\beta_1 = - \tilde{\beta}_1$ is independent of
$\alpha$~\cite{aprop,Thesis}. For generic $\alpha$, we expect
the coefficient $\beta_2(\alpha)$ to disagree with the standard
value; as we will explicitly confirm in~\cite{mgierg2}, agreement
is reached for $\beta_2(0)$.\footnote{We note that whilst
we expect $\beta_2(0) = \tilde{\beta}_2(0)$, there is no
reason to generically expect $\tilde{\beta}_2(1/\alpha) = \beta_2(\alpha)$
since $g$ and $g_2$ are not treated symmetrically in the flow equation.}

Utilising eqns.~\eqs{eq:WCE:flow-g}{eq:WCE:flow-g_2}
and~\eq{eq:alpha-defn}, it is apparent
that $\flow \alpha$ has the following weak coupling expansion:
\begin{equation}
	\flow \alpha = \sum_{i=1}^{\infty} \gamma_i g^{2i},
\label{eq:flow-alpha}
\end{equation}
where
\[
	\gamma_i = -2 \alpha \left(\beta_i(\alpha) - \alpha^i \tilde{\beta}_i(1/\alpha) \right).
\]
Writing\footnote{We avoid writing $\partial / \partial \alpha$ as $\partial_\alpha$
to avoid confusion later, when we will have momentum derivatives which are written \eg
$\partial_\alpha^k$.}
\[
	\flow S = \flowConstAl S + \flow \alpha \ \pder{S}{\alpha},
\]
the weak coupling flow equations are given by
\begin{equation}
	\dot{S}_n = \sum_{r=1}^n \left[ 2(n-r-1) \beta_r S_{n-r} + \gamma_r \pder{S_{n-r}}{\alpha} \right]
		+ \sum_{r=0}^n a_0 \left[ S_{n-r}, \Sigma_r \right] -  a_1 \left[\Sigma_{n-1} \right],
\label{eq:WeakCouplingExpansion-NewFlow}
\end{equation}
where we have now changed notation slightly such
that $\dot{X}$ is redefined to mean $-\flowConstAl X$.
Incidentally, we note that the kernels $\dd^{ff}$ appearing in the 
flow equation are defined
according to this new definition of $\dot{X}$.
This is a choice we are free to make about the flow equation and do so, 
since it makes life easier.

The diagrammatics for the new weak coupling flow equation follow, directly.
However, we note that the classical term can be brought 
into a more symmetrical form. This follows from the invariance of
$a_0[S_{n-r}, \Sigma_r] + a_0[S_{r}, \Sigma_{n-r}]$ 
under $r \rightarrow n-r$. We exploit this by recasting
the classical term as follows:
\begin{eqnarray}
	a_0[\bar{S}_{n-r}, \bar{S}_r] 	& \equiv 	& a_0[S_{n-r}, S_r] - a_0[S_{n-r}, \hat{S}_r] - a_0[\hat{S}_{n-r}, S_r].
\label{eq:NFE:BarNotation}
\\
									& =			& 	\left\{
														\begin{array}{cc}
															\frac{1}{2} \left(a_0[S_{n-r}, \Sigma_r] + a_0[S_{r}, \Sigma_{n-r}]\right) 	& n-r \neq r
														\\
															a_0[S_r, \Sigma_r]														& n-r =r.
														\end{array}
													\right.
\nonumber
\end{eqnarray}

Hence, we can rewrite the flow equation in the following form:
\begin{equation}
	\dot{S}_n = \sum_{r=1}^n \left[ 2(n-r-1) \beta_r S_{n-r} + \gamma_r \pder{S_{n-r}}{\alpha} \right]
		+ \sum_{r=0}^n a_0 \left[\bar{S}_{n-r}, \bar{S}_r \right] -  a_1 \left[\Sigma_{n-1} \right].
\label{eq:WeakCouplingExpansion-NewFlow-B}
\end{equation}
The diagrammatic version is shown in \fig{fig:NFE:NewDiagFE}.
A vertex whose argument is a letter, say $n$, represents
$S_n$. We define  $n_r := n-r$ and $n_\pm := n \pm 1$.
\begin{center}
\begin{figure}[h]
	\[
	\dec{
		\ensuremath{\begin{array}{c}\input{pstex/Vertex-n-LdL.pstex_t} \end{array}} 
	}{\{f\}}
	= 
	\dec{
		2 \sum_{r=1}^n \left[\left(n_r -1 \right) \beta_r +\gamma_r \pder{}{\alpha} \right]\ensuremath{\begin{array}{c}\input{pstex/Vertex-n_r-B.pstex_t} \end{array}} 
		+ \frac{1}{2} \sum_{r=0}^n \ensuremath{\begin{array}{c}\input{pstex/Dumbbell-n_r-r.pstex_t} \end{array}} -\frac{1}{2} \ensuremath{\begin{array}{c}\input{pstex/Vertex-Sigma_n_-B.pstex_t} \end{array}}
	}{\{f\}}
	\]
\caption{The new diagrammatic form for the weak coupling flow equation.}
\label{fig:NFE:NewDiagFE}
\end{figure}
\end{center}

Terms like the one on the \lhs, in which the entire diagram is struck by
$\flowConstAl$, are referred to as $\Lambda$-derivative terms. On the \rhs,
in addition to the usual classical and quantum terms, we have 
what we call the $\beta$ and $\alpha$ terms.

\subsubsection{The effective propagator relation}

The tree level flow equations are
obtained by specialising eqn.~\eq{eq:WeakCouplingExpansion-NewFlow}
or~\eq{eq:WeakCouplingExpansion-NewFlow-B} to $n=0$:
\begin{equation}
	\dot{S}_0 = a_0[S_0, \Sigma_0].
\label{eq:TreeLevelFlow}
\end{equation}
We now further specialise, to consider the flow of all two-point vertices,
as shown in \fig{fig:TLTP-flow}. Recall that the solid
lines represent dummy fields, which we choose to be instances
of $A^1$, $A^2$, $C^1$, $C^2$, $\bar{F}$ and $F$.
\begin{center}
\begin{figure}[h]
	\begin{equation}
		\ensuremath{\begin{array}{c}\input{pstex/Vertex-TLTP-LdL.pstex_t} \end{array}} = \ensuremath{\begin{array}{c}\input{pstex/Dumbbell-S_0-Sigma_0.pstex_t} \end{array}}
	\label{eq:TLTP-flow}
	\end{equation}
\caption{Flow of all possible two-point, tree level vertices.}
\label{fig:TLTP-flow}
\end{figure}
\end{center}

Eqn.~\eq{eq:TLTP-flow} is analogous to eqn.~\eq{twoflo},
but there are a number of things to note. First,
there is no possibility
of embedding components of $\C$ at the ends  of the kernels and
so the $\dd^{AA}_\sigma$ kernel does not appear.\footnote{Up to
instances which have been used to map is into the $A^{1,2}$ basis,
in the first place.}
Secondly, diagrams containing one-point, tree level vertices 
have not been drawn, since these vertices do not exist in any sector,
for either action.\footnote{Recall that one-point, seed action vertices
exist only from the one-loop level.} Lastly, if we work in the
$C$-sector, then each of the vertices can possess more than one
supertrace. We can and do consistently set~\cite{Thesis}
\[
	\hS^{\ C^{1,2},C^{1,2}}_0(p), \ S^{\ C^{1,2},C^{1,2}}_0(p) = 0.
\]

As described in \sec{Review}, our strategy is now to
set the two-point, tree level, seed action vertices
equal to their Wilsonian effective action counterparts
and integrate up, choosing the integration constants
appropriately~\cite{Thesis}. Aided by the Ward identities, it is 
straightforward to obtain the effective propagator relation~\cite{aprop,Thesis}:
\begin{equation}
		S_{0 MS}^{\ \, X \, Y}(p) \EP{Y Z}{SN}(p) = \delta_{MN} - p'_M p_N
\label{eq:EffPropReln}
\end{equation}
where the fields $X,Z$ are any broken phase fields, the field
$Y$ is summed over and
we identify the components of the \rhs\ according to table~\ref{tab:NFE:k,k'}.
(Note that the field $Z$ must, in fact, be the same as $Y$, since `mixed'
effective propagators do not exist.)
\renewcommand{\arraystretch}{1.5}
\begin{center}
\begin{table}[h]
	\[
	\begin{array}{c|ccc}
					& \delta_{MN}		& p_M'							& p_N
	\\ \hline 
		F,\bar{F}	& \delta_{MN}		& (f_p p_\mu / \Lambda^2, g_p)	& (p_\nu, 2)
	\\
		A^i			& \delta_{\mu \nu}	& p_\mu / p^2					& p_\nu
	\\
		C^i			& \one				& \mbox{---}					& \mbox{---}
	\end{array}
	\]
\caption{Prescription for interpreting eqn.~\eq{eq:EffPropReln}.}
\label{tab:NFE:k,k'}
\end{table}
\end{center}
\renewcommand{\arraystretch}{1}

The functions $f(k^2/\Lambda^2)$ and $g(k^2/\Lambda^2)$ 
need never be exactly determined (though for an explicit
algebraic realisation for arbitrary $\alpha$ see~\cite{Thesis});
rather, they must satisfy general constraints enforced
by the requirements of
proper UV regularisation of the physical $SU(N)$ theory
and gauge invariance. We will see the effect induced by the
latter shortly.

The object  $p'_M p_N$ is, of course, a gauge remainder, 
which we represent diagrammatically by \FullGR;
the constituent components are furnished with the following
diagrammatic representations:
\begin{eqnarray}
	\ensuremath{\begin{array}{c}\input{pstex/GaugeRemainder-ppr.pstex_t} \end{array}} & \equiv & p'_M,
\label{eq:NFE:Diag:p'}
\\[1ex]
	\ensuremath{\begin{array}{c}\input{pstex/GaugeRemainder-p.pstex_t} \end{array}} & \equiv & p_N.
\label{eq:NFE:Diag:p}
\end{eqnarray}

\subsubsection{Diagrammatic Identities}
\label{sec:DiagrammaticIdentities}

We conclude this section with a set of diagrammatic
identities. First, denoting the effective 
propagator by a long, solid line,
we cast
the effective propagator relation~\eq{eq:EffPropReln}
in a particularly appealing diagrammatic form. Note
that we have attached the effective propagator, which only
ever appears as an internal line, to an arbitrary structure.
\begin{equation}
	\ensuremath{\begin{array}{c}\input{pstex/EffPropReln.pstex_t} \end{array}}
	:= \ensuremath{\begin{array}{c}\begin{picture}(0,0)%
\includegraphics{pstex/K-Delta.pstex}%
\end{picture}%
\setlength{\unitlength}{3947sp}%
\begingroup\makeatletter\ifx\SetFigFont\undefined%
\gdef\SetFigFont#1#2#3#4#5{%
  \reset@font\fontsize{#1}{#2pt}%
  \fontfamily{#3}\fontseries{#4}\fontshape{#5}%
  \selectfont}%
\fi\endgroup%
\begin{picture}(374,395)(1791,-1006)
\put(1791,-843){\makebox(0,0)[lb]{\smash{\SetFigFont{8}{9.6}{\rmdefault}{\mddefault}{\updefault}{\color[rgb]{0,0,0}$M$}%
}}}
\end{picture}
 \end{array}} - \ensuremath{\begin{array}{c}\begin{picture}(0,0)%
\includegraphics{pstex/FullGaugeRemainder.pstex}%
\end{picture}%
\setlength{\unitlength}{3947sp}%
\begingroup\makeatletter\ifx\SetFigFont\undefined%
\gdef\SetFigFont#1#2#3#4#5{%
  \reset@font\fontsize{#1}{#2pt}%
  \fontfamily{#3}\fontseries{#4}\fontshape{#5}%
  \selectfont}%
\fi\endgroup%
\begin{picture}(424,395)(2053,-930)
\put(2053,-773){\makebox(0,0)[lb]{\smash{\SetFigFont{8}{9.6}{\rmdefault}{\mddefault}{\updefault}{\color[rgb]{0,0,0}$M$}%
}}}
\end{picture}
 \end{array}} 
	\equiv \ensuremath{\begin{array}{c}\begin{picture}(0,0)%
\includegraphics{pstex/K-Delta.pstex}%
\end{picture}%
\setlength{\unitlength}{3947sp}%
\begingroup\makeatletter\ifx\SetFigFont\undefined%
\gdef\SetFigFont#1#2#3#4#5{%
  \reset@font\fontsize{#1}{#2pt}%
  \fontfamily{#3}\fontseries{#4}\fontshape{#5}%
  \selectfont}%
\fi\endgroup%
\begin{picture}(374,395)(1791,-1006)
\put(1791,-843){\makebox(0,0)[lb]{\smash{\SetFigFont{8}{9.6}{\rmdefault}{\mddefault}{\updefault}{\color[rgb]{0,0,0}$M$}%
}}}
\end{picture}
 \end{array}} - \ensuremath{\begin{array}{c}\begin{picture}(0,0)%
\includegraphics{pstex/DecomposedGR.pstex}%
\end{picture}%
\setlength{\unitlength}{3947sp}%
\begingroup\makeatletter\ifx\SetFigFont\undefined%
\gdef\SetFigFont#1#2#3#4#5{%
  \reset@font\fontsize{#1}{#2pt}%
  \fontfamily{#3}\fontseries{#4}\fontshape{#5}%
  \selectfont}%
\fi\endgroup%
\begin{picture}(540,395)(1936,-925)
\put(1936,-776){\makebox(0,0)[lb]{\smash{\SetFigFont{8}{9.6}{\rmdefault}{\mddefault}{\updefault}{\color[rgb]{0,0,0}$M$}%
}}}
\end{picture}
 \end{array}}
\label{eq:NFE:EffPProp}
\end{equation}

It is important to note that we have defined the diagrammatics
in eqn.~\eq{eq:NFE:EffPProp} such that there are no $1/N$
corrections where the effective propagator attaches
to the two-point, tree level vertex. We do this because, when the composite
object on the \lhs\ of eqn.~\eq{eq:NFE:EffPProp} appears
in actual calculations, it always occurs inside some larger diagram.
It is straightforward to show that, in this case, the aforementioned attachment 
corrections always vanish~\cite{Thesis}. 

The next diagrammatic identity follows
from gauge invariance and the constraint
placed on the vertices of the Wilsonian effective action by
the requirement that the minimum of the Higgs potential is
not shifted by quantum corrections:
\be
	\ensuremath{\begin{array}{c}\begin{picture}(0,0)%
\includegraphics{pstex/GR-TLTP.pstex}%
\end{picture}%
\setlength{\unitlength}{3947sp}%
\begingroup\makeatletter\ifx\SetFigFont\undefined%
\gdef\SetFigFont#1#2#3#4#5{%
  \reset@font\fontsize{#1}{#2pt}%
  \fontfamily{#3}\fontseries{#4}\fontshape{#5}%
  \selectfont}%
\fi\endgroup%
\begin{picture}(757,318)(1880,-963)
\put(2296,-857){\makebox(0,0)[lb]{\smash{\SetFigFont{11}{13.2}{\rmdefault}{\mddefault}{\updefault}{\color[rgb]{0,0,0}$0$}%
}}}
\end{picture}
 \end{array}} = 0.
\label{eq:GR-TLTP}
\ee
This follows directly in the $A$-sector, since 
one-point $A$-vertices do not exist.
In the $F$-sector, though,
we are left with one-point $C^1$ and $C^2$-vertices, but
these are constrained to be zero. 

From the effective propagator relation and~\eq{eq:GR-TLTP},
two further diagrammatic identities follow.
First, consider attaching
an effective propagator to the right-hand field in~\eq{eq:GR-TLTP}
and applying
the effective propagator before $\GRk$ has acted. Diagrammatically,
this gives
\[
	\ensuremath{\begin{array}{c}\begin{picture}(0,0)%
\includegraphics{pstex/GR-TLTP-EP.pstex}%
\end{picture}%
\setlength{\unitlength}{3947sp}%
\begingroup\makeatletter\ifx\SetFigFont\undefined%
\gdef\SetFigFont#1#2#3#4#5{%
  \reset@font\fontsize{#1}{#2pt}%
  \fontfamily{#3}\fontseries{#4}\fontshape{#5}%
  \selectfont}%
\fi\endgroup%
\begin{picture}(1081,306)(2490,-1356)
\put(2776,-1250){\makebox(0,0)[lb]{\smash{\SetFigFont{11}{13.2}{\rmdefault}{\mddefault}{\updefault}{\color[rgb]{0,0,0}0}%
}}}
\end{picture}
 \end{array}} = 0 = \ensuremath{\begin{array}{c}\begin{picture}(0,0)%
\includegraphics{pstex/k.pstex}%
\end{picture}%
\setlength{\unitlength}{3947sp}%
\begingroup\makeatletter\ifx\SetFigFont\undefined%
\gdef\SetFigFont#1#2#3#4#5{%
  \reset@font\fontsize{#1}{#2pt}%
  \fontfamily{#3}\fontseries{#4}\fontshape{#5}%
  \selectfont}%
\fi\endgroup%
\begin{picture}(174,174)(2239,-821)
\end{picture}
 \end{array}} - \ensuremath{\begin{array}{c}\begin{picture}(0,0)%
\includegraphics{pstex/kkprk.pstex}%
\end{picture}%
\setlength{\unitlength}{3947sp}%
\begingroup\makeatletter\ifx\SetFigFont\undefined%
\gdef\SetFigFont#1#2#3#4#5{%
  \reset@font\fontsize{#1}{#2pt}%
  \fontfamily{#3}\fontseries{#4}\fontshape{#5}%
  \selectfont}%
\fi\endgroup%
\begin{picture}(499,174)(2239,-820)
\end{picture}
 \end{array}},
\]
which implies the following diagrammatic identity:
\be
	\ensuremath{\begin{array}{c}\begin{picture}(0,0)%
\includegraphics{pstex/GR-relation.pstex}%
\end{picture}%
\setlength{\unitlength}{3947sp}%
\begingroup\makeatletter\ifx\SetFigFont\undefined%
\gdef\SetFigFont#1#2#3#4#5{%
  \reset@font\fontsize{#1}{#2pt}%
  \fontfamily{#3}\fontseries{#4}\fontshape{#5}%
  \selectfont}%
\fi\endgroup%
\begin{picture}(314,174)(2239,-821)
\end{picture}
 \end{array}} = 1.
\label{eq:GR-relation}
\ee

The effective propagator relation, together
with~\eq{eq:GR-relation} implies that
\[
	\ensuremath{\begin{array}{c}\begin{picture}(0,0)%
\includegraphics{pstex/TLTP-EP-GR.pstex}%
\end{picture}%
\setlength{\unitlength}{3947sp}%
\begingroup\makeatletter\ifx\SetFigFont\undefined%
\gdef\SetFigFont#1#2#3#4#5{%
  \reset@font\fontsize{#1}{#2pt}%
  \fontfamily{#3}\fontseries{#4}\fontshape{#5}%
  \selectfont}%
\fi\endgroup%
\begin{picture}(1059,306)(2512,-1356)
\put(2776,-1250){\makebox(0,0)[lb]{\smash{\SetFigFont{11}{13.2}{\rmdefault}{\mddefault}{\updefault}{\color[rgb]{0,0,0}0}%
}}}
\end{picture}
 \end{array}} = \ensuremath{\begin{array}{c}\begin{picture}(0,0)%
\includegraphics{pstex/kpr.pstex}%
\end{picture}%
\setlength{\unitlength}{3947sp}%
\begingroup\makeatletter\ifx\SetFigFont\undefined%
\gdef\SetFigFont#1#2#3#4#5{%
  \reset@font\fontsize{#1}{#2pt}%
  \fontfamily{#3}\fontseries{#4}\fontshape{#5}%
  \selectfont}%
\fi\endgroup%
\begin{picture}(174,174)(2379,-821)
\end{picture}
 \end{array}} - \ensuremath{\begin{array}{c}\begin{picture}(0,0)%
\includegraphics{pstex/kprkkpr.pstex}%
\end{picture}%
\setlength{\unitlength}{3947sp}%
\begingroup\makeatletter\ifx\SetFigFont\undefined%
\gdef\SetFigFont#1#2#3#4#5{%
  \reset@font\fontsize{#1}{#2pt}%
  \fontfamily{#3}\fontseries{#4}\fontshape{#5}%
  \selectfont}%
\fi\endgroup%
\begin{picture}(508,174)(2379,-817)
\end{picture}
 \end{array}} = 0.
\]
In other words, the (non-zero) structure $\ensuremath{\begin{array}{c}\begin{picture}(0,0)%
\includegraphics{pstex/EP-GR.pstex}%
\end{picture}%
\setlength{\unitlength}{3947sp}%
\begingroup\makeatletter\ifx\SetFigFont\undefined%
\gdef\SetFigFont#1#2#3#4#5{%
  \reset@font\fontsize{#1}{#2pt}%
  \fontfamily{#3}\fontseries{#4}\fontshape{#5}%
  \selectfont}%
\fi\endgroup%
\begin{picture}(406,118)(3165,-1234)
\end{picture}
 \end{array}}$ kills
a two-point, tree level vertex. But, by~\eq{eq:GR-TLTP}, 
this suggests that the structure $\ensuremath{\begin{array}{c}\begin{picture}(0,0)%
\includegraphics{pstex/EP-GR.pstex}%
\end{picture}%
\setlength{\unitlength}{3947sp}%
\begingroup\makeatletter\ifx\SetFigFont\undefined%
\gdef\SetFigFont#1#2#3#4#5{%
  \reset@font\fontsize{#1}{#2pt}%
  \fontfamily{#3}\fontseries{#4}\fontshape{#5}%
  \selectfont}%
\fi\endgroup%
\begin{picture}(406,118)(3165,-1234)
\end{picture}
 \end{array}}$
must be equal, up to some factor, to $\lhd$. Indeed,
\be
	\ensuremath{\begin{array}{c}\begin{picture}(0,0)%
\includegraphics{pstex/EP-GRpr.pstex}%
\end{picture}%
\setlength{\unitlength}{3947sp}%
\begingroup\makeatletter\ifx\SetFigFont\undefined%
\gdef\SetFigFont#1#2#3#4#5{%
  \reset@font\fontsize{#1}{#2pt}%
  \fontfamily{#3}\fontseries{#4}\fontshape{#5}%
  \selectfont}%
\fi\endgroup%
\begin{picture}(628,174)(1926,-821)
\end{picture}
 \end{array}} \equiv \ensuremath{\begin{array}{c}\begin{picture}(0,0)%
\includegraphics{pstex/GR-PEP.pstex}%
\end{picture}%
\setlength{\unitlength}{3947sp}%
\begingroup\makeatletter\ifx\SetFigFont\undefined%
\gdef\SetFigFont#1#2#3#4#5{%
  \reset@font\fontsize{#1}{#2pt}%
  \fontfamily{#3}\fontseries{#4}\fontshape{#5}%
  \selectfont}%
\fi\endgroup%
\begin{picture}(786,174)(2089,-821)
\end{picture}
 \end{array}},
\label{eq:PseudoEP}
\ee
where the dot-dash line represents `pseudo effective propagators'. The
exact form of these is not required, though
a particular algebraic representation is given in~\cite{Thesis}.

The final diagrammatic identity we require 
follows directly from
the independence of $\GRk$ on $\Lambda$ (see table~\ref{tab:NFE:k,k'}):
\be
	\stackrel{\bullet}{\GRk} = 0.
\label{eq:LdL-GRk}
\ee

\section{Further diagrammatic techniques}
\label{sec:FurtherDiagrammatics}

The effective propagator relation allows us to replace a two-point vertex 
connected to
an effective propagator with a Kronecker $\delta$ and a 
gauge remainder. In \sec{sec:GRs}, 
we will see how we can deal with these remainders, diagrammatically. This will
require that we broaden our understanding of both the Ward identities 
and no-$\A^0$ symmetry.

In \sec{sec:MomentumExpansions}, we utilise the insights 
gained from the treatment
of the gauge remainders to develop a diagrammatic 
technique for Taylor expanding vertices and kernels in momenta.

\subsection{Gauge Remainders} 
\label{sec:GRs}

Up until now, we have referred to the composite
object $\GRkpr \!\! \GRk$ as a gauge remainder.
Henceforth, we will often loosely refer to the
individual components as gauge remainders. To
make an unambiguous reference, we call $\GRk$
an active gauge remainder, $\GRkpr$ a processed
gauge remainder and $\GRkpr \!\! \GRk$ a full
gauge remainder.

\subsubsection{Action Vertices}

We begin by considering an arbitrary action vertex, which is contracted with
the momentum carried by one of its fields, $X$, as shown on the \lhs\ 
of \fig{fig:GR:Defining}.
All of the fields shown are wildcards, though the field $X$ has no support 
in the $C$-sector.
 To proceed, we use
the  Ward identities~\eq{eq:WID-U} and~\eq{eq:WID-B}, which
tell us that we either push forward or pull back (with a minus sign) the 
momentum of $X$ to the next
field on the vertex. We recall from \sec{sec:Diags:Action} 
that fields are read off
a vertex in the counterclockwise sense; hence, we push forward
counterclockwise and pull back clockwise.

Since the vertex contains all possible (cyclically independent) 
orderings of the fields, spread 
over all (legal) combinations of supertraces, any of the fields 
could precede or follow $X$. Hence, we must sum
over all possible pushes forward and pulls back, as shown on the 
\rhs\ of \fig{fig:GR:Defining}.
\begin{center}
\begin{figure}[h]
	\[
	\ensuremath{\begin{array}{c}\input{pstex/GR-Defining.pstex_t} \end{array}} = \ensuremath{\begin{array}{c}\input{pstex/GR-Defining-B.pstex_t} \end{array}}
	\]
\caption[Diagrammatics for the Ward Identities.]{The \lhs\ shows the contraction of an arbitrary vertex with one of its momenta. On the \rhs, the first row of diagrams
shows all possible pushes forward onto the explicitly drawn fields and the second row shows all possible pulls back.}
\label{fig:GR:Defining}
\end{figure}
\end{center}

It is clear from the Ward identities~\eq{eq:WID-U} 
and~\eq{eq:WID-B} that
the diagrams on the \rhs\ of \fig{fig:GR:Defining} have no explicit
dependence on the field $X$.
Nonetheless, to interpret the diagrams on the \rhs\ unambiguously, 
without reference to the parent,
we must retain information about $X$. This is achieved by keeping the 
line which used to represent $X$ but which is
now terminated by a half arrow, rather than entering the vertex. 
This line carries information about the flavour of $X$
and its momentum, whilst  indicating which field it is that has 
been pushed forward / pulled back onto. 
The half arrow can go on either side of the line.\footnote{This should be borne in 
mind when we encounter pseudo effective propagators, attached to $\GRk$.}

The new-style
diagrammatics we have been using has been, up until now, 
completely blind to details concerning the ordering of fields
and the supertrace structure. If we are to treat gauge remainders 
diagrammatically, we can no longer exactly 
preserve these features.
Let us suppose that we have pushed forward the momentum of 
$X$ onto the field $Y$, as depicted in the first diagram on the \rhs\ of 
\fig{fig:GR:Defining}. Clearly, it must be the case that $X$ and $Y$
are on the same supertrace and that $Y$ is immediately after $X$, 
in the counter-clockwise sense.
The other fields on the vertex---which we will call spectator fields---can 
be in any order and distributed 
amongst any  number of supertraces,
up to the requirement that they do not come between the fields 
$X$ and $Y$. To deduce the momentum flowing into the vertex
along $Y$, we simply follow the indicated momentum routing. 
Hence, momentum $r+s$ enters the vertex along $Y$, in the case that
it is the field $Y$ that has been pushed forward (pulled back) 
on to. Similarly, if we push forward onto the field carrying momentum $t$,
then momentum $r+t$ enters the vertex, along this field.
The flavour changes induced in $Y$ if $X$
is fermionic, are given by table~\ref{tab:GR:Flavour}.

However, we see that this has the capacity to re-introduce
$\A^0$s, via the fields $\tilde{A}^i$. We now describe how
we can use no-$\A^0$ symmetry to map us back into the
$A^{1,2}$ basis.

Let us suppose that we have some vertex which
is decorated by, for example, an $F$ and a $\bar{B}$ and that
these fields are on the same supertrace. Now consider
contracting the vertex with the momentum of the $F$.
There are two cases to analyse. The first is where there
are other fields on the same supertrace as the $F$ and
$\bar{B}$. For argument's sake, we will take them to be such
that the $\bar{B}$ follows the $F$ (in the counterclockwise
sense). Now the $F$ can push forward onto the $\bar{B}$,
generating an $\tilde{A}^1$. 
In this case,
the vertex \emph{coefficient function} is blind to whether its argument
involves $\tilde{A}^1$ or $A^1$: starting with a vertex containing $\tilde{A}^1$,
we can always remove the $\A^0$ part by no-$\A^0$ symmetry.

The second case to look at is where there are no other fields
on the same supertrace as $F$ and $\bar{B}$. Now the $F$ can
both push forward and pull back onto the $\bar{B}$ generating,
respectively, $\tilde{A}^1$ and $\tilde{A}^2$. However,
we cannot rewrite $\tilde{A}^{1,2}$ as $A^{1,2}$ since,
whereas $\str \tilde{A}^{1,2} \neq 0$, $\str A^{1,2} =0$.
Our strategy is to rewrite vertices involving a separate $\str \tilde{A}^{1,2}$
factor via no-$\A^0$ symmetry.

We have encountered a no-$\A^0$ relation already---see 
eqn.~\eq{eq:No-A0-Ex}.
Now we will generalise this relationship, which is most 
readily done by example.
Consider the following part of the action, where we remember 
that all position arguments
are integrated over:
\begin{eqnarray*}
	&& \cdots + \frac{1}{3} S^{1 \, 1 \, 1}_{\alpha \beta \gamma}(x,y,z) \str \tilde{A}^1_\alpha (x) \tilde{A}^1_\beta (y) \tilde{A}^1_\gamma (z)
\\
	&&
	+\frac{1}{2}  S^{1 \, 1, A \sigma}_{\alpha \beta,  \gamma}(x,y;z) \str \tilde{A}^1_\alpha (x) \tilde{A}^1_\beta (y) \; \str A_\gamma (z) \sigma
	+\cdots
\end{eqnarray*}
We note that, in the second term, we have combined $S^{AA,A}$ with $S^{A,AA}$, 
thereby killing the factor of $1/2!$
associated with each of these vertices.

To determine the no-$\A^0$ relationship between these vertices, 
we shift  $A$: $\delta A_\mu(x) = \lambda_\mu(x) \one$,
and collect together terms with the same supertrace structure 
and the same dependence on $\lambda_\mu$.
By restricting ourselves to the portion of the action shown, 
we only find common terms which
depend on a single power of $\lambda_\mu$. By
using a larger portion of the action we can, of course, 
obtain higher order relationships, although it can
be shown that these all follow from 
the first order relations~\cite{Antonio'sThesis}. In the single
supertrace vertex, this operation simply kills the factor 
of $1/3$; in the double supertrace term, we
focus on shifting the lone $A$ which yields:
\begin{eqnarray*}
	&&  \cdots + \lambda_\gamma(z)  S^{1 \, 1 \, 1}_{\alpha \beta \gamma}(x,y,z) \str \tilde{A}^1_\alpha (x) \tilde{A}^1_\beta (y)
\\
	&&
	+\frac{1}{2}  \lambda_\gamma(z) S^{1 \, 1, A \sigma}_{\alpha \beta, \gamma}(x,y;z) \str \tilde{A}^1_\alpha (x) \tilde{A}^1_\beta (y) \; \str \sigma
	+\cdots
\end{eqnarray*}

Taking into account invariance under the cylic permutation of
the arguments of the supertrace in the first term, no-$\A^0$ symmetry requires:
\[
	S^{1 \, 1 \, 1}_{\alpha \beta \gamma}(x,y,z) + S^{1 \, 1 \, 1}_{\alpha \gamma \beta}(x,z,y) + 2N S^{1 \, 1, A \sigma}_{\alpha \beta, \gamma}(x,y;z) = 0.
\]
We now recast the final term, so that we work with $\tilde{A}^1$s and $\tilde{A}^2$s,
rather than $A \sigma$. This is a little counterintuitive. At first, we recognise
that $A \sigma = \tilde{A}^1 - \tilde{A}^2$. However, $\tilde{A}^1$ and
$\tilde{A}^2$ are not independent. Specifically,
\[
	\str \tilde{A}^1 = N\A^0 = - \str \tilde{A}^2.
\]
Consequently, we need to be careful what we mean by the vertex
coefficient functions $S^{\cdots, \tilde{A}^{1,2}}$. If, as we will do,
we treat the vertex coefficient functions $S^{\cdots, \tilde{A}^1}$
and $S^{\cdots, \tilde{A}^2}$ as independent then, by recognising
that
\[
	S^{\cdots, \tilde{A}^1} (\str \cdots) \str \tilde{A}^1 + S^{\cdots, \tilde{A}^2} (\str \cdots) \str \tilde{A}^2
\]
is equivalent to
\[
	S^{\cdots, A\sigma} (\str \cdots) \str A \sigma
\]
and writing out the explicitly indicated supertraces in terms of $\A^0$ and $N$, we find
that
\[
	S^{\cdots, \tilde{A}^1} - S^{\cdots, \tilde{A}^2}  = 2S^{\cdots, A\sigma}.
\]
The factor of two on the \rhs\ is, perhaps, unexpected; we emphasise that it
comes from splitting up the variable $\A^0$ between
$\tilde{A}^1$ and $\tilde{A}^2$. 
In \fig{fig:noA0}
we give a diagrammatic form for the subset of first order 
no-$\A^0$ relations which relate single
supertrace vertices to two supertrace vertices. 
\begin{center}
\begin{figure}[h]
	\begin{eqnarray}
		&	& \ensuremath{\begin{array}{c}\input{pstex/Vertex-WW-A1-Full.pstex_t} \end{array}} \ + \cdots + \ensuremath{\begin{array}{c}\input{pstex/Vertex-A1-WW-Full.pstex_t} \end{array}} \hspace{1em} + \ensuremath{\begin{array}{c}\input{pstex/Vertex-WA1W-Full.pstex_t} \end{array}} \nonumber
	\\[1ex]
		& + & \ensuremath{\begin{array}{c}\input{pstex/Vertex-WW-A2-Full.pstex_t} \end{array}} \ + \cdots + \ensuremath{\begin{array}{c}\input{pstex/Vertex-A2-WW-Full.pstex_t} \end{array}} \hspace{1em} + \ensuremath{\begin{array}{c}\input{pstex/Vertex-WA2W-Full.pstex_t} \end{array}} \nonumber
	\\[1ex]
		& +	& N \left[ \ensuremath{\begin{array}{c}\input{pstex/Vertex-WW-Full--A1.pstex_t} \end{array}} \hspace{1em} -\hspace{0.5em} \ensuremath{\begin{array}{c}\input{pstex/Vertex-WW-Full--A2.pstex_t} \end{array}} \right] \nonumber
	\\[1ex]
		& =	& 0
	\label{eq:NoA0}
	\end{eqnarray}
\caption{Diagrammatic form of the first order no-$\A^0$ relations.}
\label{fig:noA0}
\end{figure}
\end{center}

A number of comments are in order. First,
this relationship is trivially generalised to include terms
with additional supertraces. Secondly, if we restrict the action to 
single supertrace terms,
as in~\cite{aprop,Antonio'sThesis}, then the first two lines reproduce
the no-$\A^0$ relations of~\cite{Antonio'sThesis}. Thirdly, the Feynman rules
are such that some of the diagrams of \fig{fig:noA0} can be set to zero,
for particular choices of the fields which decorate the vertex. For example,
if all decorative fields are $A^1$s or $C^1$s, then the second row
of diagrams effectively vanishes.

For the purposes of this paper,
the only place we generate $\tilde{A}^1$s
is as internal fields. 
Since internal fields are always attached to kernels (or effective propagators),
we can absorb the factors of $N$ appearing in eqn.~\eq{eq:NoA0} into our 
rules for attaching
kernels / effective propagators to vertices.
This is illustrated in
\fig{fig:GR:WinePrescription}, where the ellipsis denotes
un-drawn pushes forward and pulls back, onto the 
remaining fields, $\{f\}$, which also
decorate the vertex.
\begin{center}
\begin{figure}[h]
	\begin{eqnarray*}
		\ensuremath{\begin{array}{c}\input{pstex/Struc-W-F-FGR.pstex_t} \end{array}}	& =		& \ensuremath{\begin{array}{c}\input{pstex/Struc-W-F-FGR-M.pstex_t} \end{array}} - \ensuremath{\begin{array}{c}\input{pstex/Struc-W-F-FGR-M-B.pstex_t} \end{array}} + \cdots
	\\
							&\equiv	& \left[\ensuremath{\begin{array}{c}\input{pstex/Struc-Thick-W-F-FGR-M.pstex_t} \end{array}} - \frac{1}{N} \ensuremath{\begin{array}{c}\input{pstex/Struc-Thick-W-F-FGR-M-Corr.pstex_t} \end{array}}\right] 
									-\left[ \ensuremath{\begin{array}{c}\input{pstex/Struc-Thick-W-F-FGR-M-B.pstex_t} \end{array}} + \frac{1}{N} \ensuremath{\begin{array}{c}\input{pstex/Struc-Thick-W-F-FGR-M-Corr-B.pstex_t} \end{array}} \right] 
	\\[1ex]
							&	+	& \cdots
	\end{eqnarray*}
\caption{Prescription adopted for internal fermionic fields decorating a vertex
struck by
a gauge remainder.
}
\label{fig:GR:WinePrescription}
\end{figure}
\end{center}

There are a number of things to note. First, the gauge remainder strikes the
vertex and not the base of the kernel; this is ambiguous from the way
in which we have drawn the diagrams though this ambiguity is, in fact, 
deliberate, as we will discuss shortly. Thus, whilst the vertex now possesses
an $A^{1,2}$ field, the kernel is still labelled by $\bar{B}$. Secondly---and this
is the whole point of our prescription---the field struck by the gauge remainder becomes
an $A^{1,2}$, and not an $\tilde{A}^{1,2}$; the missing contributions to the vertex
have been effectively absorbed into the kernel Feynman rule
(\cf \fig{fig:Attachement-A1-B}). Thirdly, we implicitly
sum over all possible ways in which the un-drawn fields, $\{f\}$, decorate the 
vertex in all diagrams (including those with the old-style notation). This ensures
that all diagrams in the no-$\A^0$ relationship~\eq{eq:NoA0} are included.
Lastly, we note once more that the Feynman rules are such that certain
terms can be set to zero when we look at particular realisations of $\{f\}$.

We conclude our discussion of the effect of gauge remainders
on vertices by considering diagrams generated
by the new terms in the flow equation in which a component of $\A$ decorating
a vertex is replaced by a component of $\C$.\footnote{
Though not necessary for the following analysis, we recall from 
\sec{sec:BrokenPhase}
that, in this case, in the $A^1$, $A^2$
basis, components of $\A$ must be replaced by \emph{dynamical} components of $\C$.
}
The situation is illustrated in \fig{fig:GR:Replace-A-w-C} where, 
for reasons that
will become apparent, we schematically indicate the type of vertex 
whose flow generates the terms we
are interested in.
\begin{center}
\begin{figure}[h]
	\[
	\begin{array}{ccc}
		\LD{GR-Replace-AwC-Ex-pre}					& 		& \LD{GR-Replace-AwC-Ex} 
	\\[1ex]
		\dec{\ensuremath{\begin{array}{c}\input{pstex/GR-Replace-AwC-Ex-pre.pstex_t} \end{array}}}{\bullet}	&\sim 	& \ensuremath{\begin{array}{c}\input{pstex/GR-Replace-AwC-Ex.pstex_t} \end{array}} + \cdots
	\end{array}
	\]
\caption{A gauge remainder strikes a vertex in which a component of $\A$ has been 
replaced by a component of $\C$.}
\label{fig:GR:Replace-A-w-C}
\end{figure}
\end{center}

The effect of the gauge remainder requires a little thought.
In diagram~\ref{GR-Replace-AwC-Ex-pre}, the gauge remainder can
clearly strike the $\C$. However, in diagram~\ref{GR-Replace-AwC-Ex}
the $\C$ is not part of the vertex coefficient function and so
is blind to the effects of the gauge remainder.

Allowing the $\C$ of diagram~\ref{GR-Replace-AwC-Ex}
to strike the $\A$, which labels the vertex coefficient function,
the vertex coefficient 
function changes. Now we have a strange situation:
looking just at the coefficient functions of the diagrams (\ie ignoring
the implied supertrace structure), diagrams~\ref{GR-Replace-AwC-Ex-pre}
and~\ref{GR-Replace-AwC-Ex}  are consistent, after the action of the gauge
remainder. However, the implied supertrace structures of the two diagrams
seems to differ, because the $\C$ of diagram~\ref{GR-Replace-AwC-Ex} is
blind to the gauge remainder.

The solution is simple: we allow the effect of the gauge remainder
striking the $\A$ in diagram~\ref{GR-Replace-AwC-Ex} to induce
a similar change in the $\C$. This amounts to a diagrammatic prescription
which ensures that all our diagrams continue to represent both
numerical coefficients and implied supertrace structure. The key point
is that we are free to do this with the $\C$ since, not being part
of a vertex  it does not contribute to the
numerical value of the diagram but serves only to keep track of the
supertraces which have been implicitly stripped off from the vertex whose
flow we are computing. 

Finally, we should take account of attachment corrections, if we
are to work in the $A^1$, $A^2$ basis. Attachment corrections
effectively detach the embedded component of $\C$ from the 
vertex, causing it to become an isolated $\str \C$. 
In diagram~\ref{GR-Replace-AwC-Ex-pre}, this field cannot now be
struck by the gauge remainder. In diagram~\ref{GR-Replace-AwC-Ex},
when the gauge remainder acts, it no longer induces a change in the embedded 
component of $\C$.
In fact, such contributions must cancel against other terms formed by
the action of the gauge remainder; this is discussed further in
\sec{sec:Diags:GI}.

\subsubsection{Kernels} \label{sec:GR-Wines}

Thus far, we have been considering the effects of gauge remainders on 
vertices of the actions. It is straightforward to generalise this
analysis to the effect on vertices of the kernels; the generic case is 
shown in  \fig{fig:GR:kernels}.
\begin{center}
\begin{figure}[h]
	\[
	\ensuremath{\begin{array}{c}\input{pstex/GR-kernels.pstex_t} \end{array}} = \ensuremath{\begin{array}{c}\input{pstex/GR-kernels-M.pstex_t} \end{array}}
	\]
\caption[Contraction of a vertex of an 
arbitrary kernel with one of its momenta.]{Contraction of 
a vertex of an arbitrary kernel with one of its momenta. The 
sense in which we will take pushes forward and pulls back
is as in \fig{fig:NewKernel-Ex}.}
\label{fig:GR:kernels}
\end{figure}
\end{center}

If the field whose momentum is contracted into the kernel is fermionic, then pushes 
forward and pulls back will involve flavour changes.
Let us begin by supposing that one of the fields hit decorates the kernel (as opposed
to being a derivative sitting at the end); in this case, the
flavour changes are just given by table~\ref{tab:GR:Flavour}. 
Note that instances of $\C$
embedded at the ends of the kernel behave like normal kernel decorations, as 
far as gauge remainders are concerned. This follows from the gauge invariance 
of the flow equation and is natural if
we view these embedded fields as behaving just like multi-supertrace
components of the kernel. Of course, if the gauge remainder does strike a 
component of $\C$ which is really an embedded $\C$ then it must be that
this component of $\C$ is forced to be on the same portion of
supertrace as the rest of the kernel. This is just a manifestation of
the statement that the action of gauge remainders necessitates
partial specification of the supertrace structure (\cf our 
treatment of vertices).

When we generate internal $\tilde{A}^{1,2}$s, we would like
to attach to them according to the prescription of 
\fig{fig:GR:WinePrescription}
\ie we wish to extend this prescription such that the structure to which the
kernel attaches is generic, as opposed to being just a vertex of an action. 
We can and will do
this, though note that whether or not the $1/N$ corrections actually survive
depends on whether or not we endow our kernels with completely general 
supertrace structure.
If we do allow completely general kernel decorations then the $1/N$ corrections 
arise---as they did before---by combining terms with a lone 
$\str \tilde{A}^1$ (or $\str \tilde{A}^2$)
with those without. If, however, we take the only multi-supertrace terms of
the kernel to be those involving embedded $\C$s, then 
the fact that the kernel satisifies no-$\A^0$ relations
on its own
causes the $1/N$ corrections to vanish. Since these corrections
are to be hidden in our Feynman rules, it does not matter which scheme we
employ.

The next task is to consider what happens when we push forward (pull back) onto
the end of a kernel. This is done explicitly in~\cite{Thesis}. However,
we can deduce what the answer must be by gauge invariance
considerations, as discussed in the next section.

\subsubsection{Gauge Invariance}	\label{sec:Diags:GI}

We mentioned under \fig{fig:GR:WinePrescription} that
the diagrams on the \rhs\ of the figure are ambiguous: if
we ignore the \lhs, it is
not clear whether we have pushed forward / pulled back around the bottom structure
or pulled back / pushed forward down the kernel. In this section we will argue that
the two must be equivalent, by gauge invariance. This is
explicitly demonstrated to be true in~\cite{Thesis}.

Consider the flow of some vertex decorated by the fields $f_1 \cdots f_n$.
Using the form of the flow equation given in \fig{fig:New-Diags-1},
we explicitly decorate with $f_1$, but leave the other fields as
unrealised decorations (see~\cite{Thesis,oliver2} for much more
detail on this procedure). 
This yields the diagrams of \fig{fig:GR:GaugeInvariance}.
\begin{center}
\begin{figure}[h]
	\[
	-\flow 
	\dec{
		\ensuremath{\begin{array}{c}\input{pstex/Vertex-S-f_1.pstex_t} \end{array}}
	}{f_2 \cdots f_n}
	=
	\frac{1}{2}
	\dec{
		\ensuremath{\begin{array}{c}\input{pstex/Dumbbell-S-f_1-W-Sigma_g.pstex_t} \end{array}} + \ensuremath{\begin{array}{c}\input{pstex/Dumbbell-S-W-f_1-Sigma_g.pstex_t} \end{array}} \hspace{0.75em} +  \ensuremath{\begin{array}{c}\input{pstex/Dumbbell-S-W-Sigma_g-f_1.pstex_t} \end{array}}  -  \ensuremath{\begin{array}{c}\input{pstex/Vertex-Sigma_g-f1-W.pstex_t} \end{array}} - \ensuremath{\begin{array}{c}\input{pstex/Vertex-Sigma_g-W-f1.pstex_t} \end{array}}
	}{f_2 \cdots f_n}
	\]
\caption{Flow of a vertex decorated by the fields $f_1 \cdots f_n$.}
\label{fig:GR:GaugeInvariance}
\end{figure}
\end{center}

Now consider contracting each of the diagrams of \fig{fig:GR:GaugeInvariance}
with the momentum of $f_1$. On the \lhs, this generates the flow of a set of 
vertices decorated by $m-1$ fields. Amongst the diagrams generated on the \rhs\
are those for which we push forward / pull back onto fields to which the kernel
attaches. For each of these diagrams, there is then a corresponding diagram
(with opposite sign) where we have pulled back / pushed forward onto the 
end of the kernel.
Such diagrams, in which we push forward / pull back onto an internal field cannot be
generated by the \lhs; thus as a consequence of gauge invariance, it must be
that they cancel amongst themselves.
\begin{Prop}

Consider the set of diagrams generated by the flow equation,
each of which
necessarily possesses a single kernel that we will take
to attach to the fields $X_1$ and $X_2$.
Suppose that we contract each of these diagrams with
the momentum of one of the (external) fields, $Y$.

Of the resultant diagrams, we collect together those for which the momentum of $Y$
is pushed forward and / or pulled back round a vertex, onto $X_1$ ($X_2$).
We add to this set of diagrams all those for which the momentum of
$Y$ is pushed forward and / or pulled back along the kernel onto
the end attaching to $X_1$ ($X_2$).

We now split these sets into subsets, where the elements of each subset have
exactly the same supertrace structure. The elements of each of these
subsets cancel, amongst themselves.

\label{Prop-GI}
\end{Prop}

In $\beta$ function calculations, where active
gauge remainders arise in a different context from that
above, diagrams in which the ends of kernels are
pushed forward and / or pulled back onto can survive. We can
use the above relationship to always re-interpret such
terms as diagrams in which the gauge remainder has
instead pushed forward and / or pulled back onto the field
on the vertex to which the appropriate end of the
kernel attaches. The attachment corrections should be deduced
according to the flavour of the field on the \emph{vertex} 
to which the kernel
attaches \emph{after} any gauge remainders have acted.

\subsubsection{Cancellations Between Pushes forward / Pulls back} 
\label{sec:GRs:bosonic-cancellations}

Referring back to \fig{fig:GR:Defining}, we now ask when it 
is possible for the pulls back of
the second row to cancel the pushes forward of the first row (this argument can
be repeated for kernels). It is clear that, if the field structure of
corresponding terms is exactly the same, then they will cancel, 
due to the relative minus sign. 
For the purposes of this section, we wish to consider the case where 
any cancellations occur independently of the
spectator fields. In other words, we will not consider cancellations 
which involve changing the ordering, flavour or indices
of the spectator fields; this is delayed until the next section. 
Furthermore, whilst all the wildcard fields we are considering
include all possible field choices, we do not sum over these choices, 
but consider each independently.

Let us temporarily
suppose that $X$  is in the $A$-sector and focus on the case where its 
momentum is pushed forward onto field $Y$. If both $X$ and $Y$ are
bosonic, then the flavours of $X$ and $Y$ are independent of which 
field precedes the other. Moreover, for a given field arrangement,
the flavours of the other fields will not change if the order of $X$ 
and $Y$ is swapped. In this case, the push forward
onto $Y$ will be exactly cancelled by the corresponding pull back.

However, if either $X$ or $Y$ is fermionic, then interchanging their 
order will necessarily change the field content of the vertex.
This follows because a bosonic field in the 1-sector precedes an $F$ 
and follows an $\bar{F}$, whereas a bosonic field 
in the 2-sector follows an $F$ and precedes an $\bar{F}$.
As an example, consider $p_\mu S^{1F\bar{F},\ldots}_{\mu R S\ldots}(p,r,s,\ldots)$. 
To cancel the push forward onto $F_R$
would require us to change the flavour of $X$ to $A^2$: 
$p_\mu S^{F 2 \bar{F}\ldots}_{R \mu S\ldots}(r,p,s,\ldots)$. 
Instead, we could try and cancel the
push forward onto $F_R$ by constructing the term $p_\mu 
S^{F\bar{F}1\ldots}_{S R \mu\ldots}(s,r,p,\ldots)$, 
but now it is the $\bar{F}$ carrying the 
index $R$, rather than the $F$. As we will see in the next 
section such a term can, in general, either cancel or double the original
push forward. However, for the purposes of this section, 
we note that the spectator field $\bar{F}_S$ has suffered a change
and so we do not consider this further.

Similarly, if both $X$ and $Y$ are fermionic, then interchanging them will alter the field content of the vertex,  if other fields
are present on the same supertrace.
Then, we have the choice of altering the spectators or letting
 $X,Y\rightarrow \bar{X}, \bar{Y}$. The former case will be dealt with in the next section. In the latter case, we note from
table~\ref{tab:GR:Flavour} that pushing forward the momentum of
$F_R$ onto $\bar{F}_S$ yields $(A^1,C^1)_S$ whereas the pulling back the momentum of 
$\bar{F}_R$ into $F_S$ yields $-(A^1,-C^1)_S$. These contribution do
produce a cancellation over the first four indices, but the fifth index contributions add.

In conclusion, when dealing with a single vertex, a push forward can only completely 
cancel a pull back, independently of
the spectator fields, when both fields
involved are bosonic. When we generalise this analysis to 
full diagrams, rather than individual vertices, we might 
expect this constraint to be relaxed: all internal fields will be 
summed over and we have seen how, for example, pushing forward 
the momentum of an $A^1$ onto an $F$ could be can be cancelled by 
pulling back the momentum of an $A^2$. Thus, if the $A$ field is internal,
then we will be including both cases, automatically.
However, when dealing with full diagrams,
we must be aware that interchanging fields can alter the supertrace 
structure of the diagram and so we will actually find
that the conditions for cancellation between pushes forward and 
pulls back are even more stringent (see \sec{sec:GR:CompleteDiagrams}).

\subsubsection{Charge Conjugation}	\label{sec:Diagrammatics:CC}

In the previous section we looked at whether pushes forward could cancel 
pulls back, independently of the spectator fields.
If the properties of the spectator fields are allowed to change, 
then we find that every push forward is related to a pull back,
by \CC.

The 
diagrammatic recipe for \CC\ is~\cite{ym,ymi,aprop}:
\begin{enumerate}
	\item	reverse the sense in which we read fields off from the vertices / kernels,

	\item 	pick up a minus sign for each field in the $A$-sector;

	\item	let $\bar{F} \leftrightarrow -F$;
\end{enumerate}
where we must remember that fields pushed forward / pulled back onto 
may have changed flavour and that the wildcard fields 
should be interpreted according to table~\ref{tab:GR:Flavour}.
Rather than having to
specify the sense in which fields are to be read off, we can instead replace a 
given diagram with its mirror image, whilst obeying
point two and three above.

Now let us return to \fig{fig:GR:Defining} and consider 
taking the mirror image of the bottom row of diagrams.
Since the location and order of the spectator fields is unspecified 
we see that, up to a possible sign, the first and 
second rows are actually identical! However, whether corresponding 
entries in the two rows add or cancel, depends
on the whether the original vertex is even or odd under \CC. 
In the former case, pushes forward and pulls back will add;
in the latter case they will cancel.

\subsubsection{Complete Diagrams} \label{sec:GR:CompleteDiagrams}

So far, we have just been concerned with isolated vertices
and so now turn to full diagrams. We still wish to combine
pushes forward and pulls back using \CC\ but, to do so, we 
must look at the \CC\ properties of whole diagrams,
rather than the properties of individual vertices.

We begin by looking at the
example illustrated in \fig{fig:GR:d1:1.6}. 
Each diagram has two external fields, which we will choose to be $A^1$s,
carrying indices $\alpha$ and $\beta$ and momenta $p$ and $-p$. 
By Bose symmetry, the diagrams are symmetric under $p_\alpha \leftrightarrow -p_\beta$.
We have abbreviated the vertex argument $\hat{0} \equiv \hS_0$ to just the hat.
\begin{center}
\begin{figure}[h]
	\[
	\begin{array}{cccccc}
			& \LD{ExD:d1:1.6}	&	& \LD{ExD:d1:2.1}	&	&\LD{ExD:d1:2.2}
	\\[1ex]
		- 	& \ensuremath{\begin{array}{c}\input{pstex/GR-CC-1.6.pstex_t} \end{array}}	& =4& \ensuremath{\begin{array}{c}\input{pstex/Beta1-2.1-lab.pstex_t} \end{array}}& -2&\ensuremath{\begin{array}{c}\input{pstex/Beta1-2.2-lab.pstex_t} \end{array}}
	\end{array} 
	\]
\caption[Example of a gauge remainders in a complete diagram.
]{Example of a gauge remainders in a complete diagram. The dummy index $R$ is given by $\rho$, if restricted to the first four indices.}
\label{fig:GR:d1:1.6}
\end{figure}
\end{center}

The first comment to make is that the diagrammatics is 
slightly different from the previous case.
Rather than terminating the pushed forward / pulled back 
field-line with a half arrow, we just utilise the fact
that the corresponding field line already ends in a $\GRkpr$ 
and use this to indicate the field hit.

Returning to the diagrams of \fig{fig:GR:d1:1.6}
we see that, not only can we collect pushes forward and pulls back, 
but we can also exploit any symmetries of
the diagrams to collect terms. Looking at diagram~\ref{ExD:d1:1.6}, 
it makes no difference whether the gauge remainder hits the field carrying
$\alpha$ or the field  carrying $\beta$. Since we can push forward or 
pull back onto either of these fields, this accounts for the
factor of four multiplying diagram~\ref{ExD:d1:2.1}. 

Diagram~\ref{ExD:d1:2.2} is interesting. Having used \CC\ to collect the 
push forward
and pull back, let us now suppose that all fields leaving the 
three-point vertex are in the $A$-sector. 
We note that the field struck by the gauge remainder has an $\A^0$ component,
but suppose that this has been absorbed into an attachment correction. In this
picture, we cannot have an $A^1$ alone on a supertrace and so all three fields
must be on the same supertrace. However, we are still free to interchange
 $A_\alpha(p)$ and $A_\beta(-p)$ and, summing over the two possible locations
of these fields, we have:
\[
\hat{S}^{1\, 1\, 1}_{\alpha \beta \rho}(p,-p,0) + \hat{S}^{1\, 1\, 1}_{\beta \alpha \rho}(-p,p,0).
\]
These two terms cancel, as a result of \CC.

We might wonder if \CC\ causes components of diagram~\ref{ExD:d1:2.1} to cancel. However, 
attachment corrections aside, the index structure $\beta \alpha R$
corresponds to a different supertrace structure from the index structure $\alpha \beta R$. In the former case, the two $A^1$s are
on different supertraces whereas, in the latter case, they are on the same supertrace. This is illustrated
in \fig{fig:SupertraceStrucure-Ex}.
\begin{center}
\begin{figure}[h]
	\[
		\ensuremath{\begin{array}{c}\input{pstex/Beta1-2.1-lab-B.pstex_t} \end{array}} \mathrm{FAS} \ \ \neq \ \ \ensuremath{\begin{array}{c}\input{pstex/Beta1-2.1-lab.pstex_t} \end{array}} \mathrm{FAS} 
	\]
\caption{Two components of diagram~\ref{ExD:d1:2.1} which are not equal, due to their differing
supertrace structure.}
\label{fig:SupertraceStrucure-Ex}
\end{figure}
\end{center}

If we include  attachment corrections (see \fig{fig:Attachement-A1-B}), 
then the kernel
of diagram~\ref{ExD:d1:2.1}
can attach to the vertex via a false kernel. In this case, there is only 
one supertrace, and so
all fields are necessarily on it. Such components of 
diagram~\ref{ExD:d1:2.1} do cancel
amongst themselves. However, these cancellations are 
generally hidden by our notation
and are of no practical importance anyway, until we come 
to extracting numerical contributions to
$\beta$-function coefficients from $\Lambda$-derivative terms.

Returning to diagram~\ref{ExD:d1:2.2}, for the diagram to survive, 
the field carrying momentum $k$ must be in the $F$-sector. In this case, 
the gauge remainder
can produce a $C$-sector field. Under interchange of $\alpha$ and $\beta$, 
such a vertex is even and so survives.

 This
serves to illustrate a general feature of these diagrammatics, 
alluded to at the end of 
\sec{sec:GRs:bosonic-cancellations}. Suppose that we are pushing 
forward the momentum of a field $X$ onto the 
field $Y$. If we can rearrange the diagram such that, leaving 
all other fields alone, we can place $X$ on the other side of
$Y$, then the resulting pull back onto $Y$ will cancel the push 
forward, so long as no flavours  or indices have changed in the
rearrangement and the supertrace structure is still the same. 

Here is how this applies to our examples.
In the case of diagram~\ref{ExD:d1:2.1}, to convert a pull back 
onto $A_\beta^1$ into a push forward, we must change the location
of $A_\alpha^1$, to maintain the same supertrace structure 
(up to attachment corrections). 
The resulting term can then just be collected with the pull back, by
\CC. Hence the push forward onto $A_\beta^1$ can never be 
completely cancelled by a pull back.

In the case of diagram~\ref{ExD:d1:2.2} we can convert a 
push forward into a pull back without changing the locations
of the spectator fields and without having to change the 
supertrace structure.
If the fields carrying momentum $k$ are in the $A$-sector,
then interchanging them does not result in any flavour
changes, and so the push forward cancels the pull back. 
However, if the fields carrying momentum $k$ are fermionic, then interchanging
them requires us to replace $\bar{F} \leftrightarrow F$. 
This constitutes a change of flavour and we find that the push forward does not
completely cancel the pullback, since we are left with a 
contribution arising from the $C$-sector.

We note that, just as we can use \CC\ to
redraw vertices struck by gauge remainders, so too can we use \CC\ to
redraw entire diagrams. Given some diagram, the diagrammatic effect of \CC\
is to replace a diagram by its mirror image, letting 
$\bar{F} \leftrightarrow F$ (\sic)
and picking up a minus sign for each gauge remainder that has been 
performed.\footnote{So far, we have only encountered diagrams in which a single
gauge remainder has acted, but we will come across more general cases later.}
Picking up a sign in this manner automatically keeps track
of the signs associated with the rules of \sec{sec:Diagrammatics:CC}.

Let us now examine a second example of gauge remainders
in complete diagrams, as shown in \fig{fig:GR:ExD:1.10}.
\begin{center}
\begin{figure}[h]
	\[
	\begin{array}{cccccc}
			& \LD{ExD:d1:1.10}	&	& \LD{ExD:d1:2.3(S)}&	&\LD{ExD:d1:2.4(S)}
	\\[1ex]
		-	& \ensuremath{\begin{array}{c}\input{pstex/GR-CC-1.10.pstex_t} \end{array}} 	&=-2& \ensuremath{\begin{array}{c}\input{pstex/Beta1-2.3S.pstex_t} \end{array}} &+2 & \ensuremath{\begin{array}{c}\input{pstex/Beta1-2.2.pstex_t} \end{array}}
	\end{array}
	\]
\caption{Example of a gauge remainder on a kernel, in a full diagram.}
\label{fig:GR:ExD:1.10}
\end{figure}
\end{center}

It is crucially important to recognise that, whilst diagram~\ref{ExD:d1:1.10}
may superficially look like a diagram in which the kernel bites its own tail, it is
very different. The difference arises due to the gauge remainder, and means that
such diagrams cannot be discarded on account of~\eq{noATailBiting}. (We can view
the gauge remainders as being some non-trivial kernel $K(x,y)$ sitting between
the functional derivatives in~\eq{noATailBiting}---which we take to carry position
argument $x$---and $\{W\}$, which we take to carry position argument $y$. Only if
the kernel reduces to $\delta(x-y)$, which it does not, 
can the constraint~\eq{noATailBiting}
contribute with non-zero measure.)

Comparing with \fig{fig:GR:d1:1.6}, 
we see that diagram~\ref{ExD:d1:2.4(S)} has exactly the same 
structure as diagram~\ref{ExD:d1:2.2}. Although the former diagram
involves a pull back along the kernel and the latter case involves a 
push forward around a vertex,
we know from \sec{sec:Diags:GI} that these two diagrams are 
identical. Taking into
account the relative sign, it is clear that they cancel.

It is worth making some comments
about the structure at the top of
diagram~\ref{ExD:d1:2.3(S)}. The line segment which joins the top
of the kernel to the $\GRkpr$---thereby forming a `hook'---performs no \role\ other 
than to make this join. In other
words, it is neither a section of kernel nor an effective propagator. We could imagine
deforming this line segment so that the hook becomes arbitrarily large. Despite
appearances, we must always remember that this line segment simply performs the role
of a Kronecker~$\delta$. When part of a complete diagram, this line
segment can always be distinguished from an effective propagator,
to which it can be made to look identical, by the context. This
follows because hooks in which the line segment is a Kronecker~$\delta$
only ever attach to effective propagators or kernels, whereas hook-like structures
made out of an effective propagator only ever attach to vertices
(see 
diagram~\ref{ExD:d1:2.4(S)} for an example but note that, in
this case, the effective propagator is differentiated).
When viewed in isolation, we will always take the hook structure to
comprise just a line segment and so will draw the hook as tightly as
possible.

We conclude this section by discussing a particular scenario---which will
crop up repeatedly in our computation of $\beta$-function coefficients---in
which it is possible to neglect attachment corrections.

Consider some complete diagram possessing
a three-point vertex which is decorated by an external $A^1$
and is struck by a gauge remainder. 
The type of
diagram we are
considering is represented in \fig{fig:Diags:GR-SuperCorrs},
where the fields $\{f\}$ can attach anywhere except the bottom vertex, which must
be three-point (hence the superscript three). If any of these fields are internal fields, then we take
pairs of them to be connected by an effective propagator.
\begin{center}
\begin{figure}[h]
	\[
	\dec{
		\ensuremath{\begin{array}{c}\input{pstex/GR-SuperCorrs.pstex_t} \end{array}}
	}{\{f\}}
	\]
\caption{A diagram for which attachment corrections are restricted.}
\label{fig:Diags:GR-SuperCorrs}
\end{figure}
\end{center}

We now argue that we can 
forget about any attachment corrections.
Let us start by supposing that the gauge remainder
is in the $F$-sector. If the gauge remainder strikes
the internal field, then it can generate an effective
attachment correction. However this correction isolates the newly
formed two-point vertex from the rest of the diagram,
leaving us with $\str A^1 = 0$.

Next, suppose that the gauge remainder is in the $A$-sector.
Since two
of the three fields entering the vertex are now in the $A$-sector,
the third must also be bosonic. Moreover, 
the final field must be in the $A$-sector also, else the action
of the gauge remainder will produce an $AC$ vertex, which does not
exist. Now, any attachment corrections 
would mean that all fields on the vertex are guaranteed
to be on the same portion of supertrace, \wrt\ the diagram
as a whole, irrespective of location. Summing over the independent
locations of the fields causes the diagram to vanish, by \CC.

\begin{center}
\begin{figure}[h]
	\[
	\LDi{GR-CC-1.20}{ExD:d1:1.20}
	= 4
	\left[
		\begin{array}{l}
			\begin{array}{cccc}
					& \LD{ExD:d1:2.10,11}	&	& \LD{ExD:d1:2.12}
			\\[1ex]
				-	& \ensuremath{\begin{array}{c}\input{pstex/Beta1-2.10pre.pstex_t} \end{array}}	& +	& \ensuremath{\begin{array}{c}\input{pstex/Beta1-2.12.pstex_t} \end{array}}
			\end{array}
		\\[1ex]	
			\hspace{2.5em} \cdeps{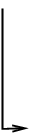}
			\begin{array}{cccc}
					& \LD{ExD:d1:2.10}	&	& \LD{ExD:d1:2.11}
			\\[1ex]
				-	& \ensuremath{\begin{array}{c}\input{pstex/Beta1-2.10.pstex_t} \end{array}}	& +	& \ensuremath{\begin{array}{c}\input{pstex/Beta1-2.11.pstex_t} \end{array}}
			\end{array}
		\end{array}
	\right]
	\]
\caption{Gauge remainder which produces a diagram in which the effective propagator relation can be applied.
}
\label{fig:GR:nested}
\end{figure}
\end{center}

Henceforth, whenever we deal with a three-point vertex
decorated by an external field and struck by a gauge remainder,
we will automatically discard all attachment corrections. Let us
now apply this to a specific case, illustrated in \fig{fig:GR:nested}.
Referring also to 
\fig{fig:GR:d1:1.6}, 
diagram~\ref{ExD:d1:2.10} straightforwardly cancels diagram~\ref{ExD:d1:2.1}.

\subsubsection{Nested Contributions} \label{sec:GRs-Nested}

The basic methodology for nested gauge remainders is  similar to the 
methodology just presented, but
we must take account of the fact that the supertrace structure is now 
partially specified. In particular,
this will generally mean that we cannot use \CC\ to collect nested pushes 
forward and pulls back (the exception 
being if a gauge remainder produced in one factorisable 
sub-diagram hits a separate factorisable sub-diagram)
and so must count them separately. Indeed, nested pushes 
forward and pulls back can have different supertrace 
structures as illustrated by considering the result of 
processing diagram~\ref{ExD:d1:2.11}, as shown in \fig{fig:GR:nested-M}.
\begin{center}
\begin{figure}[h]
	\[
	4
	\left[
		\begin{array}{ccccccc}
			\LD{ExD:d1:2.17a}	&	& \LD{ExD:d1:2.17b}		&	& \LD{ExD:d1:2.18a}		&	& \LD{ExD:d1:2.18b}
		\\[1ex]
			\ensuremath{\begin{array}{c}\input{pstex/GR-CC-1.20M-B1.pstex_t} \end{array}}	& -	& \ensuremath{\begin{array}{c}\input{pstex/GR-CC-1.20M-B2.pstex_t} \end{array}}	& -	& \ensuremath{\begin{array}{c}\input{pstex/GR-CC-1.20M-B3.pstex_t} \end{array}}	& -	& \ensuremath{\begin{array}{c}\input{pstex/GR-CC-1.20M-B4.pstex_t} \end{array}}
		\end{array}
	\right]
	\]
\caption{Result of processing diagram~\ref{ExD:d1:2.11}.}
\label{fig:GR:nested-M}
\end{figure}
\end{center}

We begin our analysis of 
the diagrams of \fig{fig:GR:nested-M}
by noting that, from our
discussion at the end of \sec{sec:GR:CompleteDiagrams},
there are no attachment corrections. 

Diagrams~\ref{ExD:d1:2.17a} and~\ref{ExD:d1:2.18a} have supertrace structure $\pm N\str{A^1_\alpha A^1_\beta}$, with
the plus or minus depending on the sector of the wildcard fields. On the other hand, diagrams~\ref{ExD:d1:2.17b} and~\ref{ExD:d1:2.18b}
have supertrace structure $\str{A^1_\alpha}\str{A^1_\beta} =0$.
Note in the latter case that the particular supertrace structure puts constraints
on the field content of the diagram. Specifically, the kernel in 
diagrams~\ref{ExD:d1:2.17b} and~\ref{ExD:d1:2.18b} cannot be fermionic. Let us suppose
that it is. Then, the end which attaches to the vertex must be an $F$ and so
the end which attaches to the $\wedge$ must be an $\bar{F}$. However, referring
to table~\ref{tab:GR:Flavour} we see that an $\bar{F}$ cannot pull back onto bosonic
fields in the 1-sector. There is an inconsistency in such a diagram and so our
original supposition that it exists must be wrong.

\subsection{Momentum Expansions} 
\label{sec:MomentumExpansions}

The computation of $\beta$-function coefficients involves working at 
fixed order in external momentum. 
If a diagram contains a structure that is already manifestly of the 
desired order, then it is useful to Taylor expand 
at least some of the remaining structures in the external momentum. 
Vertices can always be expanded in momentum,
as it is a requirement of the setup that such a step is
 possible~\cite{ymi,ymii}. 
Whereas kernels, too, can always be Taylor expanded in momentum it is 
not necessarily
possible to do so with effective propagators that form part of a 
diagram, as this step can introduce IR divergences, at intermediate
stages.
This will be discussed in detail in~\cite{mgierg2}.

The key idea in what follows is that, if an $A$-field decorating 
a vertex decorated by $n$ other fields
carries zero momentum, then we can relate
this vertex to the momentum derivative of a set of vertices
decorated just by the $n$ fields.
This relation arises as a consequence of
the Ward identity~(\ref{eq:WID-U}) and so it is no 
surprise that the diagrammatics of this section is very similar to 
those we employed for the gauge remainders.

\subsubsection{Basics}

Consider the structure general vertex, $U$, 
the decorations of which include an $A^i$-field carrying zero momentum.
Let us begin by supposing that this $A^i$-field is sandwiched between the 
fields $X$ and $Y$. By the Ward Identity~\eq{eq:WID-U}, we have:
\begin{eqnarray}
\lefteqn{
\epsilon_\mu U^{\cdots X A^i Y \cdots}_{\cdots R \, \mu \ S \cdots}(\ldots,r,\epsilon,s-\epsilon,\ldots) =} \nonumber \\
& &
U^{\cdots X Y \cdots}_{\cdots R \, S \cdots}(\ldots,r,s,\ldots) - U^{\cdots X Y \cdots}_{\cdots R \, S \cdots}(\ldots,r+\epsilon,s-\epsilon,\ldots).
\end{eqnarray}
Taylor expanding both sides in $\epsilon$ and equating the $\Oep$ terms yields:
\begin{equation}
 U^{\cdots X A^i Y \cdots}_{\cdots R \, \mu \ S \cdots}(\ldots,r,0,s,\ldots) = \left.\left(\partial_\mu^{s'} - \partial_\mu^{r'} \right)
 U^{\cdots X Y \cdots}_{\cdots R \, S \cdots}(\ldots,r',s',\ldots)\right|_{r'=r,s'=s}.
\label{eq:MomeExp:Defining}
\end{equation}
This equation, for the case of a vertex, is represented diagrammatically in \fig{fig:MomExp:dummy}, which highlights the similarity between
the  momentum expansions and gauge remainders. The top row on the \rhs\ correspond to `push forward like' terms, whereas those
on the second row correspond to `pull back like' terms. (As with the gauge remainders, pushes forward are performed
in the counterclockwise sense.)
\begin{center}
\begin{figure}[h]
\[
	\ensuremath{\begin{array}{c}\input{pstex/MomExp-dummy.pstex_t} \end{array}} =
	\left[
	\ensuremath{\begin{array}{c}\input{pstex/MomExp-dummyM.pstex_t} \end{array}}
	\hspace{0.3in}
\right]
\]
\caption[Diagrammatics expression for a vertex decorated by an $A$-field carrying zero momentum.
]{Diagrammatics expression for a vertex decorated by an $A$-field carrying zero momentum. The filled circle attached to 
the $A$-field line tells us to first replace all momenta with dummy momenta; then
to differentiate \wrt\ the dummy momenta of the field hit, holding
all other momenta constant and finally to replace the dummy momenta with the original momenta.}
\label{fig:MomExp:dummy}
\end{figure}
\end{center}

As with the gauge remainders, we must consider all possible 
independent locations of the $A$-field \wrt\ the other fields.
Hence, terms between the first and second rows can cancel, 
if the field hit is bosonic.

We now want to convert derivatives \wrt\ the dummy momenta to 
derivatives \wrt\ the original momenta. There are two cases to deal with.
The first---in which we shall say that the momenta are 
paired---is where there are a pair of fields, carrying equal and opposite momentum. 
The second---in which we shall say that the momenta are 
coupled---is where there are three fields carrying, say, $(r, s, -s-r)$.

\paragraph{Paired Momenta}

This is the simplest case to deal with. If the momentum $r$ has been replaced with dummy momentum $r'$ and $-r$ has been
replaced with dummy momentum $s'$ then
\[
\left(\partial_\mu^{r'} - \partial_\mu^{s'} \right) \rightarrow \partial_\mu^r.
\]
Hence, we can collect together a push forward like diagram with a pull back like diagram to give a derivative \wrt\ one of 
the original momenta. An example of this is shown in \fig{fig:MomExp:Defining}, for a field-ordered three-point vertex.
\begin{center}
\begin{figure}[h]
\[
	\ensuremath{\begin{array}{c}\input{pstex/MomExp-Defining.pstex_t} \end{array}}
\]
\caption[Re-expressing a three-point vertex with zero momentum entering along an $A$-field.]
{A field ordered three-point vertex with zero momentum entering along an $A$-field can be expressed as 
the momentum derivative of a two-point vertex. The open circle attached to the $A$-field line represents a derivative
\wrt\ the momentum entering the vertex along the field hit.}
\label{fig:MomExp:Defining}
\end{figure}
\end{center}

\paragraph{Coupled Momenta}

The structures in this section contain momentum arguments of the form $(r,s,-s-r)$.
Referring back to eqn.~(\ref{eq:MomeExp:Defining}), we will denote the dummy momenta
by $(r',s',t')$. We can make progress by noting that:
\begin{eqnarray}
\left(\partial_\mu^{r'} - \partial_\mu^{s'} \right) & \rightarrow & \left(\left.\partial_\mu^{r}\right|_s - 
\left.\partial_\mu^{s}\right|_r\right) \label{eq:dummy-original}
\end{eqnarray}
and likewise, for all other combinations of $(r',s',t')$.
Thus, as with the previous case, we need to combine a pair of terms differentiated \wrt\ dummy momenta
to obtain a structure which is differentiated \wrt\ its original momenta. The difference is that, whilst in the
previous case the pair combined into one diagram, in this case they remain as a pair. An example is
shown in \fig{fig:MomExp:Coupled}.
\begin{center}
\begin{figure}[h]
\[
	\ensuremath{\begin{array}{c}\input{pstex/MomExp-Coupled.pstex_t} \end{array}}
\]
\caption[Re-expressing a four-point vertex with zero momentum entering along an $A$-field.]
{A field ordered four-point vertex with zero momentum entering along an $A$-field can be expressed as 
the momentum derivative of two three-point vertices. 
}
\label{fig:MomExp:Coupled}
\end{figure}
\end{center}

The open circle attached to the $A$-field line represents a derivative
\wrt\ the momentum entering the vertex along the field hit. However, this derivative is performed holding the
momentum of the field hit in the partner diagram constant. Hence, the final two diagrams 
of \fig{fig:MomExp:Coupled} must be interpreted as a pair.
The difference between this
and the paired momentum case highlights the care that
must be taken interpreting the new diagrammatics.

\subsubsection{Kernels}

When we come to deal with kernels, we must adapt the diagrammatic notation slightly.
If the momentum derivative strikes a field decorating a kernel, then we just use the
current notation. However, it is desirable to change the notation when the
momentum derivative strikes one of the ends a kernel.
In complete diagrams, placing the
diagrammatic object representing a momentum derivative at the end of the 
kernel becomes confusing; rather we place the object in middle 
and use an arrow to indicate which end of the kernel it acts on, as 
shown in \fig{fig:MomExp:wine}.

\begin{center}
\begin{figure}[h]
\[
	\ensuremath{\begin{array}{c}\input{pstex/MomExp-wine.pstex_t} \end{array}}
\]
\caption{A field ordered one-point kernel with zero momentum entering along a decorative $A$-field can be expressed as
the momentum derivative of a zero-point kernel.}
\label{fig:MomExp:wine}
\end{figure}
\end{center}

Hence, the second diagram denotes a derivative \wrt\ $+k$, whereas the third diagram denotes a derivative \wrt\ $-k$.

\subsubsection{Complete Diagrams} \label{sec:MomExp:CompleteDiagrams}

We illustrate the application to complete
diagrams by showing how to manipulate
diagram~\ref{ExD:d1:2.12}.
\begin{center}
\begin{figure}[h]
	\[
	\begin{array}{ccccccccc}
							&			&\LD{ExD:d1:2.12M}	&			& \LD{ExD:d1:2.12Mb}&		&\LD{ExD:d1:2.12Mc}	&	&\LD{ExD:d1:2.12Md}
	\\[1ex]
		\ensuremath{\begin{array}{c}\input{pstex/Beta1-2.12-lab.pstex_t} \end{array}}	&\rightarrow&\ensuremath{\begin{array}{c}\input{pstex/ExD-d12.12M.pstex_t} \end{array}}	&\rightarrow&\ensuremath{\begin{array}{c}\input{pstex/ExD-d12.12Mb.pstex_t} \end{array}}	&\equiv	& \ensuremath{\begin{array}{c}\input{pstex/ExD-d12.12Mc.pstex_t} \end{array}}	&=-	&\ensuremath{\begin{array}{c}\input{pstex/ExD-d12.12Md.pstex_t} \end{array}}
	\end{array}
	\]
\caption{Manipulation of a diagram at $\Op{2}$. Discontinuities in momentum flow are indicated by a bar.}
\label{fig:MomExp:Ex}
\end{figure}
\end{center}

Taking the external momentum of the parent diagram to be $p$,
we note that the two-point vertex at the base of the 
diagram is $\Op{2}$, which is the order in $p$ to
which we wish to work. We call this base structure an `$\Op{2}$ stub'.
The first step is to Taylor expand the three-point vertex to zeroth order in $p$,
as shown in diagram~\ref{ExD:d1:2.12M}. 
There is now a discontinuity in momentum arguments, since
although momentum $l$ flows into and out of the 
differentiated two-point vertex, this vertex is attached to an
effective propagator 
carrying momentum $l$ and a kernel carrying momentum $l-p$. 
This discontinuity is indicated by the bar between the vertex and the kernel.
We can Taylor expand the kernel to zeroth order in momentum, too, and this
is done in diagram~\ref{ExD:d1:2.12Mb}. Since the discontinuity in 
momentum has now vanished, the bar is
removed. 

In diagram~\ref{ExD:d1:2.12Mc} we have introduced an arrow on 
the diagrammatic representation
of the derivative. We have come across this arrow already the 
context of kernel but have not
yet required it for vertices. Indeed, in the current example, 
it is effectively redundant notation.
We note, though, that we can reverse the direction of the arrow, 
at the expense of a minus sign,
as in diagram~\ref{ExD:d1:2.12Md}. By reversing the 
direction of the arrow, we are now differentiating \wrt\
the momentum leaving the vertex along the struck field, 
rather than the momentum thus entering. 

We conclude our discussion of momentum expansions by commenting on how
we can redraw a diagram using \CC. For any diagram, we use the following recipe:
\begin{enumerate}
	\item	take the mirror image (this includes reflecting
			any arrows accompanying derivative symbols);

	\item	pick up a minus sign for each performed gauge remainder;

	\item	pick up a minus sign for each derivative symbol.
\end{enumerate}

\section{One Loop Diagrammatics}
\label{sec:beta1}

In this section, we present the entire diagrammatics 
for the computation of $\beta_1$,
arriving at a manifestly gauge invariant, diagrammatic expression, 
from which the universal value in $D=4$ can be 
immediately extracted. 
This computation of $\beta_1$ not only serves as an 
illustration of the diagrammatic techniques 
of secs.~\ref{New} and~\ref{sec:FurtherDiagrammatics}, 
but is a necessary
intermediate step in the computation of $\beta_2$.
Much of the work presented in this chapter overlaps with the 
computation
of $\beta_1$ presented in~\cite{aprop}. However, there are a 
number of important differences, which we now outline.

First, the computation here is done for an unrestricted Wilsonian 
effective action. Previously, the action was restricted to
just single supertrace terms; a consequence of which is that the 
single supertrace terms $S^{\ AAC}_{0 \mu \nu}(p,q,r)$ 
and $S^{\ AAC\sigma}_{0 \mu \nu}(p,q,r)$
can be set to zero~\cite{aprop}. The second major difference is 
that the diagrammatics are no longer terminated after the use of the
effective propagator relation: gauge remainders and 
$\Op{2}$ manipulations are dealt with in an entirely diagrammatic fashion.

We also choose to use a completely general $\hat{S}$, thereby 
demonstrating complete scheme independence. In fact, the inclusion
of $\hat{S}_1$ (higher loop vertices do not occur in the calculation) 
actually leads only to a trivial extension of the scheme
independence. The instances of $\hat{S}$ beyond tree level are (at $\Op{2}$)
restricted to those of the form $\hat{S}_1^C$,
 and are only ever involved
in cancellations via the weak coupling expansion of the
constraint 
eqn.~\eq{eq:NPC-sec-Constraint}.
Nonetheless, it is instructive to see this occurring and to confirm that
$\beta_1$ is universal.

\subsection{A Diagrammatic Expression for $\beta_1$}

\subsubsection{The Starting Point}

The key to extracting $\beta$-function coefficients from
the weak coupling flow equations~\eq{eq:WeakCouplingExpansion-NewFlow-B} 
is to use the renormalisation condition~\eq{defg}, which places 
a constraint on  the vertex $S^{1\, 1}_{\mu \nu}(p)$. From 
eqns.~\eqs{gcondn}{eq:WeakCouplingExpansion-Action} we see that,
apart from the required $\Box_{\mu \nu}(p)$,
the $\Op{2}$ part of $S_{0\mu\nu}^{\ 1 \, 1}(p)$ is just a number
(two) and that
$S_{n\geq 1 \mu \nu}^{\ \ \ \ 1\, 1}(p) = \Op{4}$.

To utilise this information, we
begin by specialising eqn.~\eq{eq:WeakCouplingExpansion-NewFlow-B} 
to compute the flow of $S_{1 \mu \nu}^{\ 1 \, 1}(p)$:
\begin{equation}
	\flow{S_{1 \mu \nu}^{\ 1 \, 1}(p)} = 2\beta_1 S_{0 \mu \nu}^{\ 1 \, 1}(p) - \gamma_1 \pder{S_{0 \mu \nu}^{\ 1 \, 1}(p)}{\alpha}
	- \sum_{r=0}^1 a_0[\bar{S}_{1-r},\bar{S}_r]^{1\, 1}_{\mu \nu}(p) +  a_1[\Sigma_0]^{1\, 1}_{\mu \nu}(p).
\label{eq:Beta1-Defining-pre}
\end{equation}

The $a_0$ term can be simplified. 
Defining $\Pi_{RS}^{XY}(k) = S_{RS}^{XY}(k) - \hat{S}_{RS}^{XY}(k)$
and using the definition of the barred vertices, \eq{eq:NFE:BarNotation}, 
we can write
\[
	- \sum_{r=0}^1 a_0[\bar{S}_{1-r},\bar{S}_r]^{1\, 1}_{\mu \nu}(p) = -2a_0[\Pi_0, S_1]^{1\, 1}_{\mu \nu}(p) + 2a_0[S_0, \hat{S}_1]^{1\, 1}_{\mu \nu}(p).
\]

All the $a_0$ terms generate two vertices, joined by a kernel. 
Unless one of these vertices is decorated by a single field, both
vertices must be decorated by an internal field and one of the external fields. 
Now, one-point $\Pi_0$ vertices do
not exist and two-point $\Pi_0$ vertices vanish, since we have identified 
the two-point, tree level Wilsonian effective
action vertices with the corresponding seed action vertices. 
We choose to discard one-point $S_1^C$ vertices at this
stage of the calculation\footnote{We could instead process
these terms, which would essentially amount to 
enforcing the constraint~\eq{eq:NPC-sec-Constraint} on
the fly~\cite{Thesis}.} and so 
$a_0[\Pi_0, S_1]^{1\, 1}_{\mu \nu}(p)$ does not contribute.

The next step is to focus on the $\Op{2}$ part 
of eqn.~\eq{eq:Beta1-Defining-pre}. Noting that $S_{1 \mu \nu}^{\ 1 \, 1}(p)$
is at least $\Op{4}$, and that the $\Op{2}$ part
of   $S_{0\mu\nu}^{\ 1 \, 1}(p)$ is independent of $\alpha$,
we arrive at
an algebraic expression for $\beta_1$:
\begin{equation}
-4 \beta_1 \Box_{\mu \nu}(p) + \Op{4} = a_1[\Sigma_0]^{1\, 1}_{\mu \nu}(p) + 2a_0[S_0,\hat{S}_1]^{1\, 1}_{\mu \nu}(p),
\label{eq:Beta1-Defining}
\end{equation}
which is shown diagrammatically in \fig{fig:Beta1-LevelZero}. It is implicit 
in all that follows that, unless otherwise stated,
the external indices are $\mu$ and $\nu$ and
we are working at $\Op{2}$.
\begin{center}
\begin{figure}[h]
	\[
	\begin{array}{c}
		-4 \beta_1 \Box_{\mu \nu}(p) + \Op{4} = 
	\\[2ex]
		\ds
		\frac{1}{2}
		\left[
			\begin{array}{c}
				\begin{array}{ccccc}
					\PDCD{d1:1.2(W)}{fig:Beta1:FirstManip}{d1:1.2(S)}{d1:1.5}	&	& \CDCD{d1:1.3(W)}{d1:1.22}{d1:1.3(S)}{d1:1.8}	&	& \CDBl{d1:1.4}{d1:1.12}
				\\[1ex]
					\ensuremath{\begin{array}{c}\input{pstex/d1-1.2.pstex_t} \end{array}}													& +2& \ensuremath{\begin{array}{c}\input{pstex/d1-1.3.pstex_t} \end{array}}									&-	& \ensuremath{\begin{array}{c}\input{pstex/d1-1.4.pstex_t} \end{array}}
				\end{array}
			\\
				\begin{array}{cccc}
						& \DCD{d1:C-1}{d1:1.16}{d1:1.16b}	&	&\Discard{d1:C-2}
				\\[1ex]
					+2	& \ensuremath{\begin{array}{c}\input{pstex/d1-C-1.pstex_t} \end{array}}						&+4	& \ensuremath{\begin{array}{c}\input{pstex/d1-C-2.pstex_t} \end{array}}
				\end{array}
			\end{array}
		\right]	
	\end{array}
	\]
\caption[A diagrammatic representation of the equation for $\beta_1$.]{
A diagrammatic representation of the equation for $\beta_1$. On the \rhs, we implicitly take the indices to be $\mu$ and $\nu$ and
work at $\Op{2}$.}
\label{fig:Beta1-LevelZero}
\end{figure}
\end{center}

The diagrams are labelled in boldface as follows: 
any diagram containing $\Sigma$ as a vertex
argument has two labels: the first corresponds to the 
Wilsonian effective action part and
the second corresponds to the seed action part. Diagrams with only a Wilsonian
effective action vertex or a seed action vertex have a single label.
If the reference number of a diagram
is followed by an arrow, it can mean one of two things:
\begin{enumerate}
	\item 	$\rightarrow 0$ denotes that the corresponding 
			diagram can be set to zero, for some reason;

	\item 	$\rightarrow$ followed by a number (other than zero)
			indicates the number of the figure in which the corresponding
			diagram is processed.
\end{enumerate}
If a diagram is cancelled, then its reference number is enclosed in curly braces,
together with the reference number of the diagram against which it cancels.
A diagram will not be taken as cancelled until the diagram against
which it cancels has been explicitly generated. Thus, at various stages
of the calculation where we collate surviving terms, we include those diagrams
whose cancelling partner does not yet exist.

Returning to \fig{fig:Beta1-LevelZero}, 
the first three diagrams  are formed by the $a_1[\Sigma_0]$ term and 
the last two are formed by the 
$a_0[S_0,\hat{S}_1]$ term.  
We do not draw any diagrams possessing either a one-point, tree level vertex
or a kernel which bites its own tail.
In the third diagram, we have used the equality between 
Wilsonian effective action and seed action two-point, tree
level vertices
to replace $\Sigma_{0 RS}^{\ XX}(k)$ with $-S_{0 RS}^{\ XX}(k)$. 
The final diagram vanishes: the one-point
vertex must be in the $C$-sector but, 
since an $\dot{\Delta}^{AC,A}$ kernel does not exist, it is not possible to form a
legal diagram.

Note that we have not included the diagram which can be obtained from 
diagram~\ref{d1:C-2}
by taking the field on the kernel and placing it on the top-most vertex, 
since such a term vanishes
at $\Op{2}$: the vertex $S_{0 \mu \alpha}^{\ 1 \, 1}(p)$ is, as we know 
already, at least $\Op{2}$;
the same too applies to $\hat{S}_{1 \mu \alpha}^{\ 1 \, 1}(p)$, as 
a consequence of the
Ward identity~\eq{eq:WID-U}.\footnote{There is no argument that the $\Op{2}$ 
part of $\hat{S}_{n\geq 1 \mu \nu}^{\ \ \ \ 1\, 1}(p)$ vanishes,
since the renormalisation condition does not apply to the seed action.}

\subsubsection{Diagrammatic Manipulations}

As it stands, we cannot directly extract a value for $\beta_1$ 
from eqn.~\eq{eq:Beta1-Defining}.
The \rhs\ is phrased in terms of non-universal objects. 
Whilst one approach would be to choose
particular schemes in which these objects are defined
algebraically (up to a choice of cutoff functions)~\cite{ymii} we 
know from~\cite{aprop}
that this is unnecessary: owing to the universality of $\beta_1$, all
non-universalities must somehow cancel out. To proceed, we utilise the flow equations.

Our aim is to try and reduce the expression for $\beta_1$ to a 
set of $\Lambda$-derivative terms---terms where
the entire diagram is hit by $\Lambda \partial_\Lambda|_\alpha$---since, 
such terms either vanish directly or 
combine to give only universal contributions 
(in the limit that $D \rightarrow 4$)~\cite{aprop,Thesis,mgierg2}.

The approach we use is to start with the term containing the highest point Wilsonian
effective action vertex.
By focusing on the term with the highest point vertex,
we guarantee that the kernel in the diagram is un-decorated. Now,
we know that an un-decorated kernel is $-\flowConstAl$ of an effective propagator.
Hence, up to a term in which the entire diagram is hit by $-\flowConstAl$,
we can move the $-\flowConstAl$ from the effective propagator to the vertex.
This step is only useful if the vertex is a Wilsonian effective action vertex,
for now it can be processed, using the flow equations.

From \fig{fig:Beta1-LevelZero} it is clear that the highest 
point Wilsonian effective
action vertex in our calculation of $\beta_1$
is the four-point, tree level vertex contained in 
diagram~\ref{d1:1.2(W)}. 
The manipulation of this diagram is shown in \fig{fig:Beta1:FirstManip}. 
For the time being,
we will always take the $\Lambda$-derivative to act before we 
integrate over loop momenta (this is fully 
discussed in~\cite{Thesis,mgierg2}).
\begin{center}
\begin{figure}[h]
	\[
	\frac{1}{2} \ensuremath{\begin{array}{c}\input{pstex/Beta1-1.2.pstex_t} \end{array}} 
	= 
	\frac{1}{2}
	\dec{
		\PDi{Beta1-LdL-A}{d1:1.4b}{fig:beta1}
	}{\bullet}	
	-\frac{1}{2}
	\PDi{Beta1-1.2c}{d1:1.2(W):Manip}{fig:beta1:LevelOne}
	\]
\caption[The manipulation of diagram~\ref{d1:1.2(W)}.]{
The manipulation of diagram~\ref{d1:1.2(W)}. 
In the final diagram, the $\Lambda$-derivative operates 
on just the four-point vertex.}
\label{fig:Beta1:FirstManip}
\end{figure}
\end{center}

\begin{center}
\begin{figure}
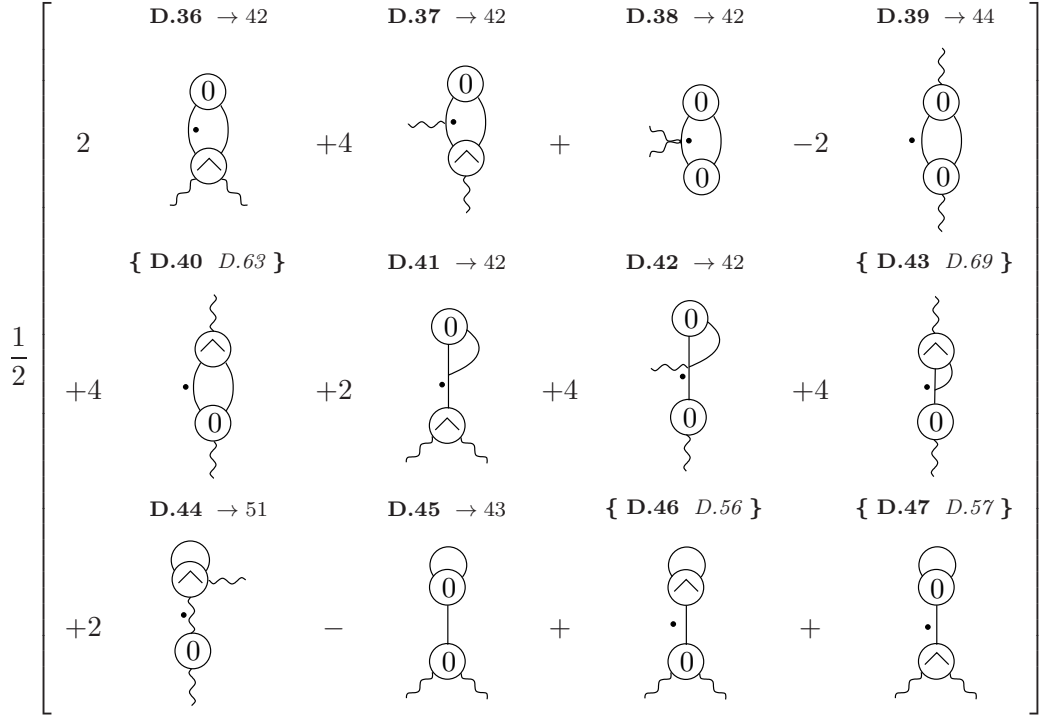

	\[
	\frac{1}{2}
	\left[
		\begin{array}{cccccccc}
				&\PD{d1:1.5,6}{fig:beta1:LevelOne:Manip}&	&\PD{d1:1.8,9}{fig:beta1:LevelOne:Manip}&	&\PD{d1:1.12:pre}{fig:beta1:LevelOne:Manip}	
				&	&\PD{d1:1.15}{fig:beta1:LevelTwo-B}
		\\[1ex]
			2	& \ensuremath{\begin{array}{c}\input{pstex/d1-1.5,6.pstex_t} \end{array}}							&+4	&\ensuremath{\begin{array}{c}\input{pstex/d1-1.8,9.pstex_t} \end{array}}							&+	&\ensuremath{\begin{array}{c}\input{pstex/d1-1.12-pre.pstex_t} \end{array}}							
				&-2	&\ensuremath{\begin{array}{c}\input{pstex/d1-1.15.pstex_t} \end{array}}
		\\[1ex]
				&\CD{d1:1.15b}{d1:1.19b}&	&\PD{d1:1.10:pre}{fig:beta1:LevelOne:Manip}	&	&\PD{d1:1.13:pre}{fig:beta1:LevelOne:Manip}	&	&\CD{d1:1.11}{d1:1.24}
		\\[1ex]
			+4	&\ensuremath{\begin{array}{c}\input{pstex/d1-1.15b.pstex_t} \end{array}}			&+2	&\ensuremath{\begin{array}{c}\input{pstex/d1-1.10-pre.pstex_t} \end{array}}							&+4	&\ensuremath{\begin{array}{c}\input{pstex/d1-1.13-pre.pstex_t} \end{array}}							&+4	&\ensuremath{\begin{array}{c}\input{pstex/d1-1.11.pstex_t} \end{array}}
		\\[1ex]
				&\PD{d1:1.7}{fig:beta1:c0}	&	&\PD{d1:1.14}{fig:beta1:LevelTwo-A}	&	&\CD{d1:1.14b}{d1:1.15d}&	&\CD{d1:1.14c}{d1:1.15e}
		\\[1ex]
			+2	&\ensuremath{\begin{array}{c}\input{pstex/d1-1.7.pstex_t} \end{array}}				&-	&\ensuremath{\begin{array}{c}\input{pstex/d1-1.14.pstex_t} \end{array}}						&+	&\ensuremath{\begin{array}{c}\input{pstex/d1-1.14b.pstex_t} \end{array}}			&+	&\ensuremath{\begin{array}{c}\input{pstex/d1-1.14c.pstex_t} \end{array}}
		\end{array}
	\right]
	\]
\caption{The result of manipulating diagram~\ref{d1:1.2(W):Manip}, using the tree-level flow equation.}
\label{fig:beta1:LevelOne}
\end{figure}
\end{center}

We can now use the tree-level flow eqn.~\eq{eq:TreeLevelFlow} 
to process the $\Lambda$-derivative of the four-point vertex. 
The flow of a four-point vertex with two $A^1$ fields and two wildcards 
is shown in \fig{fig:App:TLFP-A1A1XX} (appendix~\ref{app:Ex-ClassicalFlows}). Throwing
away all terms which vanish at $\Op{2}$ and joining the wildcards
together with an effective propagator, we arrive at 
\fig{fig:beta1:LevelOne}.

We now arrive at a key juncture in the diagrammatic
procedure: Diagrams~\ref{d1:1.5,6}--\ref{d1:1.12:pre}, 
\ref{d1:1.10:pre} and~\ref{d1:1.13:pre}
 can be 
further manipulated using the effective propagator relation.
The results of this procedure are shown in 
\fig{fig:beta1:LevelOne:Manip}. 
Diagrams in which a kernel bites its own tail have been discarded as 
have those in which 
a gauge remainder strikes a two-point vertex (see eqn.~\eq{eq:GR-TLTP}). 
Henceforth, we will assume that such terms have always
been discarded.
\begin{center}
\begin{figure}[h]
	\[
	\frac{1}{2}
	\left[
		\begin{array}{c}
			\begin{array}{cccccccc}
					&\CD{d1:1.5}{d1:1.2(S)}	&	&\PD{d1:1.6}{fig:beta1:GR1-A}	&	&\CD{d1:1.8}{d1:1.3(S)}	&	& \CD{d1:1.9}{d1:1.23}
			\\[1ex]
				2	& \ensuremath{\begin{array}{c}\input{pstex/d1-1.5.pstex_t} \end{array}}			&-2	&\ensuremath{\begin{array}{c}\input{pstex/d1-1.6.pstex_t} \end{array}}				&+4	&\ensuremath{\begin{array}{c}\input{pstex/d1-1.8.pstex_t} \end{array}}			&-4	&\ensuremath{\begin{array}{c}\input{pstex/d1-1.9.pstex_t} \end{array}}
			\end{array}
		\\[1ex]
			\begin{array}{cccccc}
					&\CD{d1:1.12}{d1:1.4}	&	&\PD{d1:1.10}{fig:beta1:GR1-A}&	&\PD{d1:1.13}{fig:beta1:GR1-B}
			\\[1ex]
				+	&\ensuremath{\begin{array}{c}\input{pstex/d1-1.4.pstex_t} \end{array}}			&-2	&\ensuremath{\begin{array}{c}\input{pstex/d1-1.10.pstex_t} \end{array}}				&-4	&\ensuremath{\begin{array}{c}\input{pstex/d1-1.13.pstex_t} \end{array}}
			\end{array}
		\end{array}
	\right]	
	\]
\caption{Manipulation of diagrams~\ref{d1:1.5,6}--\ref{d1:1.12:pre}, 
\ref{d1:1.10:pre} and~\ref{d1:1.13:pre}.}
\label{fig:beta1:LevelOne:Manip}
\end{figure}
\end{center}

Three of the diagrams generated exactly cancel
the contributions in the first row of 
\fig{fig:Beta1-LevelZero} 
containing seed action vertices 
(or Wilsonian effective action two-point, tree level vertices).

\Cancel{d1:1.5}{d1:1.2(S)}
\Cancel{d1:1.8}{d1:1.3(S)}
\Cancel{d1:1.12}{d1:1.4}

Other than the $\Lambda$-derivative term, diagram~\ref{d1:1.4b}, 
there are now only two diagrams left which contain four-point vertices.
The first of these, diagram~\ref{d1:1.7}, 
can be manipulated at $\Op{2}$ since the bottom vertex is at least $\Op{2}$ and
so the rest of the diagram
can be Taylor expanded to zeroth order in  $p$.
Note, in particular, that given the effective propagator
$\Delta^{11}(p) \sim \mathsf{B}(p^2/\Lambda^2) / p^2$~\cite{aprop,Thesis}, 
the differentiated kernel which attaches to the two-point
vertex contributes the non-universal factor $2\BPZ/ \Lambda^2$, where
the prime denotes differentiation \wrt\ the argument.
We henceforth call terms of this type \BPZ\ terms.
If we were to
explicitly perform this Taylor expansion, we would
reduce the four-point vertex to the (double) momentum
derivative of a two-point vertex~\cite{aprop}.\footnote{
As we will see in \sec{sec:beta1:A'(0)} this manipulation
is not necessary as there is a more elegant way to deal with the diagram.}
In the second of the diagrams containing a four-point vertex, 
diagram~\ref{d1:1.6}, the vertex is hit by a gauge 
remainder and so will automatically be reduced
to a three-point vertex. Thus the effect of our manipulations 
is to ensure that all occurrences of the highest-point vertex in 
the calculation occur only in a $\Lambda$-derivative term.

Before moving on to the next stage of the calculation, 
we compare our current expression to that of reference~\cite{aprop}. 
Ignoring the
multiple supertrace terms contained within each of our diagrams, 
the two expressions are superficially
the same, up to diagrams~\ref{d1:1.10:pre} and~\ref{d1:1.14}--\ref{d1:1.14c}.
In each of these terms, the internal field joining the two 
three-point vertices must be in the $C$-sector. If it is in the $F$-sector, then
each diagram vanishes because net fermionic vertices vanish. 
If it is in the $A$-sector, then charge conjugation invariance 
causes the diagrams to 
vanish when we sum over permutations of the 
bottom vertex. Looking a little
harder, we see a related difference between the 
current expression and that of reference~\cite{aprop}:
amongst the components of diagrams~\ref{d1:1.15} and~\ref{d1:1.15b} 
are diagrams possessing $AAC$ vertices.

Now, with the aim of removing all three-point vertices 
from the calculation (up to $\Lambda$-derivative terms), we iterate the procedure.
Referring to \fig{fig:beta1:LevelOne}, 
only diagrams~\ref{d1:1.14} and~\ref{d1:1.15} 
possess exclusively Wilsonian effective
action vertices and an un-decorated kernel
and so it is these which we manipulate.

\Fig{fig:beta1:LevelTwo-A} shows the 
manipulation of diagram~\ref{d1:1.14} which proceeds
along exactly the same lines as the manipulations 
of \fig{fig:Beta1:FirstManip}. 
This time, however, we utilise \fig{fig:App:TLThP-WildCardx3}
for the flow of a three-point vertex with three wildcard 
fields and \fig{fig:App-TLThP-A1A1X} for the flow of
a three-point vertex containing two 
external $A^1$s (which carry moment $p$ and $-p$). In this 
latter case, we discard all terms which
vanish at $\Op{2}$.
\begin{center}
\begin{figure}[h]
	\[
	\begin{array}{c}
	\vspace{2ex}
		\ds 
		-\frac{1}{2}
		\dec{
			\PDi{Beta1-LdL-D}{d1:1.15c}{fig:beta1}
		}{\bullet}
		-\frac{1}{2}
		\left[
			\begin{array}{ccc}
				\CD{d1:1.15d}{d1:1.14b}	&	& \CD{d1:1.15e}{d1:1.14c}
			\\[1ex]
				\ensuremath{\begin{array}{c}\input{pstex/d1-1.15d.pstex_t} \end{array}}			& +	& \ensuremath{\begin{array}{c}\input{pstex/d1-1.15e.pstex_t} \end{array}}
			\end{array}
		\right]
	\\
		\ds
		+\frac{1}{2}
		\left[
			\begin{array}{ccccccc}
				\CDB{d1:1.16}{d1:C-1}	&	&\CDC{d1:1.16b}{d1:C-1}	&	&\PD{d1:1.17}{fig:beta1:GR1-A}&	&\PD{d1:1.18}{fig:beta1:GR1-A}
			\\[1ex]
				\ensuremath{\begin{array}{c}\input{pstex/d1-1.16.pstex_t} \end{array}}			&-	&\ensuremath{\begin{array}{c}\input{pstex/d1-1.16b.pstex_t} \end{array}}			&+2	&\ensuremath{\begin{array}{c}\input{pstex/d1-1.17.pstex_t} \end{array}}				&+2	&\ensuremath{\begin{array}{c}\input{pstex/d1-1.18.pstex_t} \end{array}}
			\end{array}
		\right]
	\end{array}
	\]
\caption{Result of the manipulation of diagram~\ref{d1:1.14} using the tree level flow equations.}
\label{fig:beta1:LevelTwo-A}
\end{figure}
\end{center}

As with diagrams~\ref{d1:1.2(S)}, \ref{d1:1.3(S)} and~\ref{d1:1.4},
we find that diagrams of the same structure as the parent but 
containing a seed action
vertex are cancelled. 
\Cancel{d1:1.15d}{d1:1.14b}
\Cancel{d1:1.15e}{d1:1.14c}

We also find, as promised, that the sole instance of a 
one-point, seed action vertex is cancelled.
\begin{cancel}
Diagrams~\ref{d1:1.16}  and~\ref{d1:1.16b} exactly cancel diagram~\ref{d1:C-1}
by virtue of the weak coupling expansion of the
constraint~\eq{eq:NPC-sec-Constraint}.
\label{cancel:d1:1.16}
\end{cancel}

In this context, the notation used in \fig{fig:beta1:LevelTwo-A} to 
describe the cancellation of diagram~\ref{d1:C-1} has an obvious
interpretation. Note that 
the only surviving terms from \fig{fig:beta1:LevelTwo-A} 
both contain gauge remainders.

\begin{center}
\begin{figure}[h]
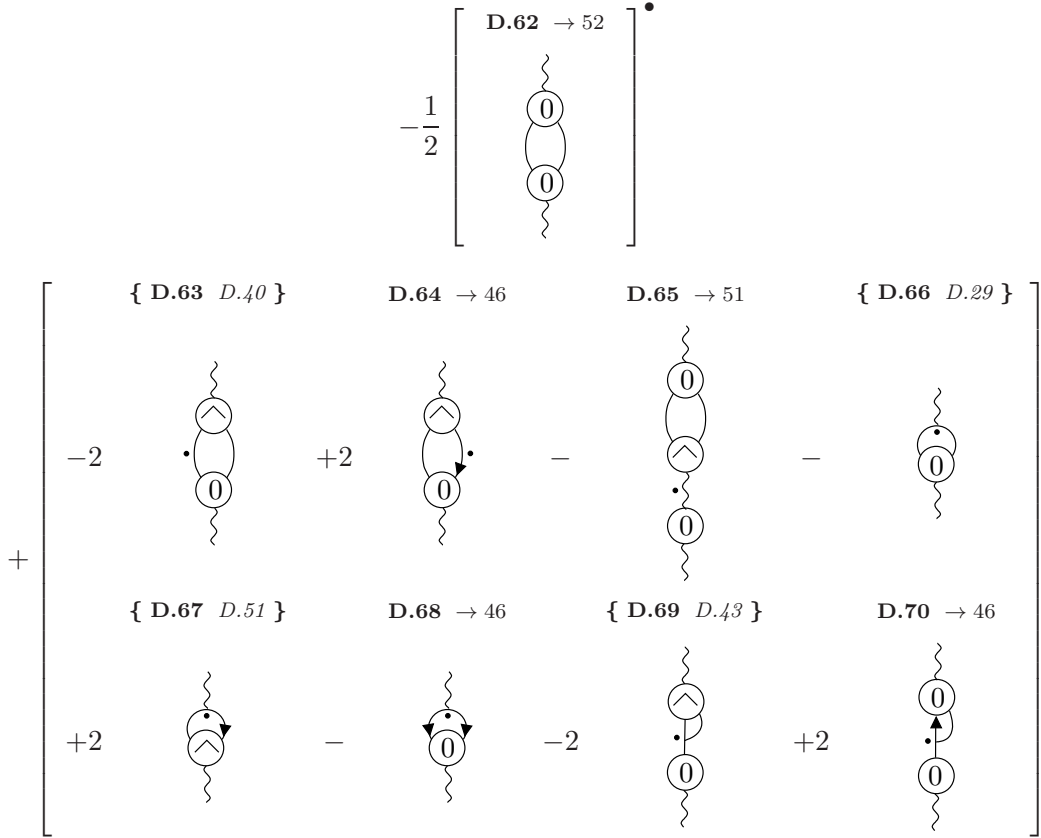

	\[
	\begin{array}{c}
	\vspace{2ex}
		\ds
		-\frac{1}{2}
		\dec{
			\PDi{Beta1-LdL-B}{d1:1.19}{fig:beta1}
		}{\bullet}
	\\
		+
		\left[
			\begin{array}{cccccccc}
					&\CD{d1:1.19b}{d1:1.15b}&	&\PD{d1:1.20}{fig:beta1:GR1-B}&	&\PD{d1:1.21}{fig:beta1:c0}	&	&\CD{d1:1.22}{d1:1.3(W)}
			\\[1ex]
				-2	&\ensuremath{\begin{array}{c}\input{pstex/d1-1.15b.pstex_t} \end{array}}			&+2	&\ensuremath{\begin{array}{c}\input{pstex/d1-1.20.pstex_t} \end{array}}				&-	&\ensuremath{\begin{array}{c}\input{pstex/d1-1.21.pstex_t} \end{array}}				&-	&\ensuremath{\begin{array}{c}\input{pstex/d1-1.22.pstex_t} \end{array}}
			\\[1ex]
					&\CD{d1:1.23}{d1:1.9}	&	&\PD{d1:1.24b}{fig:beta1:GR1-B}	&	&\CD{d1:1.24}{d1:1.11}	&	&\PD{d1:1.25}{fig:beta1:GR1-B}
			\\[1ex]
				+2	&\ensuremath{\begin{array}{c}\input{pstex/d1-1.9.pstex_t} \end{array}}			&-	&\ensuremath{\begin{array}{c}\input{pstex/d1-1.24b.pstex_t} \end{array}}					&-2	&\ensuremath{\begin{array}{c}\input{pstex/d1-1.11.pstex_t} \end{array}}			&+2	&\ensuremath{\begin{array}{c}\input{pstex/d1-1.25.pstex_t} \end{array}}
			\end{array}
		\right]
	\end{array}
	\]
\caption{Result of the manipulation of diagram~\ref{d1:1.15} using the tree level flow equation.}
\label{fig:beta1:LevelTwo-B}
\end{figure}
\end{center}

\Fig{fig:beta1:LevelTwo-B} shows the manipulation of 
diagram~\ref{d1:1.15}, where the overall 
factor of $1/2$ arises from
the symmetry of the $\Lambda$-derivative term~\ref{d1:1.19} 
under rotations by $\pi$ (alternatively, the indistinguishability of the
two internal fields). When we compute the flow of the 
vertices of diagram~\ref{d1:1.19}, we can 
utilise this symmetry to remove the
factor of $1/2$.

We find a number of cancellations. The first of these is the 
expected cancellation of
the partner of the parent diagram, possessing a seed action vertex.
\Cancel{d1:1.19b}{d1:1.15b}

The next cancellation completes the removal of all
terms from \fig{fig:Beta1-LevelZero}
formed by the action of $a_0[\Sigma_0]$.

\Cancel{d1:1.22}{d1:1.3(W)}

Of the remaining cancellations, one involves two 
diagrams, each possessing active gauge remainders,
which we notice can be cancelled without the need to perform the
gauge remainders. The final cancellation occurs only at $\Op{2}$.

\CancelCom{d1:1.23}{d1:1.9}{. When a three-point tree level 
vertex is struck by 
a gauge remainder it is reduced to a two-point,
tree level vertex. Since Wilsonian effective action two-point, tree
level vertices are equal to the corresponding seed action vertices,
 it thus makes no difference whether the original three-point
vertex is a Wilsonian effective
action vertex or a seed action vertex.}

\CancelOpCom{d1:1.24}{d1:1.11}{. In each case, we Taylor 
expand the three-point, tree level vertex
to zeroth order in external momentum, reducing it to the momentum derivative of
a two-point, tree level vertex.
It is then of no consequence that one of the three-point, 
tree level vertices was a 
seed action vertex whereas the other was a Wilsonian 
effective action vertex.}

At this stage, up to diagrams in which
the sole three-point vertex is hit by a gauge remainder, 
we have removed all three-point, tree level vertices
from the calculation with the following exceptions:
\begin{enumerate}
	\item 	the $\Lambda$-derivative terms, diagrams~\ref{d1:1.19} 
			and~\ref{d1:1.15c};

	\item 	the \BPZ\ term, diagram \ref{d1:1.21};

	\item 	three diagrams~\ref{d1:1.10}, \ref{d1:1.17} and~\ref{d1:1.18}, 
			which are each left with a three-point, tree level vertex,
			even after the action of the gauge remainders.
\end{enumerate}

The last three terms, which  all possess an
$S_{0 \mu \nu}^{\ 1 \, 1 C^{1,2}}$ vertex, 
have no analogue in
the version of the calculation presented in~\cite{aprop}.
To make further progress, we must process the gauge remainders.
In \figs{fig:beta1:GR1-A}{fig:beta1:GR1-B}, we utilise the 
techniques of \sec{sec:GRs} to 
manipulate the gauge remainders, stopping after the use of the
effective propagator relation. We discard all terms 
which vanish due to their supertrace structure
being $\str A^1_\mu \, \str A^1_\nu$ and neglect attachment corrections to
three-point vertices which are decorated by an 
external field and struck by a gauge remainder (\cf \fig{fig:Diags:GR-SuperCorrs}).
\begin{center}
\begin{figure}[h]
	\[
	\begin{array}{l}
	\vspace{0.2in}
		\ref{d1:1.6} = 2 
		\left[
			\begin{array}{cccc}
					&\CD{d1:2.1}{d1:2.10}	&	&\CD{d1:2.2}{d1:2.4(S)}
			\\[1ex]
				2	&\ensuremath{\begin{array}{c}\input{pstex/d1-2.1.pstex_t} \end{array}}			&-	&\ensuremath{\begin{array}{c}\input{pstex/d1-2.2.pstex_t} \end{array}}
			\end{array}
		\right]
	\\
	\vspace{0.2in}
		\ref{d1:1.10} + \ref{d1:1.18} = 2 
		\left[
			\begin{array}{ccc}
				\CDBl{d1:2.3}{d1:2.15b}	&	&\CDCD{d1:2.4(W)}{d1:2.8}{d1:2.4(S)}{d1:2.2}
			\\[1ex]
				\ensuremath{\begin{array}{c}\input{pstex/d1-2.3.pstex_t} \end{array}}				&-	&\ensuremath{\begin{array}{c}\input{pstex/d1-2.4.pstex_t} \end{array}}
			\end{array}
		\right]
	\\
	\vspace{0.2in}
		 \ref{d1:1.17} = 2
		\left[
			\begin{array}{ccc}
				\CD{d1:2.8}{d1:2.4(W)}	&	&\PD{d1:2.9}{fig:beta1:LdLConversion}
			\\[1ex]
				\ensuremath{\begin{array}{c}\input{pstex/d1-2.8.pstex_t} \end{array}}				&-	&\ensuremath{\begin{array}{c}\input{pstex/d1-2.9.pstex_t} \end{array}}
			\end{array}
		\right]
	\end{array}
	\]
\caption{Terms arising from processing the gauge remainders
of the diagrams in \figs{fig:beta1:LevelTwo-A}{fig:beta1:LevelTwo-B} part~I.}
\label{fig:beta1:GR1-A}
\end{figure}
\end{center}

\begin{center}
\begin{figure}
	\[
	\begin{array}{l}
	\vspace{0.2in}
		\ref{d1:1.20} = -4 
		\left[
			\begin{array}{ccccc}
				\CD{d1:2.10}{d1:2.1}&	& \PD{d1:2.11}{fig:beta1:GR2}	&	&\PD{d1:2.12}{fig:beta1:Op2}
			\\[1ex]
				\ensuremath{\begin{array}{c}\input{pstex/d1-2.1.pstex_t} \end{array}}			& -	& \ensuremath{\begin{array}{c}\input{pstex/d1-2.11.pstex_t} \end{array}}					&-	&\ensuremath{\begin{array}{c}\input{pstex/d1-2.12.pstex_t} \end{array}}
			\end{array}
		\right]
	\\
	\vspace{0.2in}
		\ref{d1:1.13} = 4 
		\left[
			\begin{array}{ccccc}
				\PD{d1:2.7-pre}{fig:beta1:Op2}	&	&\CD{d1:2.5}{d1:2.13}	&	& \Discard{d1:2.6}
			\\[1ex]
				\ensuremath{\begin{array}{c}\input{pstex/d1-2.7-pre.pstex_t} \end{array}}					& +	& \ensuremath{\begin{array}{c}\input{pstex/d1-2.5.pstex_t} \end{array}}			& - & \ensuremath{\begin{array}{c}\input{pstex/d1-2.6.pstex_t} \end{array}}
			\end{array}
		\right]
	\\
		\ref{d1:1.24b} = 2
		\left[
			\PDi{Beta1-GR-G}{d1:2.16:pre}{fig:beta1:Op2}
		\right]
	\hspace{3em}
		\ref{d1:1.25} = -4  
		\left[
			\begin{array}{ccc}
				\CD{d1:2.13}{d1:2.5}	&	&\PD{d1:2.14}{fig:beta1:GR2}
			\\[1ex]
				\ensuremath{\begin{array}{c}\input{pstex/d1-2.5.pstex_t} \end{array}}			& -	& \ensuremath{\begin{array}{c}\input{pstex/d1-2.14.pstex_t} \end{array}}
			\end{array}
		\right] 
	\end{array}
\]
\caption{Terms arising from processing the gauge remainders
of the diagrams in \figs{fig:beta1:LevelTwo-A}{fig:beta1:LevelTwo-B} part~II.}
\label{fig:beta1:GR1-B}
\end{figure}
\end{center}

Note that diagram~\ref{d1:2.16:pre}
is our first example of a diagram possessing a trapped gauge remainder.
The full gauge remainder is prevented from killing the two-point,
tree level vertex by the processed gauge remainder: the vertex
and the full gauge remainder do not carry the same momentum.
There is no corresponding diagram in which the gauge remainder
instead bites the external field because then we are left with an active
gauge remainder striking a two-point, tree level vertex.
Note also that all diagrams which either cannot be processed further
or do not contain an $S_{0 \mu \nu}^{\ 1 \, 1 C^{1,2}}$ vertex cancel,
amongst themselves.
\Cancel{d1:2.4(W)}{d1:2.8}
\Cancel{d1:2.4(S)}{d1:2.2}
\Cancel{d1:2.10}{d1:2.1}
\Cancel{d1:2.5}{d1:2.13}

\begin{center}
\begin{figure}[h]
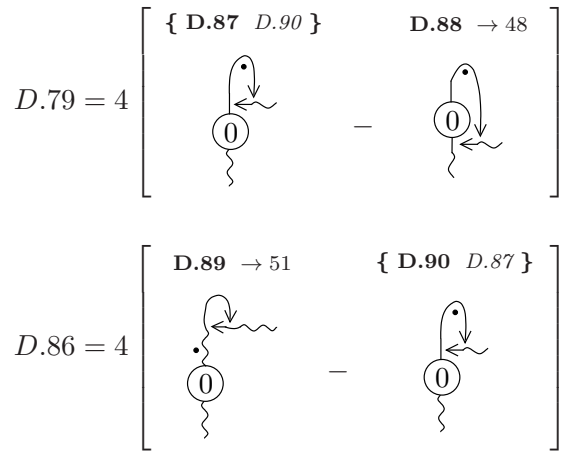

	\[
	\begin{array}{c}
		\ref{d1:2.11} = 4
		\left[
			\begin{array}{ccc}
				\CD{d1:2.17}{d1:2.20}	&	&\PD{d1:2.18}{fig:beta1:LdLConversion}
			\\[1ex]	
				\ensuremath{\begin{array}{c}\input{pstex/d1-2.17.pstex_t} \end{array}}			& -	&\ensuremath{\begin{array}{c}\input{pstex/d1-2.18.pstex_t} \end{array}}
			\end{array}
		\right] 
	\\[10ex]
		\ref{d1:2.14} = 4
		\left[
			\begin{array}{ccc}
				\PD{d1:2.19}{fig:beta1:c0}	&	&\CD{d1:2.20}{d1:2.17}
			\\[1ex]	
				\ensuremath{\begin{array}{c}\input{pstex/d1-2.19.pstex_t} \end{array}}				& -	&\ensuremath{\begin{array}{c}\input{pstex/d1-2.17.pstex_t} \end{array}}
			\end{array}
		\right]
	\end{array}
\]
\caption{Diagrams arising from processing the nested gauge 
remainders of \fig{fig:beta1:GR1-B}.}
\label{fig:beta1:GR2}
\end{figure}
\end{center}

Whilst these cancellations are very encouraging, it is not
clear that we are any closer to solving the mystery of
the diagrams containing $S_{0 \mu \nu}^{\ 1 \, 1 C^{1,2}}$ vertices.
We will, however, persevere and 
process the nested gauge remainders arising from the
previous procedure. 
The result of this is shown in \fig{fig:beta1:GR2}.

Once again, we find a cancellation between a pair of the terms
generated by this procedure.

\Cancel{d1:2.20}{d1:2.17}

This exhausts the active gauge remainders and so is a good point to pause and
collate the surviving terms. These fall into five sets:
\begin{enumerate}
	\item 	The $\Lambda$-derivative terms, diagrams~\ref{d1:1.4b}, \ref{d1:1.15c} and~\ref{d1:1.19};
	
	\item 	The \BPZ\ terms, diagrams~\ref{d1:1.7}, \ref{d1:1.21} and~\ref{d1:2.19}. Notice
			that the last of these has been formed via the action of a nested gauge
			remainder;
	
	\item	Terms possessing an $\Op{2}$ stub formed by the action of a gauge remainder,
			diagrams~\ref{d1:2.16:pre}, \ref{d1:2.7-pre} and~\ref{d1:2.12}. Notice that
			the former of these has a trapped gauge remainder;

	\item	Terms possessing a $S_{0 \mu \nu}^{\ 1 \, 1 C^{1,2}}$ vertex,
			diagrams~\ref{d1:2.3} and~\ref{d1:2.9};

	\item	Diagram~\ref{d1:2.18}.
\end{enumerate}

We will leave the first three sets of diagrams, for the time being,
and focus on the final two. Remarkably, diagram~\ref{d1:2.9} from the fourth
set and diagram~\ref{d1:2.18} share a common feature: a two-point,
tree level vertex, attached to an undecorated kernel, which terminates
in a processed gauge remainder. This two-point, tree level
vertex cannot be directly removed by the effective propagator;
however, we can use a combination of the diagrammatic
identities~\eq{eq:NFE:EffPProp}--\eq{eq:LdL-GRk} to make progress:
\bea
	\ensuremath{\begin{array}{c}\input{pstex/D-ID-VWG-A.pstex_t} \end{array}} 	& =	& \dec{\ensuremath{\begin{array}{c}\begin{picture}(0,0)%
\includegraphics{pstex/D-ID-VWG-B.pstex}%
\end{picture}%
\setlength{\unitlength}{3947sp}%
\begingroup\makeatletter\ifx\SetFigFont\undefined%
\gdef\SetFigFont#1#2#3#4#5{%
  \reset@font\fontsize{#1}{#2pt}%
  \fontfamily{#3}\fontseries{#4}\fontshape{#5}%
  \selectfont}%
\fi\endgroup%
\begin{picture}(656,238)(376,-654)
\put(537,-596){\makebox(0,0)[lb]{\smash{\SetFigFont{11}{13.2}{\rmdefault}{\mddefault}{\updefault}{\color[rgb]{0,0,0}0}%
}}}
\end{picture}
 \end{array}}}{\bullet} - \ensuremath{\begin{array}{c}\input{pstex/D-ID-VWG-C.pstex_t} \end{array}} - \ensuremath{\begin{array}{c}\input{pstex/D-ID-VWG-D.pstex_t} \end{array}}
\nonumber \\
						& = & \dec{\ensuremath{\begin{array}{c}\begin{picture}(0,0)%
\includegraphics{pstex/D-ID-VGP.pstex}%
\end{picture}%
\setlength{\unitlength}{3947sp}%
\begingroup\makeatletter\ifx\SetFigFont\undefined%
\gdef\SetFigFont#1#2#3#4#5{%
  \reset@font\fontsize{#1}{#2pt}%
  \fontfamily{#3}\fontseries{#4}\fontshape{#5}%
  \selectfont}%
\fi\endgroup%
\begin{picture}(711,238)(376,-654)
\put(540,-593){\makebox(0,0)[lb]{\smash{\SetFigFont{11}{13.2}{\rmdefault}{\mddefault}{\updefault}{\color[rgb]{0,0,0}0}%
}}}
\end{picture}
 \end{array}}}{\bullet} - \ensuremath{\begin{array}{c}\input{pstex/D-ID-DVGP.pstex_t} \end{array}} - \ensuremath{\begin{array}{c}\begin{picture}(0,0)%
\includegraphics{pstex/D-ID-VWG-E.pstex}%
\end{picture}%
\setlength{\unitlength}{3947sp}%
\begingroup\makeatletter\ifx\SetFigFont\undefined%
\gdef\SetFigFont#1#2#3#4#5{%
  \reset@font\fontsize{#1}{#2pt}%
  \fontfamily{#3}\fontseries{#4}\fontshape{#5}%
  \selectfont}%
\fi\endgroup%
\begin{picture}(155,227)(878,-583)
\put(935,-440){\makebox(0,0)[lb]{\smash{\SetFigFont{8}{9.6}{\rmdefault}{\mddefault}{\updefault}{\color[rgb]{0,0,0}$\bullet$}%
}}}
\end{picture}
 \end{array}} + \ensuremath{\begin{array}{c}\begin{picture}(0,0)%
\includegraphics{pstex/D-ID-VWG-F.pstex}%
\end{picture}%
\setlength{\unitlength}{3947sp}%
\begingroup\makeatletter\ifx\SetFigFont\undefined%
\gdef\SetFigFont#1#2#3#4#5{%
  \reset@font\fontsize{#1}{#2pt}%
  \fontfamily{#3}\fontseries{#4}\fontshape{#5}%
  \selectfont}%
\fi\endgroup%
\begin{picture}(296,229)(736,-585)
\put(935,-440){\makebox(0,0)[lb]{\smash{\SetFigFont{8}{9.6}{\rmdefault}{\mddefault}{\updefault}{\color[rgb]{0,0,0}$\bullet$}%
}}}
\end{picture}
 \end{array}}
\nonumber\\
						& =	& -\ensuremath{\begin{array}{c}\begin{picture}(0,0)%
\includegraphics{pstex/D-ID-VWG-E.pstex}%
\end{picture}%
\setlength{\unitlength}{3947sp}%
\begingroup\makeatletter\ifx\SetFigFont\undefined%
\gdef\SetFigFont#1#2#3#4#5{%
  \reset@font\fontsize{#1}{#2pt}%
  \fontfamily{#3}\fontseries{#4}\fontshape{#5}%
  \selectfont}%
\fi\endgroup%
\begin{picture}(155,227)(878,-583)
\put(935,-440){\makebox(0,0)[lb]{\smash{\SetFigFont{8}{9.6}{\rmdefault}{\mddefault}{\updefault}{\color[rgb]{0,0,0}$\bullet$}%
}}}
\end{picture}
 \end{array}}.
\label{eq:D-ID-VWG}
\eea
(Strictly, we should consider the above diagrams to occur in some larger
diagrams, \cf~\eq{eq:NFE:EffPProp}.)

To go from the first line to the second, 
we have employed~\eq{eq:PseudoEP} and the effective
propagator relation. On the second line, the first term
vanishes, courtesy of~\eq{eq:GR-TLTP}; similarly the second
term if we also employ~\eq{eq:LdL-GRk}. The final term on
the second line vanishes courtesy of~\eqs{eq:GR-relation}{eq:LdL-GRk}:
\[
	\dec{ \GRk \!\! \GRkpr}{\bullet} = 0 = \stackrel{\bullet}{\GRk} \! \GRkpr + \GRk \! \! \stackrel{\bullet}{\GRkpr} = \GRk \!\!  \stackrel{\bullet}{\GRkpr}.
\]

Redrawing diagrams~\ref{d1:2.9} and~\ref{d1:2.18} using~\eq{eq:D-ID-VWG},
we find that they can be converted into $\Lambda$-derivative terms. The
result of this procedure is shown 
in \fig{fig:beta1:LdLConversion}, where we have discarded
any terms which vanish at $\Op{2}$.
\begin{center}
\begin{figure}[h]
	\[
	\begin{array}{lcccl}
		\ref{d1:2.9}	& =	& 2 \ensuremath{\begin{array}{c}\input{pstex/Beta1-2.9.pstex_t} \end{array}} 		& = & 
		2 
		\dec{
			\ensuremath{\begin{array}{c}\input{pstex/Beta1-2.15a.pstex_t} \end{array}}
		}{\bullet} 
		-\ensuremath{\begin{array}{c}\input{pstex/d1-2.15b.pstex_t} \end{array}} - \ensuremath{\begin{array}{c}\input{pstex/d1-2.15c.pstex_t} \end{array}}
	\\[10ex]
						&	&						& =	&
		2 
		\dec{
			\PDi{Beta1-2.15a}{d1:2.15a}{fig:beta1}
		}{\bullet}
		- 2 \CDi{d1-2.3}{d1:2.15b}{d1:2.3} + \Op{4}
	\\[10ex]	
		\ref{d1:2.18}	& =	&4 \ensuremath{\begin{array}{c}\input{pstex/Beta1-2.25pre.pstex_t} \end{array}}	& =	& 
		2 
		\dec{
			\PDi{Beta1-LdL-C}{d1:2.25}{fig:beta1}
		}{\bullet}
	\end{array}
	\]
\caption{Re-drawing of diagrams~\ref{d1:2.9} and~\ref{d1:2.18}
using~\eq{eq:D-ID-VWG} and their subsequent
conversion into $\Lambda$-derivative terms.}
\label{fig:beta1:LdLConversion}
\end{figure}
\end{center}

Note how diagram~\ref{d1:2.25} has a factor of $1/2$, relative to the parent diagram.
This recognises the indistinguishability of the two processed gauge remainders
possessed by this diagram.

Finally, we find the cancellation of the remaining diagrams possessing
a $S_{0 \mu \nu}^{\ 1 \, 1 C^{1,2}}$ vertex, up to those which are cast
as $\Lambda$-derivative terms.

\Cancel{d1:2.15b}{d1:2.3}

Our next task is to analyse the surviving diagrams possessing 
an $\Op{2}$ stub formed by
the action of a gauge remainder. This is a two-step process. First, we Taylor
expand each of the sub-diagrams attached to the stub to zeroth order 
in $p$.\footnote{In these cases, this process does not
generate IR divergences and so can be safely performed.} 
We then re-draw them, if possible, using various diagrammatic identities. 
The results of
the complete procedure are shown in \fig{fig:beta1:Op2}.
\begin{center}
\begin{figure}[h]
	\[
	\begin{array}{lcl}
	\vspace{0.2in}
		\ref{d1:2.7-pre} 	&\rightarrow& 4
		\CDi[0.1]{Beta1-2.24}{d1:2.7}{d1:2.24}
	\\
	\vspace{0.2in}
		\ref{d1:2.12} 		&\rightarrow& 4 \ensuremath{\begin{array}{c}\input{pstex/Beta1-Op2-A.pstex_t} \end{array}}
		= -4 
		\left[
			\begin{array}{ccccc}
				\PD{d1:2.21}{fig:beta1:FinalManip}	&	&\PD{d1:2.16b}{fig:beta1:FinalManip}&	&\CD{d1:2.16}{d1:2.22a}
			\\[1ex]
				\ensuremath{\begin{array}{c}\input{pstex/d1-2.21.pstex_t} \end{array}}						& +	& \ensuremath{\begin{array}{c}\input{pstex/d1-2.16b.pstex_t} \end{array}}						& +	&\ensuremath{\begin{array}{c}\input{pstex/d1-2.16.pstex_t} \end{array}}
			\end{array}
		\right] 
	\\
		\ref{d1:2.16:pre} 	&\rightarrow& -2 \ensuremath{\begin{array}{c}\input{pstex/Beta1-Op2-C.pstex_t} \end{array}}
		= 2
		\left[
			\begin{array}{ccc}
				\PD{d1:2.22b}{fig:beta1:FinalManip}	&	&\CD{d1:2.22a}{d1:2.16}
			\\[1ex]
				\ensuremath{\begin{array}{c}\input{pstex/d1-2.16b.pstex_t} \end{array}}						&+2	& \ensuremath{\begin{array}{c}\input{pstex/d1-2.16.pstex_t} \end{array}}
			\end{array}
		\right]
\end{array}
\]
\caption{Manipulations at $\Op{2}$, followed by a re-expression of the resulting diagrams.}
\label{fig:beta1:Op2}
\end{figure}
\end{center}

Some comments are in order. Having Taylor expanded diagram~\ref{d1:2.12},
we obtain the final set of diagrams via the 
diagrammatic relation, 
\bea
		\ensuremath{\begin{array}{c}\begin{picture}(0,0)%
\includegraphics{pstex/D-ID-dVE.pstex}%
\end{picture}%
\setlength{\unitlength}{3947sp}%
\begingroup\makeatletter\ifx\SetFigFont\undefined%
\gdef\SetFigFont#1#2#3#4#5{%
  \reset@font\fontsize{#1}{#2pt}%
  \fontfamily{#3}\fontseries{#4}\fontshape{#5}%
  \selectfont}%
\fi\endgroup%
\begin{picture}(670,395)(335,-410)
\put(537,-351){\makebox(0,0)[lb]{\smash{\SetFigFont{11}{13.2}{\rmdefault}{\mddefault}{\updefault}{\color[rgb]{0,0,0}0}%
}}}
\end{picture}
 \end{array}}	& =	&	\ensuremath{\begin{array}{c}\begin{picture}(0,0)%
\includegraphics{pstex/D-ID-VdE.pstex}%
\end{picture}%
\setlength{\unitlength}{3947sp}%
\begingroup\makeatletter\ifx\SetFigFont\undefined%
\gdef\SetFigFont#1#2#3#4#5{%
  \reset@font\fontsize{#1}{#2pt}%
  \fontfamily{#3}\fontseries{#4}\fontshape{#5}%
  \selectfont}%
\fi\endgroup%
\begin{picture}(668,393)(928,-816)
\put(1138,-759){\makebox(0,0)[lb]{\smash{\SetFigFont{11}{13.2}{\rmdefault}{\mddefault}{\updefault}{\color[rgb]{0,0,0}0}%
}}}
\end{picture}
 \end{array}} - \cdeps{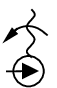},
\eea
which follows from the effective propagator
relation. The final diagram is interpreted as
the derivative \wrt\ the momentum entering the
encircled structure from the left. In diagrams~\ref{d1:2.16b}
and~\ref{d1:2.16} we have rewritten the (derivative of)
the full gauge remainder in terms of its constituent
parts.

In the case of diagram~\ref{d1:2.16:pre} the procedure after Taylor expansion
is different: we re-express it as a total momentum 
derivative---which we discard---plus
supplementary terms. These supplementary terms come with a 
relative minus sign and
correspond to the momentum derivative hitting all structures 
other than the kernel.
We then note that the two contributions in which the momentum derivative strikes
a $\GRkpr$ can be combined:
\[
	\ensuremath{\begin{array}{c}\begin{picture}(0,0)%
\includegraphics{pstex/Beta1-2.16-A.pstex}%
\end{picture}%
\setlength{\unitlength}{3947sp}%
\begingroup\makeatletter\ifx\SetFigFont\undefined%
\gdef\SetFigFont#1#2#3#4#5{%
  \reset@font\fontsize{#1}{#2pt}%
  \fontfamily{#3}\fontseries{#4}\fontshape{#5}%
  \selectfont}%
\fi\endgroup%
\begin{picture}(548,914)(632,-619)
\put(914,-316){\makebox(0,0)[lb]{\smash{\SetFigFont{11}{13.2}{\rmdefault}{\mddefault}{\updefault}{\color[rgb]{0,0,0}0}%
}}}
\end{picture}
 \end{array}} = \ensuremath{\begin{array}{c}\begin{picture}(0,0)%
\includegraphics{pstex/Beta1-2.16-B.pstex}%
\end{picture}%
\setlength{\unitlength}{3947sp}%
\begingroup\makeatletter\ifx\SetFigFont\undefined%
\gdef\SetFigFont#1#2#3#4#5{%
  \reset@font\fontsize{#1}{#2pt}%
  \fontfamily{#3}\fontseries{#4}\fontshape{#5}%
  \selectfont}%
\fi\endgroup%
\begin{picture}(586,911)(630,-619)
\put(914,-316){\makebox(0,0)[lb]{\smash{\SetFigFont{11}{13.2}{\rmdefault}{\mddefault}{\updefault}{\color[rgb]{0,0,0}0}%
}}}
\end{picture}
 \end{array}} = - \ensuremath{\begin{array}{c}\begin{picture}(0,0)%
\includegraphics{pstex/Beta1-2.16-C.pstex}%
\end{picture}%
\setlength{\unitlength}{3947sp}%
\begingroup\makeatletter\ifx\SetFigFont\undefined%
\gdef\SetFigFont#1#2#3#4#5{%
  \reset@font\fontsize{#1}{#2pt}%
  \fontfamily{#3}\fontseries{#4}\fontshape{#5}%
  \selectfont}%
\fi\endgroup%
\begin{picture}(484,868)(582,-619)
\put(914,-316){\makebox(0,0)[lb]{\smash{\SetFigFont{11}{13.2}{\rmdefault}{\mddefault}{\updefault}{\color[rgb]{0,0,0}0}%
}}}
\end{picture}
 \end{array}} = \ensuremath{\begin{array}{c}\begin{picture}(0,0)%
\includegraphics{pstex/Beta1-2.16-D.pstex}%
\end{picture}%
\setlength{\unitlength}{3947sp}%
\begingroup\makeatletter\ifx\SetFigFont\undefined%
\gdef\SetFigFont#1#2#3#4#5{%
  \reset@font\fontsize{#1}{#2pt}%
  \fontfamily{#3}\fontseries{#4}\fontshape{#5}%
  \selectfont}%
\fi\endgroup%
\begin{picture}(547,914)(826,-619)
\put(914,-316){\makebox(0,0)[lb]{\smash{\SetFigFont{11}{13.2}{\rmdefault}{\mddefault}{\updefault}{\color[rgb]{0,0,0}0}%
}}}
\end{picture}
 \end{array}},
\]
where the last step follows from first applying \CC\ and secondly
reversing the direction of the arrow on the derivative symbol,
at the expense of a minus sign.
\Cancel{d1:2.22a}{d1:2.16}
The four surviving diagrams possessing an $\Op{2}$ stub formed by the
action of a gauge remainder can now be combined into $\Lambda$-derivatives.
Diagram~\ref{d1:2.21} can be re-drawn via~\eq{eq:D-ID-VWG}  and then, 
together with diagram~\ref{d1:2.7}
converted into a $\Lambda$-derivative term.

Diagram~\ref{d1:2.22b} precisely halves the overall factor
of diagram \ref{d1:2.16b}. The resultant diagram is then re-drawn
using~\eq{eq:PseudoEP}, \eq{eq:GR-relation} and~\eq{eq:LdL-GRk}
which, upon inspection, yields a $\Lambda$-derivative term.
 This is all
shown in \fig{fig:beta1:FinalManip}. Note that we have taken
the $\Lambda$-derivative to strike entire diagrams rather than
just the sub-diagram attached to the stub. This step is
valid at $\Op{2}$.
\begin{center}
\begin{figure}[h]
	\[
	\begin{array}{l}
	\vspace{0.2in}
		\ref{d1:2.21} = 4 \ensuremath{\begin{array}{c}\input{pstex/Beta1-2.23pre.pstex_t} \end{array}}
		\rightarrow 4
		\dec{
			\PDi{Beta1-LdL-F}{d1:2.23}{fig:beta1}
		}{\bullet}
		-4 \CDi{Beta1-2.24}{d1:2.24}{d1:2.7}
	\\
		\ref{d1:2.16b} + \ref{d1:2.22b} = -2 \ensuremath{\begin{array}{c}\input{pstex/Beta1-2.26pre.pstex_t} \end{array}}
		\rightarrow 2 
		\dec{
			\PDi{Beta1-LdL-G}{d1:2.26}{fig:beta1}
		}{\bullet}
\end{array}
\]
\caption{The final conversion into $\Lambda$-derivative terms.}
\label{fig:beta1:FinalManip}
\end{figure}
\end{center}

In diagram~\ref{d1:2.26} we have moved the momentum derivative from
the $\GRk$ onto the pseudo effective propagator, discarding a total
momentum derivative, in the process.

\begin{center}
\begin{figure}[h]
\[
	\ensuremath{\begin{array}{c}\input{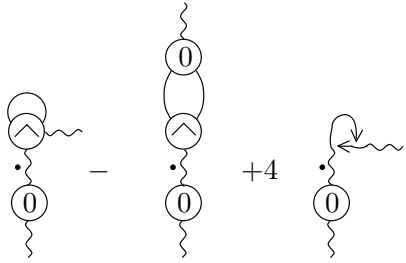} \end{array}}
\]
\caption{The set of \BPZ\ terms
(which do not manifestly vanish at $\Op{2}$).}
\label{fig:beta1:c0}
\end{figure}
\end{center}

We now find that all terms, other than the $\Lambda$-derivatives, have cancelled, with the sole exception of the \BPZ\ terms---which
have been collected together in \fig{fig:beta1:c0}.

In anticipation of the cancellation of the \BPZ\ terms, the
$\Lambda$-derivative terms have been collected together in \fig{fig:beta1} to give
 an entirely diagrammatic expression for $\beta_1$,
in terms of $\Lambda$-derivatives. 
\begin{center}
\begin{figure}[h]
	\be
	\label{eq:Beta1-Diag}
	4 \beta_1 \Box_{\mu \nu} (p) = 
	-\frac{1}{2}
	\dec{
		\begin{array}{c}
			\begin{array}{ccccccc}
				\LD{Beta1-LdL-A}	&	& \LD{Beta1-LdL-B}	&	& \LD{Beta1-LdL-C} 	&	&\LD{Beta1-LdL-D}
			\\[1ex]
				\ensuremath{\begin{array}{c}\input{pstex/Beta1-LdL-A.pstex_t} \end{array}}	& -	& \ensuremath{\begin{array}{c}\input{pstex/Beta1-LdL-B.pstex_t} \end{array}}	&+4	& \ensuremath{\begin{array}{c}\input{pstex/Beta1-LdL-C.pstex_t} \end{array}}	& -	&\ensuremath{\begin{array}{c}\input{pstex/Beta1-LdL-D.pstex_t} \end{array}}
			\end{array}
		\\[10ex]
			\begin{array}{cccccc}
					& \LD{Beta1-LdL-E}	&	& \LD{Beta1-LdL-F}	&	& \LD{Beta1-LdL-G}
			\\[1ex]
				+4	& \ensuremath{\begin{array}{c}\input{pstex/Beta1-LdL-E.pstex_t} \end{array}}	&+8	& \ensuremath{\begin{array}{c}\input{pstex/Beta1-LdL-F.pstex_t} \end{array}}	&+4	& \ensuremath{\begin{array}{c}\input{pstex/Beta1-LdL-G.pstex_t} \end{array}}
			\end{array}
		\end{array}
	}{\bullet}
	\ee
\caption{Diagrammatic, gauge invariant expression for $\beta_1$, phrased entirely in terms of $\Lambda$-derivatives.}
\label{fig:beta1}
\end{figure}
\end{center}

The diagrams in this expression will arise so many times in future 
that we will name them. 
The complete set of diagrams inside the square brackets will be 
referred to as $\OLDs$.
The first
three of these are henceforth referred to as the standard set. 
The last two will be known as the little set.

\subsection{The \BPZ\ Terms} \label{sec:beta1:A'(0)}

The first thing to notice about the \BPZ\ terms is that they are very similar
to the first three diagrams of \fig{fig:beta1}. Indeed, the \BPZ\ terms
are very nearly just the standard set joined to an $\Op{2}$ stub, via an un-decorated
kernel. The only difference is that the standard set contains exclusively Wilsonian
effective action vertices, whereas the \BPZ\ terms do not. However, we know
that the \BPZ\ terms can be manipulated, at $\Op{2}$. Doing this, we can replace the
four-point (three-point) seed action vertex with a double (single) momentum
derivative of a two-point, tree level vertex. Now, rather than making this
replacement, we use the equality of the two-point, tree level Wilsonian effective
action and seed action vertices to realise that, at $\Op{2}$, we can trade the seed
action vertices of the un-processed \BPZ\ terms for Wilsonian effective action vertices.
Now the \BPZ\ terms take the form of the standard set attached to an $\Op{2}$ stub, 
via an un-decorated kernel.

At this stage, we might wonder why the set of \BPZ\ terms does not contain diagrams
like the fourth and fifth of \fig{fig:beta1}. The answer is that we have discarded
these terms already, on the basis that they vanish at $\Op{2}$.\footnote{
\BPZ\ terms corresponding to the little set do not occur at all.}

There are several strategies to demonstrate that the \BPZ\ terms vanish. In 
reference~\cite{aprop}, it was demonstrated algebraically
that the \BPZ\ terms cancel, at $\Op{2}$: Taylor expanding the standard
set sub-diagrams to zeroth order in $p$, 
we can algebraically substitute for all constituent structures. There is, however, a much
more elegant way to proceed which minimises the algebra and is more intuitive.

Let us assume for the moment that  our calculation of $\beta_1$
is consistent (of course, part of the purpose of having performed this calculation
is to demonstrate this). Then we know that the set of diagrams contributing to $\beta_1 \Box_{\mu\nu}(p)$
must be transverse in $p$. The \BPZ\ terms are automatically transverse and so the only
diagrams not manifestly transverse are those constituting the standard set. For the calculation to be
consistent, then, the standard set must be transverse in $p$ and hence at least $\Op{2}$.
This immediately tells us that the \BPZ\ terms are at least $\Op{4}$ and so can be discarded.

Hence, our task is to demonstrate the transversality of the standard set.
\Fig{fig:SS-Transversality} shows the result of contracting one of the free indices of the standard set with its external
momentum where, as usual, we have used the techniques of \sec{sec:GRs}. 
\begin{center}
\begin{figure}
	\[
	\begin{array}{c}
		\ensuremath{\begin{array}{c}\input{pstex/SS-Transversality.pstex_t} \end{array}}
	\\[1ex]
		= 4
		\left[
			\ensuremath{\begin{array}{c}\input{pstex/SS-Transversality-B.pstex_t} \end{array}}
		\right]
	\end{array}
	\]
\caption{The result of contracting the standard set with its external momentum. The first three diagrams on the \rhs\ cancel and the
fourth vanishes by Lorentz invariance.}
\label{fig:SS-Transversality}
\end{figure}
\end{center}

Now we analyse the diagrams on the right hand side.
Algebraically, the first three terms go like:
\begin{eqnarray*}
\mbox{$A$-sector} & = &
4N\int_l \frac{l_\alpha}{l^2}\left(-1 + \frac{l\cdot(l+p)}{(l+p)^2} +  \frac{p\cdot(l+p)}{(l+p)^2} \right) \\
\mbox{$F$-sector} & = &
4(-N)
\int_l \frac{f_l l_\alpha}{\Lambda^2}\left(-1 + \frac{l\cdot(l+p)f_{l+p}}{\Lambda^2} + 2g_{l+p} + \frac{p\cdot(l+p)f_{l+p}}{\Lambda^2} \right),
\end{eqnarray*}
where the $UV$ finite sum vanishes after using the 
relationship~\eq{eq:GR-relation} and shifting momenta
(this is a special case of
diagrammatic identity~\ref{D-ID-Alg-2}).
The final term of \fig{fig:SS-Transversality} vanishes by Lorentz invariance: the $l$-integral contains a single index and
the only momentum available to carry this index, after integration over $l$, is $p$. However, this index is contracted into a vertex
transverse in $p$ and so the diagram vanishes.

Therefore, contracting the standard set with its external momenta yields zero. 
Since the standard set carries two Lorentz indices we have proven that it must
be transverse in external momenta, as predicted. This, then, guarantees that the \BPZ\
terms vanish, at $\Op{2}$, and also confirms the consistency of the calculation.

\section{Conclusions}
\label{sec:Conclusions}

The basis of this paper is the
modification of the flow equation 
of~\cite{aprop}, eqn.~\eq{sunnfl},
via the redefinition~\eq{weev}, 
thereby allowing straightforward
renormalisation at one loop and beyond.

At the heart of these changes is the
necessity to properly account for
multiple supertrace terms, in particular
lifting the restriction imposed in~\cite{aprop}
that the Wilsonian effective action comprises
only single supertrace terms. In turn, this guides us
 to the proper generalisation of the
flow equation~\eq{weev}, where now multiple
supertrace terms are effectively incorporated
into the covariantisation of the kernels. Crucially,
these generalisations must respect no-$\A^0$
symmetry, which plays a central \role\ in
properly understanding the broken phase diagrammatics.

The diagrammatic techniques of~\cite{aprop}
were essentially developed for single supertrace
calculations. With the \role\ of $\A^0$ obscured,
and $g(\Lambda) = g_2(\Lambda)$, it made sense
to work with the field $A =\tilde{A}^1 + \tilde{A}^2$,
which contains an $\A^0$ component. The
benefit of keeping $\A^0$, despite the invariance
of complete action functionals under no-$\A^0$ symmetry
was that 
the diagrammatics became particularly
simple, as both supersowing and supersplitting were
effectively exact, in all sectors.

Having modified the flow equation, our first task
was to suitably adapt the diagrammatic techniques.
Initially, in \sec{Diagrammatics}, this was done along 
the lines of~\cite{aprop}, retaining $\A^0$.
However, it rapidly became apparent that the 
benefits of exact supersowing / supersplitting were
really a red-herring. The generalisations of the
covariantised kernels to multi-supertrace objects
generates diagrams similar in structure to those
we were able to remove by working in the
exact supersowing / supersplitting picture. 
Thus, we were led to remove $\A^0$ from our picture, accepting
the corrections that this now generates, but recognising
that the overall diagrammatic structure is simplified.

Indeed, this inspired the new diagrammatic picture
of \sec{sec:NewDiags}, where we now package up 
the single and multi-supertrace terms.
In retrospect, that such a simplification is possible
is hardly surprising. Although all ingredients
in the treatment of~\cite{aprop}
were restricted to single supertrace terms, the
structure of the diagrammatic cancellations strongly
suggested that multi-supertrace terms, if included, would cancel
in the same way. Indeed, since all non-universal
contributions must cancel anyway, it is natural that
sets of them can be packaged up together and thus
removed in one go.

Moreover, our scheme now amounts to using standard
Feynman diagrammatic expansions, except that the
Feynman rules are novel and, embedded within the
diagrams, there is a prescription for automatically
evaluating the group theory factors.

An immediate consequence of these new diagrammatics
is that the calculation of~\cite{aprop} can be
essentially repeated, line for line. However, in 
the new way of doing things, multi-supertrace terms
come along for the ride, without really adding any
complication.\footnote{Effectively, the only additional terms
arise from the existence of $AAC$ vertices, which could be discarded
in the original treatment, as a consequence of the action
being restricted to single supertrace terms.}

Now, although much of the calculation of $\beta_1$
in~\cite{aprop} was done diagrammatically, these
techniques were not pushed to their limit, since 
gauge remainders and $\Op{2}$ terms were treated
algebraically.

In \sec{sec:FurtherDiagrammatics} we showed how
the gauge remainders and Taylor expansions
can be performed diagrammatically, allowing us
to reduce $\beta_1$ to a set of
$\Lambda$-derivative terms, as demonstrated in \sec{sec:beta1} .
This represents a radical
improvement over the approach in~\cite{aprop} and proves
crucial for performing calculations beyond one-loop.

Utilising the iterative approach to the one-loop
calculation, supplemented by the diagrammatic
identities of appendix~\ref{app:D-IDs}, we have reduced 
$\beta_2$ to a set of
$\Lambda$-derivative terms (and terms which vanish in the
$\alpha \rightarrow 0$ limit).\footnote{The manipulation
of a small number of the $\Op{2}$ terms is most easily
done using the subtraction techniques of~\cite{mgierg2}.}
Even so, it turns out that this iterative procedure generates far
more diagrams at two loops than we had been expecting (of order
10,000), so of course it took a long time to
complete the calculation. Nearly all of the diagrams cancel
amongst themselves  by the end of the manipulations. The few
diagrams that remain have the property that explicit dependence on
the seed action (part of the regularisation structure) and covariantisation
has disappeared, reflecting the underlying universality of the answer.

Whilst this vast number of diagrams sounds extremely
discouraging, one of us has since realised that the
underlying structure of the calculation allows for some
remarkable simplifications~\cite{Thesis,oliver1,oliver2}. 
Re-examining the $\beta_1$ diagrammatics, it becomes
clear that the same steps are repeated numerous times.
For example, cancellations~\ref{cancel:d1:1.5}--\ref{cancel:d1:1.12}
occur in parallel. Furthermore,
the conversion of diagrams~\ref{d1:1.14} and~\ref{d1:1.15} into
$\Lambda$-derivative terms mirror each other exactly 
(see figs.~\ref{fig:beta1:LevelTwo-A} and~\ref{fig:beta1:LevelTwo-B}). 
Both diagrams
possess two three-point, tree level Wilsonian effective action vertices;
upon their manipulation, the partner diagrams possessing seed action
vertices are exactly cancelled. Indeed, thinking about this more carefully,
if we take two three-point, tree level vertices, two effective propagators and
two external fields, the only $\Lambda$-derivative terms we can construct
are precisely diagrams~\ref{d1:1.15c} and~\ref{d1:1.19}. This
suggests that the generation of these $\Lambda$-derivative terms
can be done in parallel.

Indeed, this expectation is borne out~\cite{oliver1} (see also~\cite{qed})
and already represents a vast simplification of the calculational
procedure. However, even more follows: pushing these methods
to their limit, it appears that one can derive an expression for
$\beta_n$, in terms of $\Lambda$-derivatives, \emph{to all orders
in perturbation theory}~\cite{Thesis,oliver2}. At a stroke, this
removes the primary obstacle to extracting $\beta$ function coefficients
using this ERG.\footnote{Of course, we have not actually evaluated the numerical
value for $\beta_2$ in this paper---see~\cite{mgierg2}. This is
itself a subtle procedure, but there are encouraging indications
that, even here, we can expect similar simplifications to
those involved in the reduction of $\beta_n$ to $\Lambda$-derivative
terms.} We emphasise that the diagrammatic expression~\eq{eq:Beta1-Diag} 
for $\beta_1$
can now be immediately written down without the need to explicitly
perform any of the manipulations of \sec{sec:beta1} \ie $\beta_1$ can 
be directly extracted
from only seven diagrams, three of which vanish in $D=4$.
While this is a great advance on the
procedures we outline here, it also hints that there is a much
simpler, more direct framework for performing computations without
gauge fixing. This is an important direction for the future. Other
important extensions which we believe are possible and are
contemplating for the future, are the incorporation of
quarks---so that the methods are applicable to QCD, developing the
necessary techniques to compute correlators of general gauge
invariant operators, and investigating non-perturbative
approximations.

\paragraph{Acknowledgements}
TRM and OJR acknowledge financial support from PPARC Rolling Grant
PPA/G/O/2002/0468.

\appendix

\section{Examples of Classical Flows}	\label{app:Ex-ClassicalFlows}

The first vertex whose flow we will need is a three-point,
tree level vertex,
decorated by three wildcard fields labelled $R$--$T$. This
is shown in \fig{fig:App:TLThP-WildCardx3}.

\begin{center}
\begin{figure}[h]
	\[
	-\ensuremath{\begin{array}{c}\input{pstex/V0-RST-LdL.pstex_t} \end{array}}= \ensuremath{\begin{array}{c}\input{pstex/V-RSX-DEP-V-XT.pstex_t} \end{array}} + \ensuremath{\begin{array}{c}\input{pstex/V-STX-DEP-V-XR.pstex_t} \end{array}} + \ensuremath{\begin{array}{c}\input{pstex/V-TRX-DEP-V-XS.pstex_t} \end{array}} + \ensuremath{\begin{array}{c}\input{pstex/V-RX-W-S-V-TX.pstex_t} \end{array}} \hspace{1ex}
		+ \ensuremath{\begin{array}{c}\input{pstex/V-SX-W-T-V-RX.pstex_t} \end{array}} \hspace{1ex} + \ensuremath{\begin{array}{c}\input{pstex/V-TX-W-R-V-SX.pstex_t} \end{array}}
	\]
\caption{The flow of a three-point, tree level vertex decorated by three wildcard fields.}
\label{fig:App:TLThP-WildCardx3}
\end{figure}
\end{center}

We now specialise the previous example to give the flow of a three-point, tree
level vertex decorated by $A^1_\mu(p)$, $A^1_\nu(-p)$ and a wildcard field,
which we note carries zero momentum. This is shown in \fig{fig:App-TLThP-A1A1X}
where we have suppressed all labels.

\begin{center}
\begin{figure}[h]
	\[
	-\ensuremath{\begin{array}{c}\input{pstex/V0-A1A1X-LdL.pstex_t} \end{array}}= \ensuremath{\begin{array}{c}\input{pstex/V-A1A1X-DEP-V-XX.pstex_t} \end{array}} + 2 \ensuremath{\begin{array}{c}\input{pstex/V-A1XX-DEP-V-A1X.pstex_t} \end{array}} +2 \ensuremath{\begin{array}{c}\input{pstex/V-A1X-W-A1-XX.pstex_t} \end{array}} + \ensuremath{\begin{array}{c}\input{pstex/V-A1X-W-X-V-A1X.pstex_t} \end{array}} 
	\]
\caption[Flow of a three-point, tree level vertex decorated by $A^1_\mu(p)$, $A^1_\nu(-p)$
and a dummy field.]{Flow of a three-point, tree level vertex decorated by $A^1_\mu(p)$, $A^1_\nu(-p)$
and a dummy field. Lorentz indices, sub-sector labels and momentum arguments are suppressed.}
\label{fig:App-TLThP-A1A1X}
\end{figure}
\end{center}

The third diagram vanishes. First, we note that the kernel must be bosonic. Now,
it cannot be in the $C$-sector, because $AC$
vertices do not exist. If the kernel is in the $A$-sector then,
since the wildcard field carries zero momentum, this
would require a two-point $A$-vertex carrying zero
momentum, which is forbidden by gauge invariance.

The final diagram vanishes at $\Op{2}$. The kernel must be in
the $A$-sector and so each of the vertices
contributes at least $\Op{2}$, as a consequence of gauge invariance.

Note that, if the wildcard field is in the $C$-sector, then the second
diagram also vanishes at $\Op{2}$. The top vertex contributes at least $\Op{2}$.
The bottom vertex must also contribute $\Op{2}$, by gauge invariance, since
$AC$ vertices do not exist.

The last example is of the flow of a four-point, tree level vertex decorated by 
$A^1_\mu(p)$, $A^1_\nu(-p)$ and two dummy fields, as shown in \fig{fig:App:TLFP-A1A1XX}.
Summing over the flavours of the dummy fields 
and noting
that, in this current example, the dummy fields carry equal and opposite momenta,
we can treat these fields as identical.
\begin{center}
\begin{figure}
	\[
	-\ensuremath{\begin{array}{c}\input{pstex/V-A1A1XX.pstex_t} \end{array}}= 
	\begin{array}[t]{cccccccc}
		 	& 2\ensuremath{\begin{array}{c}\input{pstex/V-A1A1XX-DEP-XX.pstex_t} \end{array}}		& + & 4\ensuremath{\begin{array}{c}\input{pstex/V-A1XX-W-A1-V-XX.pstex_t} \end{array}}	& +	& \ensuremath{\begin{array}{c}\input{pstex/V-XX-W-A1A1-V-XX.pstex_t} \end{array}}		& -	& 2\ensuremath{\begin{array}{c}\input{pstex/V0-A1XX-DEP-V0-A1XX.pstex_t} \end{array}}	
	\\
		+	& 4\ensuremath{\begin{array}{c}\input{pstex/V0-A1XX-DEP-V-A1XX.pstex_t} \end{array}}	&+	& 2\ensuremath{\begin{array}{c}\input{pstex/V-A1A1X-W-X-V-XX.pstex_t} \end{array}}	& +	& 4\ensuremath{\begin{array}{c}\input{pstex/V-A1X-W-A1X-V-XX.pstex_t} \end{array}}	& +	& 4\ensuremath{\begin{array}{c}\input{pstex/V-A1X-W-X-V-A1XX.pstex_t} \end{array}}
	\\
		+	& 2\ensuremath{\begin{array}{c}\input{pstex/V-A1X-V-A1A1XX.pstex_t} \end{array}}		& -	& \ensuremath{\begin{array}{c}\input{pstex/V0-A1A1X-DEP-V0-XXX.pstex_t} \end{array}}	& +	& \ensuremath{\begin{array}{c}\input{pstex/V0-A1A1X-DEP-V-XXX.pstex_t} \end{array}}	& +& \ensuremath{\begin{array}{c}\input{pstex/V-A1A1X-DEP-V0-XXX.pstex_t} \end{array}}
	\\
		+	& 4\ensuremath{\begin{array}{c}\input{pstex/V-A1X-W-A1-VXXX.pstex_t} \end{array}}		& + & \ensuremath{\begin{array}{c}\input{pstex/VA1X-W-XX-VA1X.pstex_t} \end{array}}
	\end{array}
	\]
\caption[Flow of a four-point, tree level vertex decorated by $A^1_\mu(p)$, $A^1_\nu(-p)$
and two dummy fields.]{Flow of a four-point, tree level vertex decorated by $A^1_\mu(p)$, $A^1_\nu(-p)$
and two dummy fields. Lorentz indices and momentum arguments are suppressed.}
\label{fig:App:TLFP-A1A1XX}
\end{figure}
\end{center}

The penultimate diagram vanishes at $\Op{2}$. The kernel must be in the $A$-sector, but
then the diagram possesses an $\Op{2}$ stub. However, the kernel, which carries
three Lorentz indices, cannot have an $\Op{0}$ contribution by Lorentz invariance.
The final diagram, which possesses two $\Op{2}$ stubs, clearly vanishes at $\Op{2}$.

\section{Diagrammatic Identities}
\label{app:D-IDs}

There follows a set of diagrammatic identities
necessary for reducing $\beta_2$ to a set of
$\Lambda$-derivative terms. For more
details, see~\cite{Thesis}.

\begin{D-ID}
When attached to an arbitrary diagram,
\[
	\ensuremath{\begin{array}{c}\begin{picture}(0,0)%
\includegraphics{pstex/Struc-AR0-A.pstex}%
\end{picture}%
\setlength{\unitlength}{3947sp}%
\begingroup\makeatletter\ifx\SetFigFont\undefined%
\gdef\SetFigFont#1#2#3#4#5{%
  \reset@font\fontsize{#1}{#2pt}%
  \fontfamily{#3}\fontseries{#4}\fontshape{#5}%
  \selectfont}%
\fi\endgroup%
\begin{picture}(230,285)(197,457)
\end{picture}
 \end{array}} \rightarrow 0.
\]
\end{D-ID}

\begin{D-ID}
\label{D-ID-Alg-2}
Consider a two-point, tree level vertex, attached to an effective propagator,
joined to a nested gauge remainder, which bites the remaining
field on the vertex, in either sense. 
These sub-diagrams 
can be redrawn as shown in \fig{fig:app:D-ID-Alg-2}.
\begin{center}
\begin{figure}[h]
	\[
	\ensuremath{\begin{array}{c}\input{pstex/Struc-AR1-A+B.pstex_t} \end{array}} \equiv \cdeps{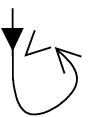}, \hspace{1em} 
	\ensuremath{\begin{array}{c}\input{pstex/Struc-AR1-A+B-2.pstex_t} \end{array}} \equiv \cdeps{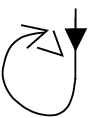}
	\]
\caption{Diagrammatic identity~\ref{D-ID-Alg-2}.}
\label{fig:app:D-ID-Alg-2}
\end{figure}
\end{center}
\end{D-ID}

\begin{D-ID}
\label{D-ID-Alg-3}
Consider the diagrams of \fig{fig:app:D-ID-Alg-3}.
This equality holds, if we change pushes forward
onto nested gauge remainders into pulls back
in all independent ways.
\begin{center}
\begin{figure}[h]
	\[
		\ensuremath{\begin{array}{c}\input{pstex/Struc-AR3-A.pstex_t} \end{array}} - \cdeps{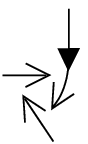} + \cdeps{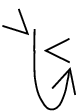} - \cdeps{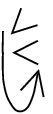} = 0
	\]
\caption{Diagrammatic identity~\ref{D-ID-Alg-3}}
\label{fig:app:D-ID-Alg-3}
\end{figure}
\end{center}
\end{D-ID}

\begin{D-ID}
It is true in all sectors for which the gauge remainder
is not null that
\[
	\ensuremath{\begin{array}{c}\input{pstex/dGRk.pstex_t} \end{array}} = \delta_{\alpha\nu}.
\]
It therefore follows that
\[
	\ensuremath{\begin{array}{c}\begin{picture}(0,0)%
\includegraphics{pstex/dGRk-Arb.pstex}%
\end{picture}%
\setlength{\unitlength}{3947sp}%
\begingroup\makeatletter\ifx\SetFigFont\undefined%
\gdef\SetFigFont#1#2#3#4#5{%
  \reset@font\fontsize{#1}{#2pt}%
  \fontfamily{#3}\fontseries{#4}\fontshape{#5}%
  \selectfont}%
\fi\endgroup%
\begin{picture}(249,608)(-110,336)
\put(-67,367){\makebox(0,0)[lb]{\smash{\SetFigFont{8}{9.6}{\rmdefault}{\mddefault}{\updefault}{\color[rgb]{0,0,0}$k$}%
}}}
\end{picture}
 \end{array}} = \ensuremath{\begin{array}{c}\begin{picture}(0,0)%
\includegraphics{pstex/TE-Ext-Arb.pstex}%
\end{picture}%
\setlength{\unitlength}{3947sp}%
\begingroup\makeatletter\ifx\SetFigFont\undefined%
\gdef\SetFigFont#1#2#3#4#5{%
  \reset@font\fontsize{#1}{#2pt}%
  \fontfamily{#3}\fontseries{#4}\fontshape{#5}%
  \selectfont}%
\fi\endgroup%
\begin{picture}(273,375)(-134,450)
\put(-79,741){\makebox(0,0)[lb]{\smash{\SetFigFont{8}{9.6}{\rmdefault}{\mddefault}{\updefault}{\color[rgb]{0,0,0}$k$}%
}}}
\end{picture}
 \end{array}}.
\]
\end{D-ID}

\end{document}